%% file: mobihoc_new_draft.tex
\documentclass[sigconf]{acmart}
\AtBeginDocument{%
  \providecommand\BibTeX{{%
    \normalfont B\kern-0.5em{\scshape i\kern-0.25em b}\kern-0.8em\TeX}}}

\usepackage{etoolbox}
\newtoggle{vtwo}

\togglefalse{vtwo}

\newcommand{\secondversion}[2]{%
\iftoggle{vtwo}{%
#1
}
{#2}
}

\toggletrue{vtwo}     

\copyrightyear{2021}
\acmYear{2021}
\setcopyright{rightsretained}
\acmConference[MobiHoc '21]{The Twenty-second International Symposium on Theory, Algorithmic Foundations, and Protocol Design for Mobile Networks and Mobile Computing}{July 26--29, 2021}{Shanghai, China}
\acmBooktitle{The Twenty-second International Symposium on Theory, Algorithmic Foundations, and Protocol Design for Mobile Networks and Mobile Computing (MobiHoc '21), July 26--29, 2021, Shanghai, China}
\acmDOI{10.1145/3466772.3467047}
\acmISBN{978-1-4503-8558-9/21/07}

\usepackage{amsmath,amsthm}
\usepackage{natbib}
\usepackage{graphicx}
\usepackage{mathtools}
\usepackage{color}
\usepackage[labelformat=simple]{subcaption}
\usepackage{pbox}
\usepackage{enumitem}
\usepackage{url}
\usepackage{bbm}
\usepackage{todonotes}
\usepackage{algpseudocode}
\usepackage{algorithm}
\usepackage{bookmark}
\usepackage{cleveref}
\usepackage{nicefrac}
\usepackage[normalem]{ulem}
\usepackage{bm}
\usepackage[export]{adjustbox}
\usepackage{float}
\usepackage{placeins}
\usepackage{thmtools}
\usepackage{thm-restate}
\makeatother
\allowdisplaybreaks

\theoremstyle{definition}
\newtheorem*{remark*}{Remark}

\newcommand{\ALOOP}[1]{\ALC@it\algorithmicloop\ #1%
  \begin{ALC@loop}}
\newcommand{\ENDALOOP}{\end{ALC@loop}\ALC@it\algorithmicendloop}

\newcommand{\arrival}{A}
\newcommand{\departure}{D}
\newcommand{\service}{S}
\newcommand{\support}{\mathcal{S}}
\newcommand{\numofserver}{K}
\newcommand{\explore}{\arrival^{(E)}}
\newcommand{\exploit}{\arrival^{(O)}}
\newcommand{\indicatorofevents}{\mathbf{1}}

\newcommand{\probability}{\mathbb{P}}
\newcommand{\expectation}{\mathbb{E}}
\newcommand{\queue}{Q}
\newcommand{\policy}{\mathcal{P}}
\newcommand{\regret}{\Psi}
\newcommand{\busy}{B}
\newcommand{\freecapacity}{r}

\newcommand{\explorerandom}{\chi}

\newcommand{\eventone}{\mathcal{E}_{\hyperref[event1]{3}}(t)}
\newcommand{\eventtwo}{\mathcal{E}_{\hyperref[event2]{4}}(t)}
\newcommand{\eventthree}{\mathcal{E}_{\hyperref[event3]{5}}(t)}
\newcommand{\eventfour}{\mathcal{E}_{\hyperref[event4]{6}}(t)}
\newcommand{\eventfive}{\mathcal{E}_{\hyperref[eqn:event5]{1}}(t)}
\newcommand{\eventsix}{\mathcal{E}_{\hyperref[eqn:event6]{2}}(t)}
\newcommand{\order}{O}

\newcommand{\dispdecision}{\sigma}
\newcommand{\mmiddle}{\;\middle|\;}
\newcommand{\eventoneone}{\mathcal{E}_{31}(t)}
\newcommand{\eventonetwo}{\mathcal{E}_{32}(t)}

\input{GJ_macros}
\begin{document}
\title{Job Dispatching Policies for Queueing Systems\\ with Unknown Service Rates} 

\date{}
\author{Tuhinangshu Choudhury}
\affiliation{%
  \institution{Carnegie Mellon University}
  \city{Pittsburgh}
  \state{PA}
}

\author{Gauri Joshi}
\affiliation{%
  \institution{Carnegie Mellon University}
  \city{Pittsburgh}
  \state{PA}
}

\author{Weina Wang}
\affiliation{%
  \institution{Carnegie Mellon University}
  \city{Pittsburgh}
  \state{PA}
}

\author{Sanjay Shakkottai}
\affiliation{%
  \institution{University of Texas - Austin}
  \city{Austin}
  \state{TX}
}

\begin{abstract}
In multi-server queueing systems where there is no central queue holding all incoming jobs, job dispatching policies are used to assign incoming jobs to the queue at one of the servers. Classic job dispatching policies such as join-the-shortest-queue and shortest expected delay assume that the service rates and queue lengths of the servers are known to the dispatcher. In this work, we tackle the problem of job dispatching without the knowledge of service rates and queue lengths, where the dispatcher can only obtain noisy estimates of the service rates by observing job departures. This problem presents a novel exploration-exploitation trade-off between sending jobs to all the servers to estimate their service rates, and exploiting the currently known fastest servers to minimize the expected queueing delay. We propose a bandit-based exploration policy that learns the service rates from observed job departures. Unlike the standard multi-armed bandit problem where only one out of a finite set of actions is optimal, here the optimal policy requires identifying the optimal fraction of incoming jobs to be sent to each server. We present a regret analysis and simulations to demonstrate the effectiveness of the proposed bandit-based exploration policy.
\end{abstract}

\ccsdesc[500]{Mathematics of computing~Queueing theory}
\ccsdesc[500]{Networks~Network performance analysis}

\maketitle

\input{newversion/intro}

\section{Problem Formulation}
\label{sec:prob_formu}
\input{newversion/prob_formu}

\section{Optimal Weighted Random Routing Policy For Known Service Rates}
\label{sec:optimal_randomized_policy}
\input{newversion/optimal-weighted-routing}

\input{newversion/algorithm}
\input{newversion/regret_analysis}

\input{newversion/simulation}

\input{newversion/conclusion}

\section*{Acknowledgments}
The authors thank Osman Ya\u{g}an for insightful initial discussions. This work was supported in part by the CMU Dean's fellowship, NSF CCF grant \#2007834, NSF CNS grant \#2007733, NSF CMMI Grant \#1826320, NSF CNS Grant \#1910112.
ONR Grant N00014-19-1-2566 and a Carnegie Bosch Institute Research Award.

\bibliographystyle{ACM-Reference-Format}
\bibliography{refs-tuhin, refs-weina}
\onecolumn
\secondversion{\include{newversion/appendix}}{}
\end{document}

%% file: GJ_macros.tex
\newtheorem{lem}{Lemma}
\newtheorem{thm}{Theorem}
\newtheorem{defn}{Definition}

\newtheorem{exple}{Example}

\let\oldexple\exple
\renewcommand{\exple}{\oldexple\normalfont}

\graphicspath{{Figures/}}

\crefname{equation}{}{}
\Crefname{equation}{}{}
\crefname{thm}{theorem}{theorems}
\Crefname{thm}{Theorem}{Theorems}
\crefname{clm}{claim}{claims}
\Crefname{clm}{Claim}{Claims}
\Crefname{coro}{Corollary}{Corollaries}
\Crefname{lem}{Lemma}{Lemmas}
\Crefname{sec}{Section}{Sections}
\crefname{app}{appendix}{appendices}
\Crefname{app}{Appendix}{Appendices}
\crefname{prop}{proposition}{propositions}
\Crefname{prop}{Proposition}{Propositions}
\Crefname{propty}{Property}{Properties}
\crefname{figure}{fig.}{figures}
\Crefname{figure}{Fig.}{Figures}
\crefname{defn}{definition}{definitions}
\Crefname{defn}{Definition}{Definitions}
\crefname{fact}{fact}{facts}
\Crefname{fact}{Fact}{Facts}
\crefname{appendix}{appendix}{appendices}
\Crefname{appendix}{Appendix}{Appendices}
\crefname{algo}{algorithm}{algorithms}
\Crefname{algo}{Algorithm}{Algorithms}
\crefname{algorithm}{algorithm}{algorithms}
\Crefname{algorithm}{Algorithm}{Algorithms}
\crefname{conj}{conjecture}{conjectures}
\Crefname{conj}{Conjecture}{Conjectures}
\crefname{obs}{observation}{observations}
\Crefname{obs}{Observation}{Observations}
\crefname{poli}{policy}{Policy}
\Crefname{poli}{Policy}{Policies}

%% file: newversion/intro.tex
\section{Introduction}
\label{sec:intro}

Traditional queueing models \citep{mor_book, srikant2014communication} such as M/M/1, M/G/k, G/G/k consist of a single central queue holding incoming jobs and one or more servers that are used to serve those jobs. However, in many applications such as supermarket or airport queues, it is more practical for each server to maintain a separate queue consisting of jobs that are assigned to it. This paradigm calls for the design of job dispatching policies such as join-the-shortest queue (JSQ), shortest expected delay (SED) and least-work-left (LWL) that seek to emulate the delay performance of systems with a single central queue by making the most efficient assignments of jobs to server queues. For example, the JSQ dispatching policy polls the queue lengths at the servers and assigns each job to the shortest queue. In large-scale systems such as computing clusters with tens of thousands of servers, an important consideration is that it can be practically infeasible to poll and maintain status information of all the queues. Therefore, alternatives to join-the-shortest-queue such as the power-of-$d$-choices (Po$d$) policy \citep{mitzenmacher1996load, mitzenmacher2001power, vvedenskaya1996queueing} obtain queue length information of only a randomly chosen subset of servers in order to reduce the communication and memory cost.

A common assumption in all the the policies described above is that the service rates at which jobs assigned to each server are served are known to the job dispatcher, or are to be homogeneous across servers, which precludes the need for the dispatcher to know them. In emerging applications such as cloud computing and crowd-sourcing, the servers may not be dedicated to jobs assigned by the dispatcher and may encounter interruptions and service slowdown due to background workload. Therefore, the service rate experienced by the assigned jobs can be unknown, highly variable across servers, and also changing over time. Traditional service-rate-agnostic policies are not effective in such systems and service-rate-aware policies such as join-the-fastest-shortest-queue (JFSQ) cannot be used due to the service rates being unknown.

\subsection{Main Contributions and Organization} In this paper, we propose a job dispatching policy that learns the unknown service rates of the servers, while simultaneously seeking to minimize the queueing delay experienced by jobs. This problem is at the 
intersection of queueing systems and online learning. It sheds light on a novel exploration-exploitation trade-off where the job dispatching policy needs to strike a balance between assigning jobs to all servers in order to better estimate their service rates (exploration) and preferentially sending jobs to the faster servers to minimize the queueing delay experienced by jobs (exploitation).

Unlike classic multi-armed bandits (MABs) where only one of the actions is optimal, in the queueing setting considered in this paper, an optimal policy would typically use several fast servers.  Therefore, it is necessary to perform exploration in order to identify the subset of servers that should continue receiving jobs asymptotically.  However, more importantly, only identifying this optimal subset of servers is \emph{not enough} for learning an optimal job dispatching policy.  We need to also accurately estimate the service rates of servers in this subset.
Interestingly, we are able to achieve this by virtue of some special properties of queueing systems. 
In particular, after we identify the optimal subset of servers, we exploit by dispatching jobs only to this subset of servers.  But meanwhile, since we keep obtaining service time samples from the jobs dispatched to these servers, we also continue to improve the learning accuracy of the service rates.  Therefore, exploitation and improvement in estimation are taking place simultaneously in this queueing system.

The rest of the paper is organized as follows. In \Cref{sec:prob_formu} we describe the system model and formulate the problem concretely. In \Cref{sec:optimal_randomized_policy}, we find the optimal weighted random routing policy, which serves as the performance baseline. In \Cref{sec:algorithm} we propose a bandit-based $\epsilon_t$-exploration algorithm. The distinction between multi-armed bandits and our queueing setting leads to very different regret analysis, presented in \Cref{sec:regret_analysis}.
In \Cref{sec:simulation} we demonstrate the effectiveness of the proposed policy via simulations. \secondversion{ }{Due to lack of space, we mainly present proof sketches here, and provide the full proofs in the technical report  \citep{choudhury2020job} available online.} 

\subsection{Related Work}
Bandits have had a rich history, both from an optimal control perspective (see \cite{mate08} for a survey) and a finite-time regret perspective (see \cite{lattimoreszepesvari2020banditbook} for a survey). Our paper focuses on bandits in queueing settings -- while this intersection has had a rich history (starting from the Klimov's model \cite{kli74} focusing on optimal control), our focus is on a finite-time regret formulation. At a high-level, the regret perspective formulates queueing problems with unknown statistics (e.g., of the service or arrival processes), with the goal of characterizing the loss/regret in performance of a resource allocation (with learning) algorithm with respect to a genie-policy that has access to the complete statistics. Such regret formulations have recently been introduced both in the adversarial setting \cite{walton13} and the stochastic setting \cite{krishnasamy2016learning}. \citet{walton13} has shown that in an adversarial setting, the queue regret (difference between the queue-length induced by a learning algorithm with respect to a static optimal policy) increases at most sublinearly in time. On the other hand, in a stochastic setting, \citet{krishnasamy2016learning} have shown that the expected regret in fact decreases with time (roughly as $\tilde{O}(1/t)$). 

Starting from the above studies, there has been increasing interest in the regret of algorithms in various queueing settings  \citep{krishnasamy2018learningcu, cayci2017learning,stahlbuhk2018learningalgorithm,liu2020pond,fbnmh20}. \citet{krishnasamy2018learningcu} studied the problem of scheduling jobs using the $c\mu$ rule with unknown service rates and showed that the cumulative queue regret (i.e., sums of queue regret over time) is $\order(1)$. \citet{liu2020pond} studied the problem of distributing different job classes across servers with unknown rewards for each job class-server pair and proves a reward regret of $\order(\sqrt{t})$. Bandits problems of similar flavour are also studied in the communication system settings, where \citet{cayci2017learning} studied channel allocation in wireless downlink systems with unknown channel statistics,
with an objective to identify the optimal number of channels to activate and proposed an UCB-based index policy that achieves $\order(\ln t)$ regret. The task of selecting the optimal channel in a wireless system with a single transmitter/receiver and multiple channels was studied by \citet{stahlbuhk2018learningalgorithm}, and they derived queue length based policies that achieve $\order(1)$ cumulative queue regret by exploiting samples acquired during the idle time of queues. Finally, \citet{fbnmh20} studied regret from an age of information viewpoint. 

Unlike these studies, our setting is one where the queues are not centrally located at the dispatcher. Instead, jobs are dispatched to individual queues based on partial information; this setting requires both learning a discrete support set and  dispatch weights.

%% file: newversion/prob_formu.tex

\begin{figure}[t]
    \centering
    \includegraphics[width=0.4\textwidth]{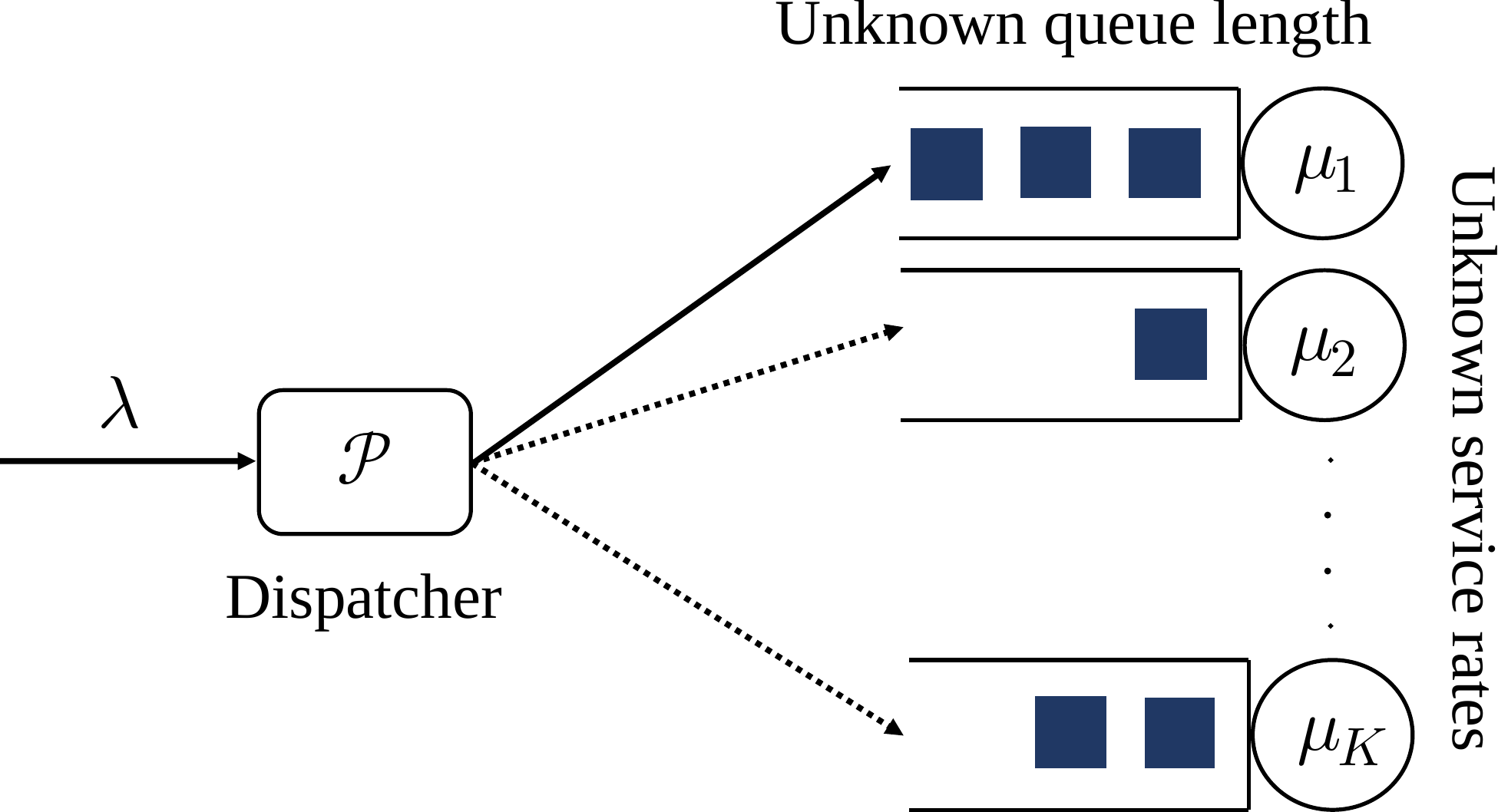}
    \caption{Our system model with job arrival rate $\lambda$ and $\numofserver$ servers with service rates $\mu_1, \dots, \mu_{\numofserver}$ respectively. The dispatching policy $\policy$ assigns each job to one of $K$ queues. The service rates and queue lengths are unknown to the dispatcher; it only observes the service times of departed jobs.
    }
    \label{fig:dia_policy}
\end{figure}

\subsection{System Model, Arrivals and Departures}\label{subsec:sysmodel_arr_dep}

We consider a multi-server discrete-time\footnote{Although we use the discrete-time assumption for the regret analysis presented in this paper, the proposed policy can be used in continuous time systems. We conjecture that the regret analysis is extendable to continuous time systems as well, but this extension is beyond the scope of this paper and is left for future work.} queueing system consisting of $\numofserver$ servers, with one queue at each server storing the unfinished jobs that are dispatched to it, as illustrated in \Cref{fig:dia_policy}. Jobs arrive into the system according to a Bernoulli process with arrival rate $\lambda$, where $0<\lambda<1$. Specifically, let $\arrival(t)$ denote the number of job arrivals at the beginning of time slot $t$.  Then $\arrival(t)=1$ with probability $\lambda$ and $\arrival(t)=0$ with probability $1-\lambda$. 
Incoming jobs are dispatched to one of the  the $\numofserver$ servers according to a scheduling policy $\policy$. Once dispatched, the job joins a first-come-first-served queue with an infinite buffer size at that server. The $\numofserver$ servers have geometrically distributed service times with parameter $\mu_i$. That is, after a job reaches the head of the queue at server $i$, it departs at the end of the next time slot with probability $\mu_i$. With the arrival rate $\lambda$ and the service of the $i$-th server being $\mu_i$, the system is stable only if $\lambda < \sum_{i=1}^{\numofserver} \mu_i$, a condition that we assume to be true. 

Let $\arrival_i(t)$ and $\departure_i(t)$ denote the number of arrivals to queue~$i$ and the number of departures from queue $i$ respectively during time slot $t$. Let $\queue_i(t)$ represent the length of queue~$i$ at the beginning of slot $t$, including the job that is currently in service. We assume that the system starts with empty queues, i.e.,  $\queue_i(0) = 0$, for all $i$. Then the queue evolution process is given by
\begin{equation}
    \queue_i(t + 1) = \queue_i(t) + \arrival_i(t) - \departure_i(t).
\end{equation}
We use $X_{i,n}$ to denote the service time of the $n$-th job that departs from server $i$. It is the time since the job reaches the head of its queue and starts service until it departs from the system. $X_{i,n}$ is geometrically distributed with success probability $\mu_i$, that is, $\Pr(X_{i,n} = x) = (1-\mu_i)^{x-1} \mu_i$ for $x \in \{1, 2, \dots \}$.


\subsection{Information Available to the Dispatcher}

\vspace{0.1cm} \textbf{Service Rates and Queue Lengths are Unknown.} Job dispatching policies for the multi-server setting described above have been extensively studied in previous literature \citep{mor_book, srikant2014communication}. However, most prior works assume that the dispatcher knows the service rates $\mu_1, \dots, \mu_K$, and it also has either full or partial information about the queue lengths $Q_i(t)$. For example, for homogeneous systems where $\mu_1 =  \dots = \mu_K = \mu$, the join-the-shortest-queue (JSQ) policy has full queue information and sends each incoming job to the server $i^* \in \arg \min Q_i(t)$, i.e., the shortest queue, with ties broken at random. Power-of-$d$-choice (Po$d$) policies \citep{mitzenmacher1996load, mitzenmacher2001power, vvedenskaya1996queueing} reduce the cost of querying queue lengths by sampling $d$ queues uniformly at random and dispatching the incoming job to the shortest queue. For heterogeneous service rates, JSQ can be generalized to the join-the-shortest-fastest-queue (JSFQ) \citep{weber1978optimal,eryilmaz2012asymptotically,weng2020optimal}, which breaks ties in favor for the queue with the fastest server. Other policies for systems with heterogeneous servers such as shortest expected delay (SED) \citep{banawan1992comparative,foschini1977heavy} also use some form of queue length information to make job dispatch decisions. In contrast, in this work, we consider that the service rates $\mu_1$, $\mu_2$, \dots,  $\mu_{\numofserver}$ of the $\numofserver$ servers are \emph{heterogeneous and unknown} to the dispatcher. Similarly, the queue lengths $Q_i(t)$ for $i = 1, \dots, \numofserver$ are also \emph{unknown} to the dispatcher.

\vspace{0.1cm} \noindent \textbf{Dispatcher Observes Service Times of Departed Jobs.} In lieu of service rates and queue lengths, we consider that the dispatcher observes service times $X_{i1}$, \dots $X_{iN_i(t)}$ of the $N_i(t)$ jobs that depart from server $i$ by time $t$. In practice, this information can be made available to the dispatcher by having the server send  the dispatcher a notification when a job reaches the head of its queue and begins service and another notification when it departs. The dispatching policy $\policy$ can use this information to estimate the service rates $\mu_1$, \dots $\mu_{\numofserver}$. For example, it can estimate the service rate vector $\hat{\bm{\mu}}(t) = \left(\hat{\mu}_1(t), \hat{\mu}_2(t), \cdots, \hat{\mu}_\numofserver(t)\right)$ at time $t$, where $\hat{\mu}_i(t)$ is given by 
\begin{equation}
    \hat{\mu}_i(t) = \frac{N_i(t)}{\sum_{j=1}^{N_i(t)} X_{i,j}} \label{eqn:mu_estimate},
\end{equation}
and use it to dispatch jobs. For instance, it can dispatch a larger fraction of jobs to server with a higher service rate estimate.

\subsection{Weighted Random Routing Policies}
\label{sec:weighted_random_routing}

The service rate estimate $\hat{\mu}_i(t)$ can be used by the dispatcher in a variety of ways to make job dispatch decisions. Among all the possible scheduling policies, we focus on the class of \textit{weighted random routing policies}, which are defined as follows.

\begin{defn}[Weighted Random Routing $\policy$]
\label{defn:weighted_random_routing}
In time slot $t$, the dispatcher associates a probability $\hat{p}_i(t)$ with server $i$, where $\hat{p}_i(t)$'s satisfy the property $\sum_{j=1}^{\numofserver} \hat{p}_j(t)=1$. We call the probability vector $\hat{\bm{p}}(t) = \left(\hat{p}_1(t), \hat{p}_2(t), \cdots, \hat{p}_\numofserver (t)\right)$ the \emph{routing vector}. A job that arrives at time $t$ is dispatched to server $i$ with probability $\hat{p}_i(t)$, independent of other jobs. The routing vector $\hat{\bm{p}}(t) = f(\lambda, \hat{\bm{\mu}}_i(t))$, where $f:[0,1]^{\numofserver+1} \rightarrow [0,1]^{\numofserver}$ is a fixed, deterministic function.
\end{defn}

The uniform random routing policy corresponds to setting $\hat{\bm{p}}(t) = \left(1/\numofserver, \cdots, 1/\numofserver \right)$. Since the routing vector $\hat{\bm{p}}(t)$ is a fixed, predetermined function of $\hat{\mu}_i(t)$, policies such as round-robin dispatching that retain a memory of where past jobs were dispatched are not included in this class of weighted random routing policies. 

\vspace{0.1cm} \noindent \textbf{Optimal Weighted Random Routing.} Consider a genie system where the dispatcher knows the service rates $\mu_1, \dots, \mu_{\numofserver}$. Then the optimal weighted routing policy $\policy^*$ is defined as the policy that chooses the optimal $\bm{p}^* = f(\lambda, \bm{\mu})$, that minimizes the expected steady-state queue length $\mathbb{E}[\sum_{i=1}^{\numofserver} Q_i(\infty)]$, which is equivalent to minimizing the mean response time experienced by incoming jobs. 
We will derive $\bm{p}^*$ in \Cref{sec:optimal_randomized_policy}.

\subsection{Measuring Performance in terms of Regret}
We seek to design a weighted random routing policy $\policy$ that starts with no knowledge of the service rates and converges to the optimal random routing policy $\policy^*$. To evaluate the transient performance of $\policy$ in terms of how quickly it learns $\policy^*$, we define a performance metric $\regret_{\policy \policy^*}(t)$, referred to as the regret of $\policy$. In \Cref{sec:regret_analysis}, we analyze the performance of our proposed dispatching policy in terms of the expected regret $\mathbb{E}[\regret_{\policy \policy^*}(t)]$.

\begin{defn}[Regret of a Dispatching Policy]
\label{defn:regret}
The regret $\regret_{\policy \policy^*}$ of a dispatching policy $\policy$ with respect to the optimal baseline $\policy^*$ is
\begin{equation}
    \regret_{\policy \policy^*}(t) \triangleq \sum_{\tau = 1}^t \sum_{i = 1}^{\numofserver}\left( \queue_i(\tau) - \queue_i^{*}(\tau)\right),
\end{equation}
where $\queue_i^*(t)$ represents the queue length at server $i$ at time $t$ when following policy $\policy^*$. 
\end{defn}

The $\regret_{\policy \policy^*}(t)$ represents the cost of using policy $\policy$ instead of $\policy^*$ in terms of cumulative queue length till time $t$. Note that the cumulative queue length is the total time spent by all jobs that arrived before time $t$, including the jobs that have departed. Hence, regret represents the additional time jobs stayed in the system when using policy $\policy$ instead of $\policy^*$. It is the penalty the policy $\policy$ has to pay for the lack of knowledge the service rates system.

\vspace{0.1cm} \noindent \textbf{Difference from the Regret used in Multi-armed Bandits.} Although similar, the regret considered in this paper and its analysis is fundamentally different from the multi-armed bandit setting \citep{bubeck2012regret,lattimoreszepesvari2020banditbook,lai1985asymptotically}. In the multi-armed bandit setting with $K$ arms, asymptotically optimal algorithms pull the best arm (with the highest mean reward) $O(t)$ times and pull all the $K-1$ sub-optimal arms $O(\log t)$ times. In our queueing setting, the optimal random routing policy generally sends $O(t)$ jobs to all servers with $p_i^*> 0$ and not just the fastest server with the highest service rate $\mu_i$. We seek fast convergence of the routing vector $\hat{\bm{p}}(t)$ to the optimal $\bm{p}^*$ so as to dispatch the optimal fraction $p_i^*$ of incoming jobs to each server $i$.





\subsection{Justification for Focusing on Weighted Random Routing Policies}

In this section, we justify why we choose to focus on the class of weighted random routing policies. First, we explain why service rates and queue lengths are unknown in the large-scale systems envisioned in this work. Furthermore, we show that even if partial or delayed queue length information is available, using it and performing join-the-shortest-queue (JSQ) or join-the-fast-shortest-queue (JFSQ) dispatching does not give a large performance improvement over weighted random routing.

\vspace{0.1cm} \noindent \textbf{Why the Service Rates are Unknown.} Traditionally, multi-server queueing systems consisted of dedicated servers and a single source of incoming jobs. However, in modern applications such as cloud data centers, a server may be receiving jobs from several different applications. For instance, it may be running background workload such as check-pointing and garbage collection, or higher priority jobs coming from other sources \citep{dean2013tail}. As a result, the effective service rate $\mu_i$ of server $i$ as seen by the dispatcher of any one application depends on the external workload. Due to privacy constraints and communication delays, it is infeasible for each dispatcher to know and keep track of the external workload at each server. Therefore, we consider that service rates $\mu_i$ are unknown to the dispatcher.
\footnote{In practice, the effective service rates may vary over time depending on the external workload. For tractability of the analysis, we do not consider time-varying service rates $\mu_i$. However, the estimates in \eqref{eqn:mu_estimate} can be modified to discount older service time observations in order to account for time-varying service rates.}


\vspace{0.1cm} \noindent \textbf{Why Queue Lengths are Unknown.} In large-scale systems with multiple job sources, each server's queue receives jobs from many dispatchers. In this setting, it is difficult to obtain queue length information due to two reasons: 1) privacy concerns -- if a server reveals its total queue length $Q_i(t)$ to one of the dispatcher, it may compromise the privacy of other dispatchers by revealing information about how many jobs they sent to that queue and 2) even if privacy is not a concern, due to large communication delays incurred when a dispatcher queries the queue length of a server, the queue length information may become stale by the time the job is dispatched. Therefore, we consider that the queue lengths $Q_i(t)$ are unknown to the dispatcher.

\begin{figure}[t]
    \centering
    \begin{subfigure}[b]{1\linewidth}
        \begin{subfigure}[b]{0.49\linewidth}
            \centering
            \renewcommand\thesubfigure{a1}
            \includegraphics[width=48mm]{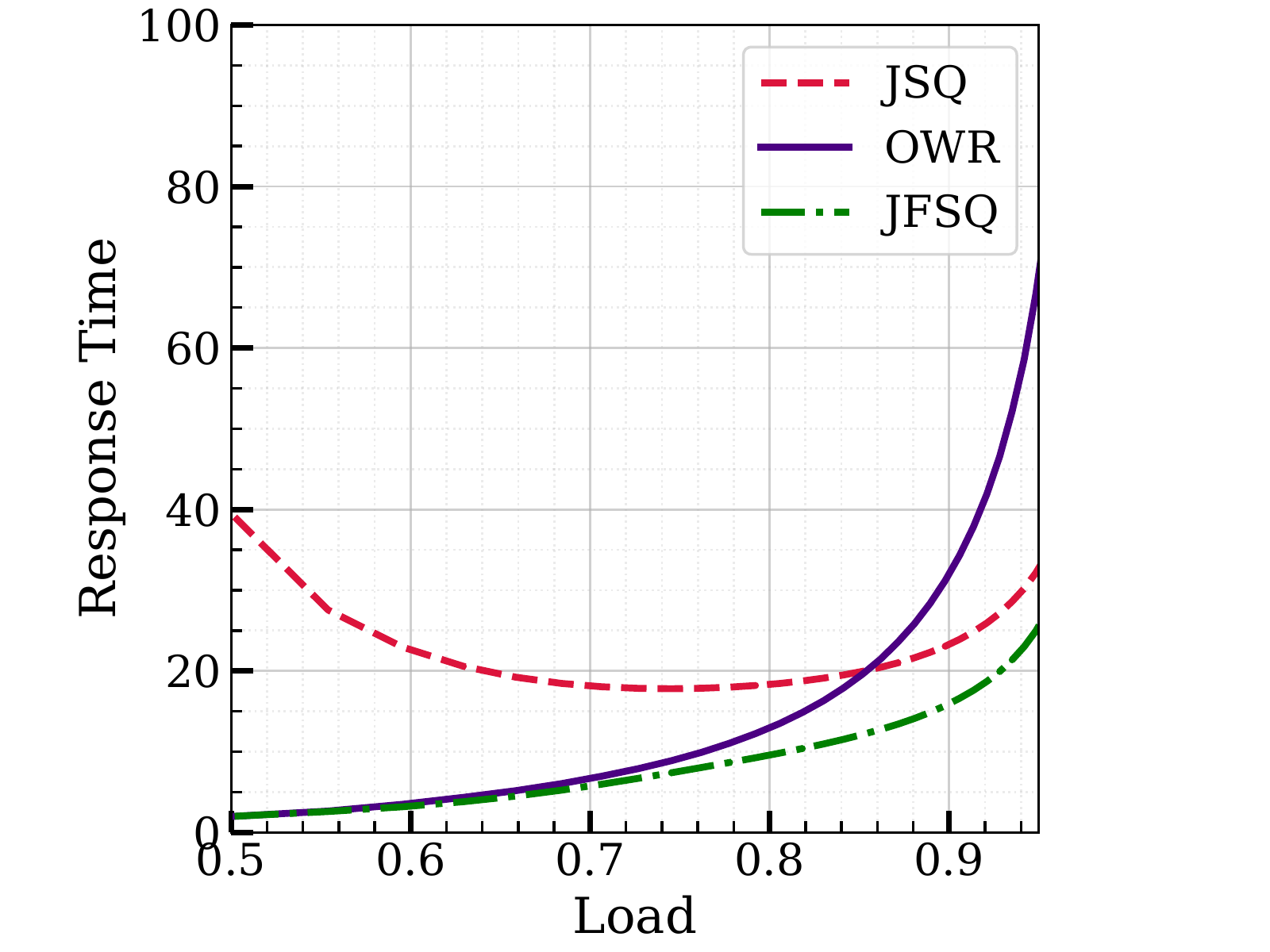}
            \caption{$\lambda_i^{ext} = \frac{\mu_i}{2}$}
            \label{fig:compare_privacy}
        \end{subfigure}
        \hfill
        \centering
        \begin{subfigure}[b]{0.49\linewidth}  
            \centering
            \addtocounter{subfigure}{-1}
            \renewcommand\thesubfigure{a2}
            \includegraphics[width=48mm]{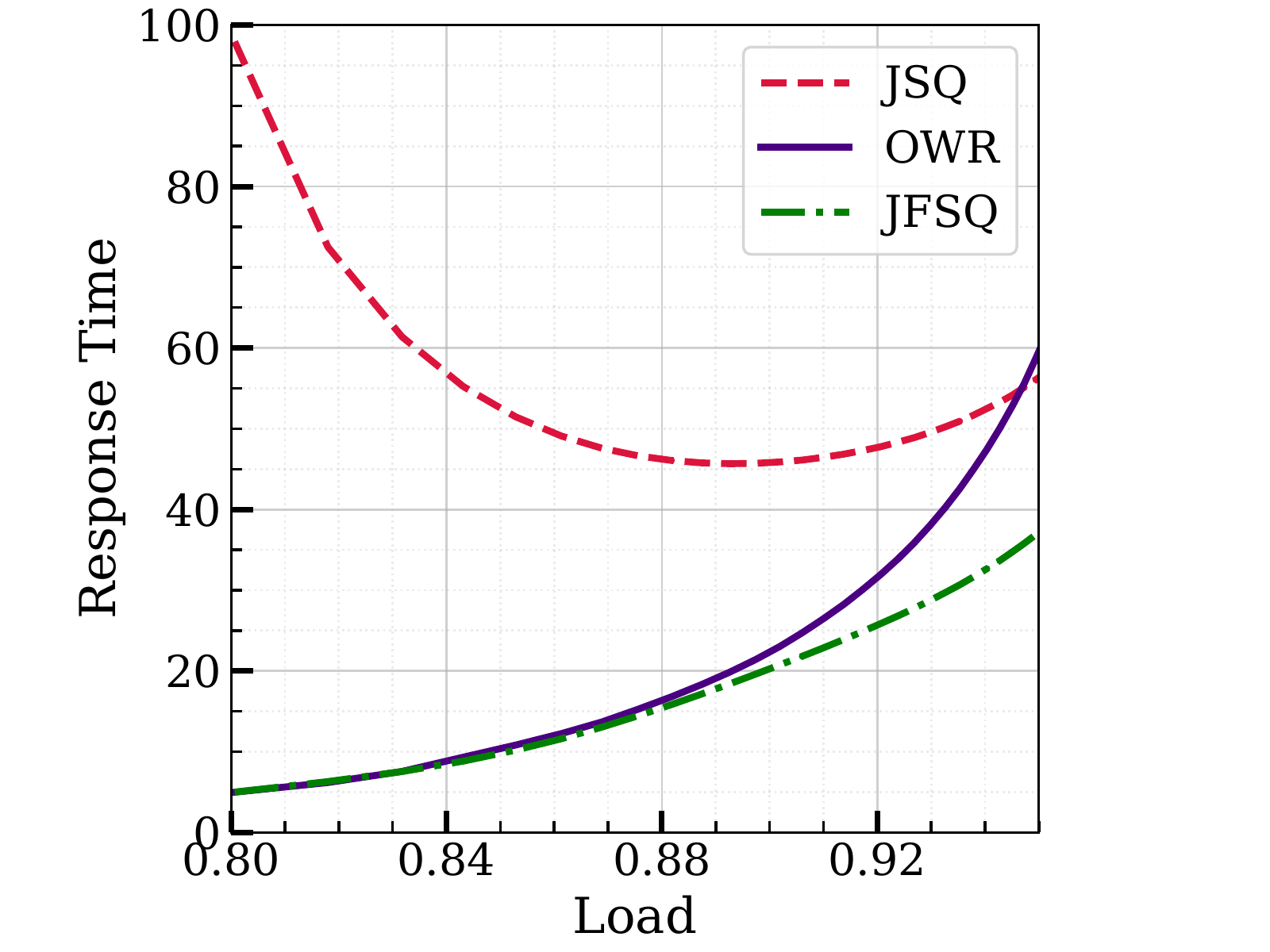}
            \caption{$\lambda_i^{ext} = \frac{4\mu_i}{5}$}
            \label{fig:compare_privacy_heavy_arrival}
        \end{subfigure}
        \addtocounter{subfigure}{-1}    
        \caption{Partial Queue Length Information (excluding external workload)}
        \label{fig:compare_privacy_combined}
    \end{subfigure}
    \hfill
    \centering
    \begin{subfigure}[b]{1\linewidth}
        \begin{subfigure}[b]{0.49\linewidth}
            \centering
            \renewcommand\thesubfigure{b1}
            \includegraphics[width=48mm]{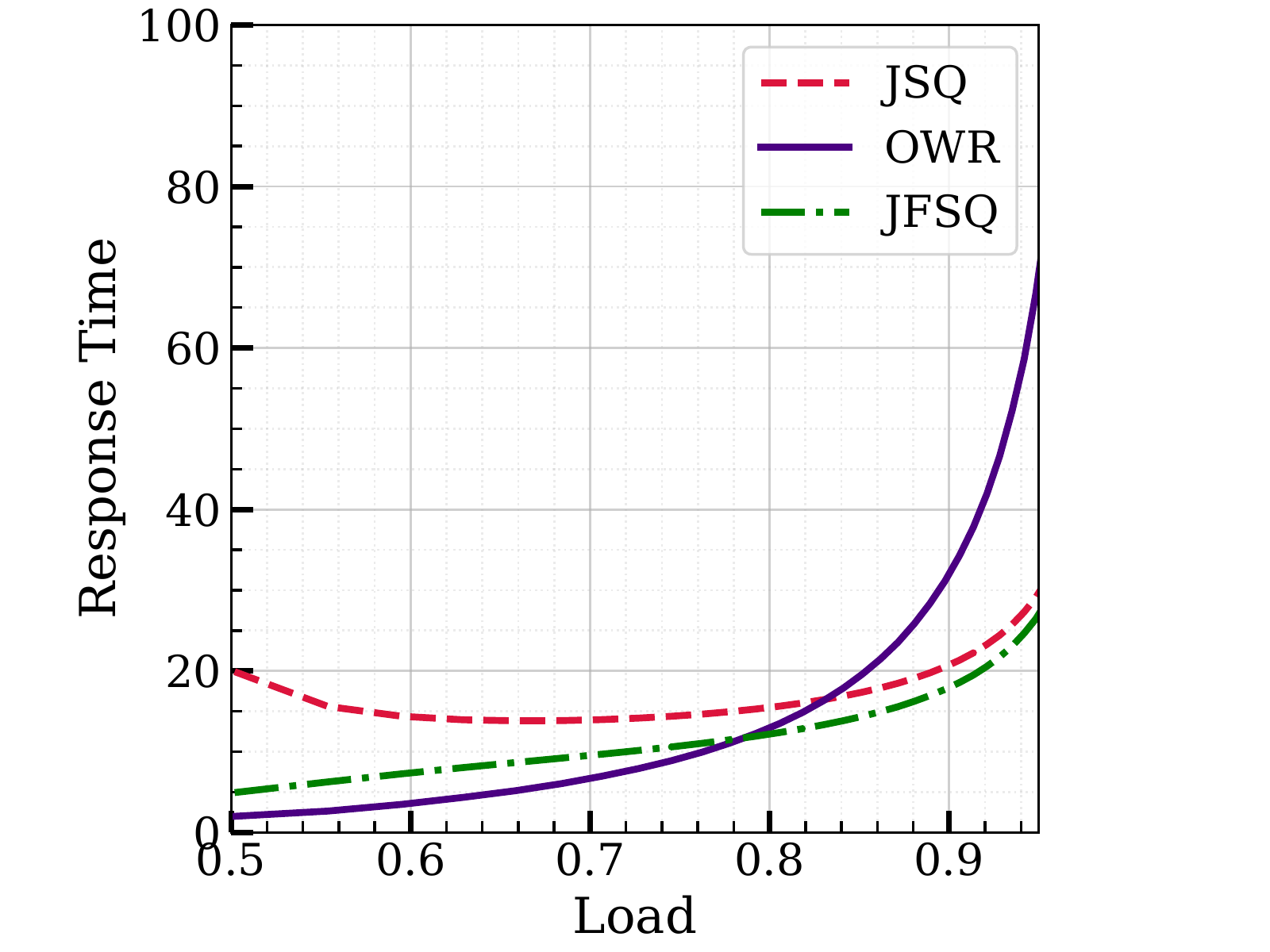}
            \caption{$\lambda_i^{ext} = \frac{\mu_i}{2}$}
            \label{fig:compare_communication}
        \end{subfigure}
        \hfill
        \centering
        \begin{subfigure}[b]{0.49\linewidth}  
            \centering
            \addtocounter{subfigure}{-1}
            \renewcommand\thesubfigure{b2}
            \includegraphics[width=48mm]{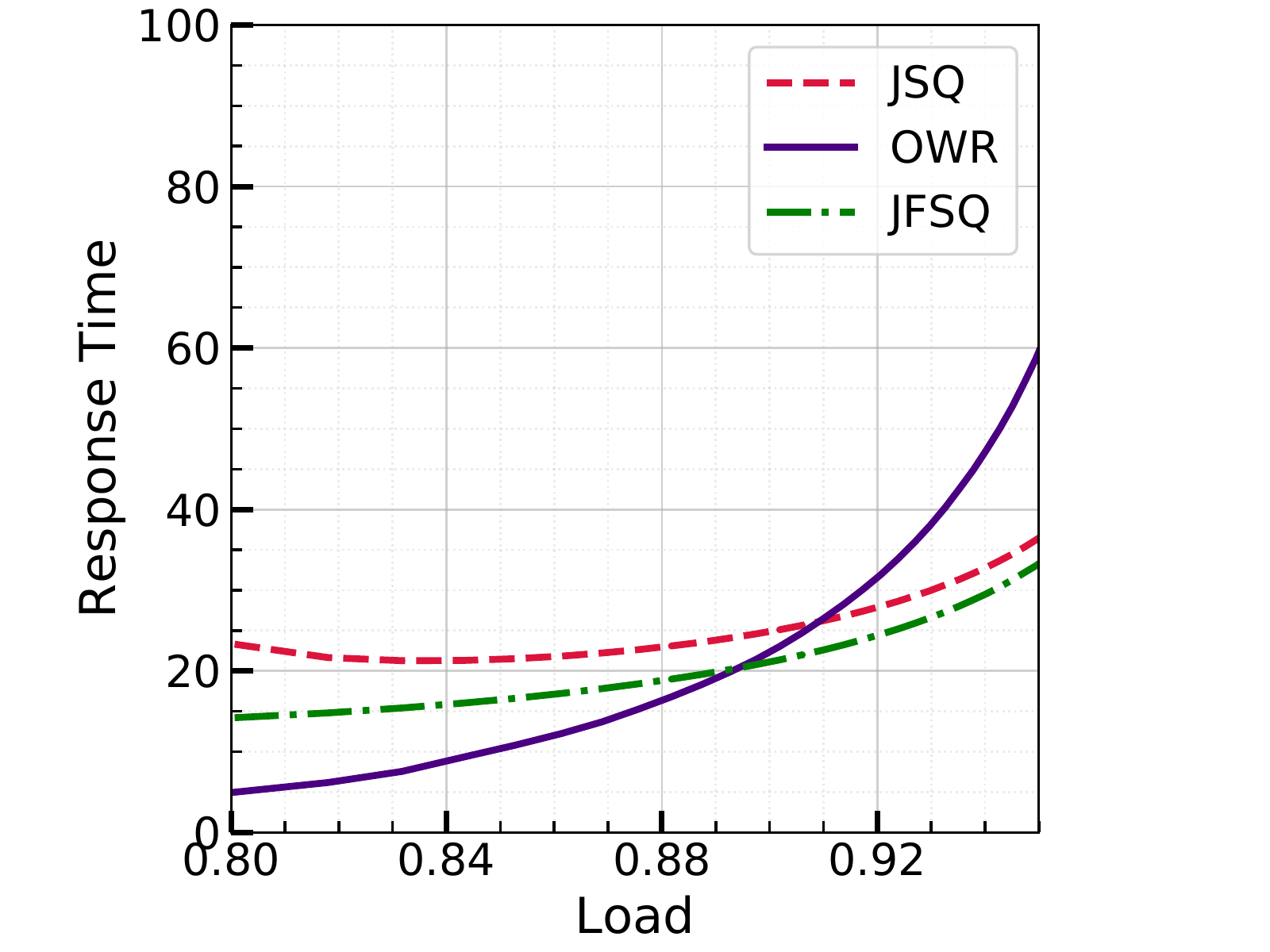}
            \caption{$\lambda_i^{ext} = \frac{4\mu_i}{5}$}
            \label{fig:compare_communication_heavy_arrival}
        \end{subfigure}
        \addtocounter{subfigure}{-1}    
        \caption{Delayed Queue Length Information (including external workload)}
            \label{fig:compare_communication_combined}
    \end{subfigure}
    \caption{
    OWR performs better than JSQ and similar to JFSQ in the middle and low traffic regime. In the heavy traffic regime, both JSQ and JFSQ performs better than OWR. OWR performs better as the external workload increases.
    \vspace{-0.5cm}
    }
    \label{fig:compare}
\end{figure}

\vspace{0.1cm}
\noindent \textbf{Limited Utility of Partial or Delayed Queue Length Information.}
In \Cref{fig:compare}, we consider a system of $6$ servers with service rates $\mu_i$ such that $\mu_i = 2^{i-1} \mu_1$, and $\sum_{i=1}^{6} \mu_i = 0.99$. We show a comparison of the mean response times (waiting time in queue plus service time) for optimal weighted random (OWR) routing, which knows the service rates $\mu_i$ but does not use queue length information, with JSQ and JFSQ, which use queue length information to make job dispatching decisions. Our goal is to demonstrate that when the queue length information is partial or delayed, a queue-length-agnostic policy such as OWR performs nearly as well as JSQ and JFSQ. 

To model partial and delayed queue length information, we consider that apart from the rate $\lambda$ job arrivals at the dispatcher, server $i$ has external Poisson arrivals at rate $\lambda_i^{ext}$ that are not visible to the dispatcher. For \Cref{fig:compare_privacy} and \Cref{fig:compare_communication}, we choose $\lambda_i^{ext} = \mu_i/2$, while for \Cref{fig:compare_privacy_heavy_arrival} and \Cref{fig:compare_communication_heavy_arrival}, we choose $\lambda_i^{ext} = 4\mu_i/5$. In \Cref{fig:compare_privacy_combined}, we consider that due to privacy concerns, the dispatcher only has queue length information about the jobs that it sent to each queue, but not about the external arrivals. In \Cref{fig:compare_communication_combined}, we consider the case of delayed queue lengths, where the dispatcher receives updated queue length information (including both its jobs and the external arrivals) with probability $1/3$. For OWR, we assumed that the policy does not know $\mu_i$ or $\lambda_i^{ext}$ but knows the difference $(\mu_i - \lambda_i^{ext})$ and uses the routing vector $f(\lambda, \bm{\mu} - \bm{\lambda}^{ext})$ to dispatch jobs. All of the simulations are averaged over $40$ trials of $10^7$ job departures each. 
For both these cases, observe in \Cref{fig:compare} that in the low to moderate load regimes (load $= (\lambda + \sum_{i=1}^{K} \lambda_i^{ext})/\sum_{i=1}^{K} \mu_i)$), the optimal weighted random routing (OWR) policy is comparable to JFSQ and better than JSQ in terms of the mean response time. 
Moreover, observe that as the external arrival rate $\lambda_i^{ext}$ increased from $\mu_i/2$ to $4\mu_i/5$, the load beyond which OWR performs worse than JFSQ shifted from $0.8$ to $0.9$ roughly. This is because at heavy load, the cost of a sub-optimal allocation of jobs using a partial or delayed information is more than using no information at all. We conjecture that OWR becomes more useful as the cross traffic load increases.


As a result of these observations, we choose to focus on the class of weighted random routing policies that do not take into account queue lengths when making job dispatching decisions, and seek to design a dispatching policy that can learn unknown service rates while simultaneously minimizing the regret (see \Cref{defn:regret}). Another reason is that for this class of policies, the optimal policy $\policy^*$ that minimizes the steady-state cumulative queue length is clearly defined, as we show in \Cref{sec:optimal_randomized_policy} below. In contrast, the optimal policy is not known in the case where queue lengths are considered for job assignment decisions. Although policies such as JSQ and JFSQ perform well in practice and in the heavy-traffic regime, it is unclear which policy is optimal in other regimes.


%% file: newversion/optimal-weighted-routing.tex


We define \textit{optimal weighted random routing (OWR) policy} as the weighted random routing policy that minimizes the mean response time of jobs in steady state or equivalently, the policy that minimizes the cumulative steady-state queue length $\mathbb{E}[\sum_{i=1}^{\numofserver} Q_i(\infty)]$. Let $\bm{p}^* = (p_1^*, p_2^*, \cdots, p_\numofserver^* )$ be the routing vector corresponding to the optimal weighted routing policy as a function of arrival rate and the service rates and we will refer to it as the \textit{optimal routing vector}.

For a $Geo/Geo/1$ queueing system with arrival rate $\lambda'$ and departure rate $\mu'$, the expected steady-state queue length, $\expectation\left[\queue(\infty)\right]$, is given by (see \citep{srikant2014communication}~Chapter~3 for the derivation)
\begin{equation}
    \expectation\left[\queue (\infty)\right] = \frac{\lambda'(1-\mu')}{\mu' - \lambda'}.
\end{equation}
Now, for our system with arrival rate $\lambda$ and service rates $\mu_i$'s, if every incoming job is assigned to server $i$ with probability $p_i$, the system can be viewed as $\numofserver$ $Geo/Geo/1$ queues each with arrival rate $p_i\lambda$ and service rate $\mu_i$ respectively. Hence, the steady state queue length of the system is given by
\begin{equation}
    \expectation\left[\sum_{i=1}^{\numofserver}\queue_i(\infty)\right] = \sum_{i=1}^{\numofserver}\expectation\left[\queue_i(\infty)\right] = \sum_{i=1}^{\numofserver} \frac{p_i\lambda(1-\mu_i)}{\mu_i-\lambda p_i}.
\end{equation}
Hence, the optimal routing vector would be the solution to the following optimization problem:
\begin{align}
\min_{p_1, p_2, \cdots, p_\numofserver} \quad &  \sum_{i = 1}^\numofserver \frac{p_i\lambda(1-\mu_i)}{\mu_i-\lambda p_i}\\
\textrm{s.t.} \quad & \sum_{i=1}^{\numofserver} p_i = 1\\
\quad  & p_i \geq 0 , 
  \quad  p_i \lambda < \mu_i,  \> \forall i,
\end{align}
where the constraint $p_i \lambda < \mu_i $ ensures stability of the queue $i$ and the remaining constraints ensure that $\mathbf{p}$ forms a valid routing vector. We show in \secondversion{\Cref{prf_optimal_p}}{\cite{choudhury2020job}} that the optimal routing vector is given by $\bm{p}^* = f(\lambda, \bm{\mu})$, where $f:[0,1]^{\numofserver+1} \rightarrow [0,1]^{\numofserver}$ is a function given by
\begin{multline}\label{optimal_p}
f(\lambda, \bm{\mu})(i) \\
=  \begin{cases} 
      \frac{\mu_i}{\lambda} - \frac{\sqrt{\mu_i(1-\mu_i)}}{\sum_{j\in \support\left(\lambda,\boldsymbol{\mu}\right)} \sqrt{\mu_j(1-\mu_j)}}\left( \frac{\sum_{j\in \support\left(\lambda,\boldsymbol{\mu}\right)} \mu_j}{\lambda} - 1\right), & \text{if }i\in \support(\lambda, \bm{\mu}), \\
      0, & \text{if }i\notin \support(\lambda, \bm{\mu}). 
   \end{cases}
\end{multline}
where $\support(\lambda, \bm{\mu})$ is a subset of servers which we refer to as the \textit{optimal support set} such that $i\in \support\left(\lambda, \bm{\mu}\right)$ if and only if $p_i^*>0$.


In \secondversion{\Cref{prf_optimal_p}}{\cite{choudhury2020job}}, we formally prove that if $\mu_i \geq \mu_j$, then $p_i^* \geq p_j^*$. Therefore, only the slowest servers are excluded from the support set $\support(\lambda, \bm{\mu})$, and the support set has to be $ [1, \dots, i ]$ for some $i \in \{ 1, 2, \dots \numofserver\}$. To find the optimal set $\support(\lambda, \bm{\mu})$ and the optimal routing vector $\bm{p}^*$ we can use the iterative algorithm given below.
\begin{enumerate}[label=(\roman*)]\label{algo:find_support}
    \item Initialize the support set $\support(\lambda, \bm{\mu}) = \{1, 2, \dots, \numofserver\}$.
    \item \label{alg:support:2} Calculate $p_i^*$ according to \eqref{optimal_p}.
    \item \label{alg:support:3} If $p_i^* < 0$ for any $i \in \{1, \dots, \numofserver\}$ or $p_i^* = 0$ for any $i\in \support(\lambda, \bm{\mu})$, then remove the slowest server $\arg \min_{i \in \support\left(\lambda,\boldsymbol{\mu}\right)} \mu_i$ from the support set and repeat from Steps \ref{alg:support:2} and \ref{alg:support:3}.
\end{enumerate}
The proof of the correctness of this algorithm is given in~\secondversion{\Cref{prf_optimal_p}.}{\cite{choudhury2020job}.} 

\begin{figure}[t]
    \centering
    \includegraphics[width=75mm]{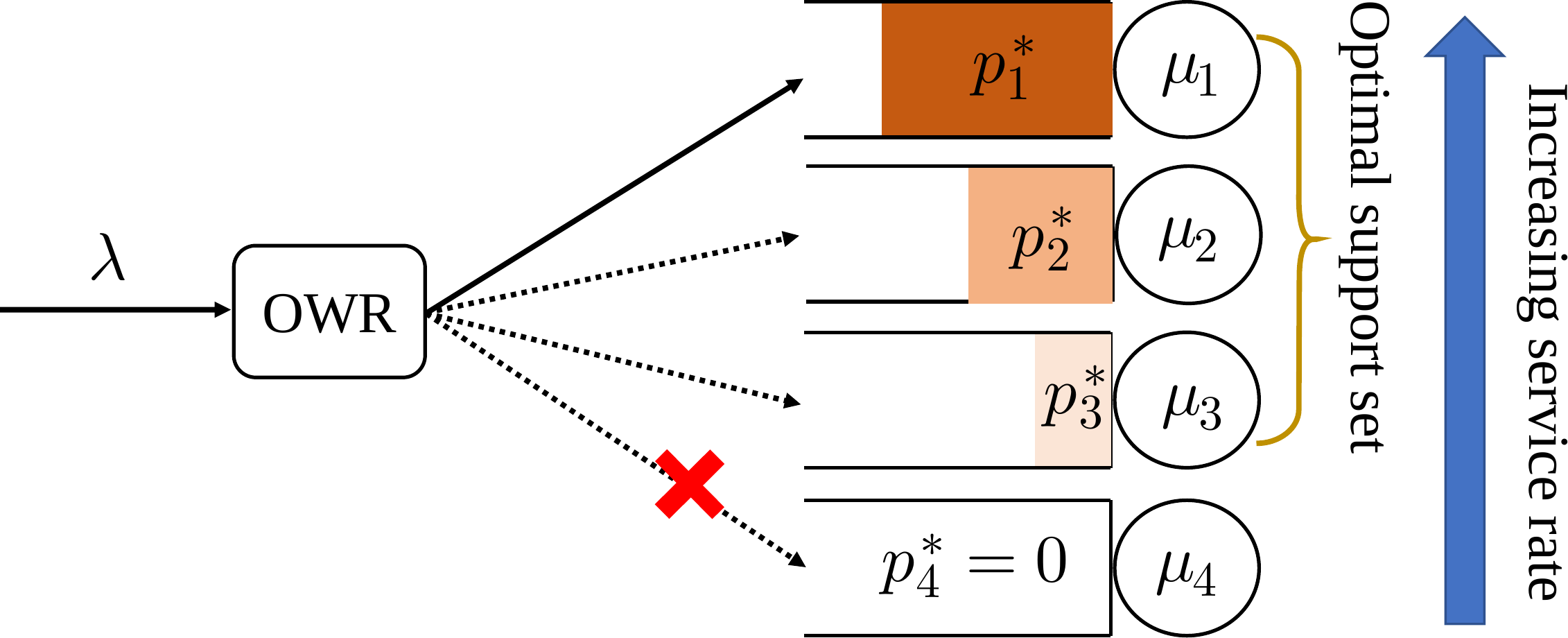}
    \caption{An illustration of the optimal weighted random routing policy. The job arrival rate is $\lambda$ is split and there are $4$ servers with service rates $\mu_1, \dots, \mu_{4}$ respectively such that $\mu_1\geq \mu_2\geq \mu_3\geq \mu_{4}$.}
    \label{fig:waterfilling_policy}
\end{figure}

%% file: newversion/algorithm.tex
\section{Proposed Job Dispatching Policy For Unknown Service Rates}
\label{sec:algorithm}

Recall that we consider a dispatcher who does not know the service rate vector $\bm{\mu}=(\mu_1,\dots,\mu_{\numofserver})$ and thus relies on the estimated service rate vector $\bm{\hat{\mu}}(t)=\left(\hat{\mu}_1(t),\dots,\hat{\mu}_{\numofserver}(t)\right)$. Our goal is to design a dispatching policy $\policy$ that minimizes the expected regret $\mathbb{E}\left[\regret_{\policy \policy^*}(t)\right]$ with respect to the optimal weighted random routing policy $\policy^*$. 

In order to asymptotically converge to $\policy^*$, it is important for the dispatching policy to correctly identify the optimal support set $\support(\lambda,\bm{\mu})$ for which $p_i^* > 0$. If a server with $p_i^* > 0$ is excluded from the estimated support set, then the dispatcher will not send any jobs to it and hence cannot converge to the optimal policy $\policy^*$. In this section we first demonstrate that it is necessary to explore (that is, send jobs to all $\numofserver$ servers) infinitely often for achieving a reasonable regret in Section~\ref{subsec:exploration}. We then present our policy in Section~\ref{subsec:algorithm}.

\subsection{The Necessity of Exploration}\label{subsec:exploration}
It is well-known that for stochastic multi-armed bandits problems, it is necessary to explore infinitely often to achieve an optimal regret.  Interestingly, there are recent results showing that no explicit exploration is needed for achieving an optimal regret in some queueing systems with unknown parameters \citep{krishnasamy2018learningcu}. There the exploration comes for free when running a stabilizing policy.  However, for the job dispatching problem we consider in this paper, we demonstrate below that infinitely often exploration is still necessary.

A naive dispatching policy may dedicate a constant amount of time at the beginning to exploration to obtain a good estimate of the service rate vector.  Then after the initial exploration phase, the policy uses the estimate $\hat{\bm{\mu}}(t)$ at every time slot $t$ to compute the routing vector and dispatches arriving jobs accordingly, while keeping updating the estimate $\hat{\bm{\mu}}(t)$ using the service times of completed jobs.  We will construct an example to demonstrate that this can lead to a situation where a server in the optimal support set $\support(\lambda,\bm{\mu})$ is forever excluded from the estimated optimal set, which will incur a linear regret.



Consider a two-server system with $\lambda = 0.2$ and  $\bm{\mu} = (0.45, 0.55)$. One can verify that the optimal routing vector is $\left(0.25, 0.75\right)$. Suppose for the first $n$ time slots, the dispatching policy assigns an arriving job to one of the two servers chosen uniformly at random. We consider an event $\mathcal{E}$ defined by the scenario below. 

Suppose that there are $k$ job arrivals during the first $n$ time slots to server $1$. Let $t'$ be the earliest time by which all the $k$ 
jobs that are assigned to server~$1$ have departed. Then we consider the scenario where
the estimated service rates satisfy that $\hat{\mu}_1(t) \leq 0.05$ for all time $t$ with $t\le t'$ and $\hat{\mu}_2(t) \geq 0.25$ for all time $t$.

We first argue that under the event $\mathcal{E}$, the regret scales linearly with time.
Note that under $\mathcal{E}$, for any time $t$ with $t\le t'$, the routing vector is $(0,1)$ since $\hat{\mu}_1(t) \leq 0.05$ and $\hat{\mu}_2(t) \geq 0.25$. 
At time $t=t'+1$, we still have that $\hat{\mu}_1(t) \leq 0.05$ since no new job is sent to server~$1$ and that $\hat{\mu}_2(t) \geq 0.25$ by the definition of event $\mathcal{E}$, which makes the routing vector remain $(0,1)$.  Repeating this argument for all the time slots after $t'+1$ we can see that for rest of the time no job will be sent to server~$1$ at all. Now the expected steady-state queue length when using only server $2$ is $9/35$, while the expected steady-state queue length using the optimal routing vector is $19/80$. Thus the regret scales roughly as $11t/560$, which is linear in $t$.

We next show that the event $\mathcal{E}$ happens with a strictly positive probability:
    \begin{align}
        \probability(\mathcal{E}) &= \left(\sum_{k = 1}^n {n \choose k}(0.5\lambda)^k(1 - 0.5\lambda)^{n - k} \probability\left(\substack{\text{all $k$ jobs to server $1$}\\ \text{have service time $\ge 20$}}\right)\right)\nonumber\\
        &\mspace{23mu}\cdot\probability\left(\cap_{t = 1}^{\infty}\left\{\hat{\mu}_2(t) \geq 0.25\right\}\right)\\
        &\geq \left(1 -  \sum_{t = 1}^{\infty}\probability\left(\hat{\mu}_2(t) < 0.25\right)\right)\nonumber\\
        &\mspace{23mu}\cdot \sum_{k = 1}^n {n \choose k}(0.5\lambda)^k (1 - 0.5\lambda)^{n - k} \left((0.45)^{20}\right)^k\\
        &\geq 0.09 \sum_{k = 1}^n {n \choose k}(0.5\lambda)^k (1 - 0.5\lambda)^{n - k} \left((0.45)^{20}\right)^k\label{eq:counter_example},
    \end{align}
which is strictly positive, where \eqref{eq:counter_example} is due to Chernoff bound.


\subsection{An \texorpdfstring{$\epsilon_t$}{et}-Exploration Policy}
\label{subsec:algorithm}


As demonstrated in Section~\ref{subsec:exploration},
exploring for a fixed amount of time can lead to a linear regret. To ensure enough exploration, we propose an $\epsilon_t$-exploration policy, which explores with probability $\epsilon_t = \frac{\numofserver \ln t}{t}$ at each time slot $t$. When not exploring, the policy treats the estimated service rates as if they were the actual service rates and calculates the optimal routing probabilities based on them; i.e., when not exploring, the policy uses $\hat{\bm{p}}(t) = (\hat{p}_1(t), \hat{p}_2(t), \cdots, \hat{p}_\numofserver(t))$ to make a routing decision at time $t$. The pseudo-code is presented in~\Cref{poli:bandit}.
\begin{algorithm}[t]
\caption{
An $\epsilon_t$-Exploration Policy for Learning Optimal Weighted Random Routing}
\label{poli:bandit}
\begin{algorithmic}[1]
\While{$t\geq 0$}
\If{\mbox{a job arrives}}
\State $\explorerandom(t)\gets$a Bernoulli sample with mean $\min\left\{1, \frac{K\ln t}{t}\right\}$
\If{$\explorerandom(t) = 1$}
\Comment{Explore}
\State{Dispatch to one of the servers uniformly at random}
\Else
\Comment{Exploit}
\State{Compute routing vector $\hat{\bm{p}}(t) = f(\lambda, \hat{\bm{\mu}}(t))$}
\State{Dispatch to server $i$ with probability $\hat{p}_i(t)$
}
\EndIf
\EndIf
\For{$i = 1, 2, \cdots K$}
\If{\mbox{a job departs from server $i$}}
\State {Update $\hat{\mu}_i(t)$ using \eqref{eqn:mu_estimate}}
\EndIf
\EndFor
\EndWhile
\end{algorithmic}
\end{algorithm}


%% file: newversion/regret_analysis.tex
\section{Regret Analysis}\label{sec:regret_analysis}
In this section, we prove an upper bound on the expected regret $\expectation\left[\regret_{\policy \policy^*} (t)\right]$ of the $\epsilon_t$-exploration policy.

To state our upper bound, we first define a quantity $\Delta_{\support}$ that we refer to as the \emph{tolerance gap}, which is analogous to the suboptimality gap for multi-armed bandits.  Specifically, let
\begin{equation}\label{eq:gap}
\begin{split}
\Delta_{\support} &:= \sup\left\{\delta \geq 0\colon \support\left(\lambda,\bm{\mu}'\right) = \support(\lambda,\bm{\mu})\right.\\
&\mspace{111mu}\left.\forall\bm{\mu}'\mathrm{ s.t.\ } \left|\mu'_i-\mu_i\right|\leq \delta,\forall i\right\},
\end{split}
\end{equation}
where recall that $\support(\lambda,\bm{\mu})$ is the optimal support set computed from arrival rate $\lambda$ and service rate vector $\bm{\mu}$.

The tolerance gap $\Delta_{\support}$ quantifies how much error in the service rates can be tolerated without incurring a discrepancy in the support set.  We can think of the $\bm{\mu}'$ in \eqref{eq:gap} as the estimated service rate vector.  If $\Delta_{\support}=0$, then even a slight imprecision in the estimated service rates would make the estimated support set deviate from the optimal support set, indicating hardness of the problem.  Therefore, we make the assumption that $\Delta_{\support}>0$ in our upper bound.  We comment that from a practical perspective, this is a very mild assumption since only a small set (with zero measure) in the parameter space $\left\{(\lambda,\bm{\mu})\in\mathbb{R}_+^{K+1}\colon \lambda<\sum_{i=1}^K \mu_i\right\}$ will violate this assumption.

\begin{thm}[Upper Bound on Expected Regret]
\label{thm:main}
Consider a system with arrival rate $\lambda$ and service rate vector $\bm{\mu}$ and assume that $\Delta_{\support}>0$.
Then there exists a constant $k_1$ and a $t_0$ such that for all $t \geq t_0$, the $\epsilon_t$-exploration policy (\Cref{poli:bandit}) has an expected regret $\expectation\left[\regret_{\policy \policy^*} (t)\right]$ that admits the following upper bound:
\begin{multline}
\expectation\left[\regret_{\policy \policy^*} (t)\right]\\
\le \sum_{\tau = t_0}^{t-1}\left(\sum_{i: p_i^* > 0}\frac{66k_1\ln \tau}{\freecapacity_i^2}\sqrt{\frac{\ln \tau}{\tau}}+\sum_{i=1}^{\numofserver}\frac{132 \ln^2 \tau}{\freecapacity_i^2 \tau}\right) + \order\left(1\right),
\end{multline}
where $\freecapacity_i$ is the residual capacity of server $i$ under the optimal weighted random routing given by $\freecapacity_i = \mu_i - \lambda p_i^*$.
\end{thm}

In the regret upper bound in Theorem~\ref{thm:main} above, the first summand $\sum_{\tau = t_0}^{t-1}\sum_{i: p_i^* > 0}\frac{66k_1\ln \tau}{\freecapacity_i^2}\sqrt{\frac{\ln \tau}{\tau}}$ is the dominant term and it comes from the estimation error in the estimated routing probability vector $\bm{\hat{p}}(t)$, and the second summand $\sum_{\tau = t_0}^{t-1}\sum_{i=1}^{\numofserver}\frac{132 \ln^2 \tau}{\freecapacity_i^2 \tau}$ results from the exploration used by the $\epsilon_t$-exploration policy.  We note that this regret bound becomes smaller in low to moderate traffic regimes where the residual capacities $\freecapacity_i$'s are large and the size of the optimal support set $\support\left(\lambda,\bm{\mu}\right)=\{i\colon p_i^*>0,i=1,2,\dots,K\}$ is small.


In the following subsections, we first couple our system with the system that runs the the optimal weighted random routing policy in Section~\ref{subsec:coupling} to facilitate the regret analysis.  We then prove Theorem~\ref{thm:main} in Section~\ref{subsec:proof-regret-bound}, using several lemmas whose proof sketches are given in Section~\ref{subsec:proof-sketch-lemmas}.

\subsection{Coupling with the Optimal Weighted Random Routing}\label{subsec:coupling}
Consider the system that runs the optimal weighted random routing policy $\policy^*$, which we refer to as the \emph{optimal system}.  We will annotate quantities in the optimal system with the superscript $^*$, e.g., $\arrival^*(t)$ denotes the total number of job arrivals at time $t$ to the optimal system, and $\queue_i^*(t)$ denotes the length of queue $i$ under $\policy^*$.  Correspondingly, recall that the regret at time $t$ is defined as
\begin{equation*}
\regret_{\policy \policy^*}(t) = \sum_{\tau = 1}^{t-1} \sum_{i = 1}^{\numofserver}\left( \queue_i(\tau) - \queue_i^*(\tau)\right).
\end{equation*}
We assume that the optimal system also starts from empty queues, i.e., $\queue_i^*(0)=0$ for all $i$.

We couple the system that runs our proposed $\epsilon_t$-exploration policy with the optimal system in the following way.

\vspace{0.1cm} \noindent {\textbf{Arrivals.}}
We couple arrivals such that $\arrival(t)=\arrival^*(t)$ for all time $t$.
    
\vspace{0.1cm} \noindent {\textbf{Service.}}
For each server~$i$, let $\service_i(t)$ for $t=0,1,2,\dots$ be i.i.d.\ Bernoulli random variables that take the value $1$ with probability $\mu_i$.  We will refer to $\service_i(t)$ as the \emph{service} offered by server $i$ at $t$.  If $\queue_i(t)+\arrival_i(t)>0$, we let $\departure_i(t)=\service_i(t)$, where recall that $\departure_i(t)$ is the number of departures from queue $i$ at $t$; otherwise it is clear that $\departure_i(t)=0$. Similarly, let $\service_i^*(t)$ denote the corresponding service offered in the optimal system.
We couple the service processes such that $\service_i(t)=\service_i^*(t)$ for all server $i$ and all time $t$.
    
\vspace{0.1cm} \noindent {\textbf{Assignment process.}}
Recall that in the $\epsilon_t$-exploration policy, for each time slot $t$, with probability $\frac{\numofserver \ln t}{t}$ we explore and otherwise we dispatch the arriving job according to the routing vector $\hat{\bm{p}}(t)$. We couple the dispatching decision generated from $\hat{\bm{p}}(t)$ with the dispatching decision generated from the optimal routing vector $\bm{p}^*$ in the optimal system as follows.

For simplicity, we can assume that for each time slot $t$, we generate a dispatching decision from $\hat{\bm{p}}(t)$, although this dispatching decision is needed only when there is a job arrival at $t$ and the $\epsilon_t$-exploration policy chooses to exploit.  Let the dispatching decision generated from $\hat{\bm{p}}(t)$ be represented by the server that an arriving job will be dispatched to, denoted as $\dispdecision(t)$.  Then $\dispdecision(t)$'s probability mass function (pmf) is $\hat{\bm{p}}(t)$.  Similarly, let $\dispdecision^*(t)$ be the dispatching decision in the optimal system, and then $\dispdecision^*(t)$'s pmf is $\bm{p}^*$. Then we couple $\dispdecision(t)$ and $\dispdecision^*(t)$ such that they have the following joint pmf:
\begin{multline}
\probability(\dispdecision(t)=i,\dispdecision^*(t)=j)\\
=
\begin{cases}
\min\{\hat{p}_i(t),p^*_i\} & \textrm{ if }i=j,\\
\frac{\left(\hat{p}_i(t)-\min\left\{\hat{p}_i(t),p^*_i\right\}\right)\left(p^*_j(t)-\min\left\{\hat{p}_j(t),p^*_j\right\}\right)}{d_{TV}(\hat{\bm{p}}(t), \bm{p}^*)} & \textrm{ if }i\neq j,
\end{cases}
\end{multline}
where $d_{TV}(\hat{\bm{p}}(t), \bm{p}^*)$ is the total variation distance between $\hat{\bm{p}}(t)$ and $\bm{p}^*$ and is given by $d_{TV}(\hat{\bm{p}}(t), \bm{p}^*) =\sum_{j=1}^K\left(p^*_j(t)-\min\left\{\hat{p}_j(t),p^*_j\right\}\right)$. This coupling is known as the \emph{maximal coupling} (see, e.g., \citep{RosPek_07}) and it guarantees that $\probability(\dispdecision(t)\neq\dispdecision^*(t))=d_{TV}(\hat{\bm{p}}(t), \bm{p}^*)$. 


With this coupling, we can quantify the probability for a mismatched dispatching decision between our system and the optimal system.  In our system, recall that $\arrival_i(t)$ denotes the number of jobs dispatched to server~$i$ at time $t$.  We now make a finer distinction between jobs dispatched through exploration and through exploitation under the $\epsilon_t$-exploration policy.  Let $\explore_i(t)$ and $\exploit_i(t)$ denote the numbers of jobs dispatched to server $i$ through exploration and exploitation, respectively.  Then $\arrival_i(t) = \explore_i(t) + \exploit_i(t)$.  Lemma~\ref{lem:mismatch} below upper-bounds the probability for the mismatch that $\exploit_i(t) = 1,  \arrival_i^*(t) = 0$ with the distance $\left|\hat{p}_i(t) - p_i^*\right|$, implying that once the estimates $\hat{p}_i(t)$'s are close to $p_i^*$'s, the probability of such a mismatch is small.
\begin{lem}\label{lem:mismatch}
For any time slot $t$ and any server $i$,
\begin{equation*}
\probability\left[\exploit_i(t) = 1,  \arrival_i^*(t) = 0 \Bigm|  \hat{p}_i(t)\right]\leq \left|\hat{p}_i(t) - p_i^*\right|.
\end{equation*}
\end{lem}
Proof of the lemma is given in \secondversion{\Cref{prf_lem:mismatch}.}{\cite{choudhury2020job}.}
\subsection{Proof of Regret Bound (Theorem~\ref{thm:main})}\label{subsec:proof-regret-bound}
In this section we prove the upper bound in Theorem~\ref{thm:main} on the expected regret $\expectation\left[\regret_{\policy \policy^*}(t)\right] = \expectation\left[\sum_{\tau = 1}^{t-1} \sum_{i = 1}^{\numofserver}\left( \queue_i(\tau) - \queue_i^*(\tau)\right)\right]$ based on several lemmas.
Proof sketches of these lemmas will be given in Section~\ref{subsec:proof-sketch-lemmas}, and the detailed proofs are presented in \secondversion{Appendices~\ref{prf_lem:inst-regret}, \ref{prf_lem:final_busy_period}, \ref{prf_lem:tight_bound_p} and \ref{prf_lem:p_wt} respectively.}{\cite{choudhury2020job}.}

We first note that the difference between $\queue_i(t)$ and $\queue_i^*(t)$ can be written in the following recursive form for any $t$ and $\tau\le t$:
\begin{align*}
    \queue_i(t) - \queue_i^*(t) &= \queue_i(\tau) - \queue_i^*(\tau)\\
    &\mspace{21mu}+\sum_{\ell = \tau}^{t-1}\left(\arrival_i(\ell) - \departure_i(\ell) - \left(\arrival_i^*(\ell) - \departure_i^*(\ell)\right)\right).
\end{align*}
In this proof, we will consider a specific $\tau$ that is the last time queue~$i$ is empty. In particular, define $\busy_i(t)$ as the length of the current busy cycle period as seen at time $t$, i.e.,
\begin{equation}
    \busy_i(t) = \min\left\{s\ge 0 : \queue_i(t - s) =0\right\}.
\end{equation}
Then it is easy to see that for $\tau = t-\busy_i(t)$, we have $\queue_i(\tau)=0$ and $\queue^*_i(\tau)\ge 0$.  In addition, for any $\ell$ with $\tau\le\ell\le t-1$, we have $\departure_i(\ell)=\service_i(\ell)$ since $\queue_i(\ell)>0$.
Based on this choice of $\tau$, the queue length difference can be bounded as follows:
\begin{align}
\mspace{21mu}\queue_i(t) - &\queue_i^*(t)\nonumber
= \queue_i(t-\busy_i(t)) - \queue_i^*(t - \busy_i(t)) + \nonumber  \\
&\mspace{21mu} \sum_{\ell = t- \busy_i(t)}^{t-1}\left(\arrival_i(\ell) - \departure_i(\ell) - \left(\arrival_i^*(\ell) - \departure_i^*(\ell)\right)\right)\nonumber\\
&\leq \sum_{\ell = t- \busy_i(t)}^{t-1}\left(\arrival_i(\ell) - \arrival_i^*(\ell)\right)  + \sum_{\ell = t- \busy_i(t)}^{t-1}\left(\departure_i^*(\ell) - \departure_i(\ell)\right)\nonumber\\
&\le \sum_{\ell = t- \busy_i(t)}^{t-1} \left(\explore_i(\ell) + \exploit_i(\ell) - \arrival_i^*(\ell)\right)\nonumber\\
&\mspace{21mu}+ \sum_{\ell = t- \busy_i(t)}^{t-1} \left(\service_i^*(\ell) - \service_i(\ell) \right)\label{eq:align_queue_a}\\
&= \sum_{\ell = t- \busy_i(t)}^{t-1} \explore_i(\ell) + \sum_{\ell = t- \busy_i(t)}^{t-1} \left(\exploit_i(\ell) - \arrival_i^*(\ell)\right)\label{eq:align_queue_b}\\
&\leq \sum_{\ell = t- \busy_i(t)}^{t-1} \explore_i(\ell) + \sum_{\ell = t- \busy_i(t)}^{t-1} \indicatorofevents_{ \left\{ \exploit_i(\ell)=1, \arrival_i^*(\ell) = 0\right\}},\label{rel:regret_equation}
\end{align}
where \eqref{eq:align_queue_a} uses the facts that $\arrival_i(\ell)=\explore_i(\ell) + \exploit_i(\ell)$, $\departure^*_i(\ell)\le \service^*_i(\ell)$, and $\departure_i(\ell)= \service_i(\ell)$; \eqref{eq:align_queue_b} is due to our coupling $\service_i(\ell)=\service^*_i(\ell)$.

In the upper bound \eqref{rel:regret_equation} on the queue length difference, the first summand comes from exploration.  Since we know that by our $\epsilon_t$-exploration, we have $\expectation[\explore_i(\ell)]=\frac{K\ln \ell}{\ell}$, this summand can be properly bounded if we obtain a suitable upper bound on $\busy_i(t)$.  The second summand in \eqref{rel:regret_equation} comes from exploitation, and it can be bounded with the estimation error through Lemma~\ref{lem:mismatch}.  To formalize the above intuition, we define the following events:
\begin{equation}\label{eqn:event5}
\eventfive := \left\{\busy_i(t) \leq v_i(t),\forall i\right\},\textrm{ where }v_i(t)=\frac{66\ln t}{\freecapacity_{i}^2}, \text{ and}
\end{equation}
\begin{equation}\label{eqn:event6}
\eventsix := \left\{\left| \hat{p}_i(\tau) - p_i^*\right| \leq k_1\min\left\{\sqrt{\frac{ \ln t}{t}}, p_i^*\right\},\forall \tau \in \left[\frac{t}{2} + 1, t\right],\forall i\right\},
\end{equation}
where $k_1$ is a properly chosen constant.
Utilizing these two events, Lemma~\ref{lem:inst-regret} below establishes an upper bound on the expected queue length difference, which enables us to further bound the regret by analyzing the busy period and the estimation error in service rates.
\begin{lem}\label{lem:inst-regret}
There exists a $t_0$ such that for any time $t\ge t_0$, the total expected queue length difference can be bounded as
\begin{multline}
\sum_{i=1}^K\expectation\left[\queue_i(t) - \queue_i^*(t)\right]
\le\sum_{i=1}^K\frac{2v_i(t)\ln t}{t} \\
+ \sum_{i:p_i^*>0} k_1v_i(t) \sqrt{\frac{\ln t}{t}} + t\probability\left(\left(\eventfive\right)^c\right) + 2t\probability\left(\left(\eventsix\right)^c\right).
\end{multline}
\end{lem}

\begin{figure}
    \centering
    \includegraphics[width=80mm]{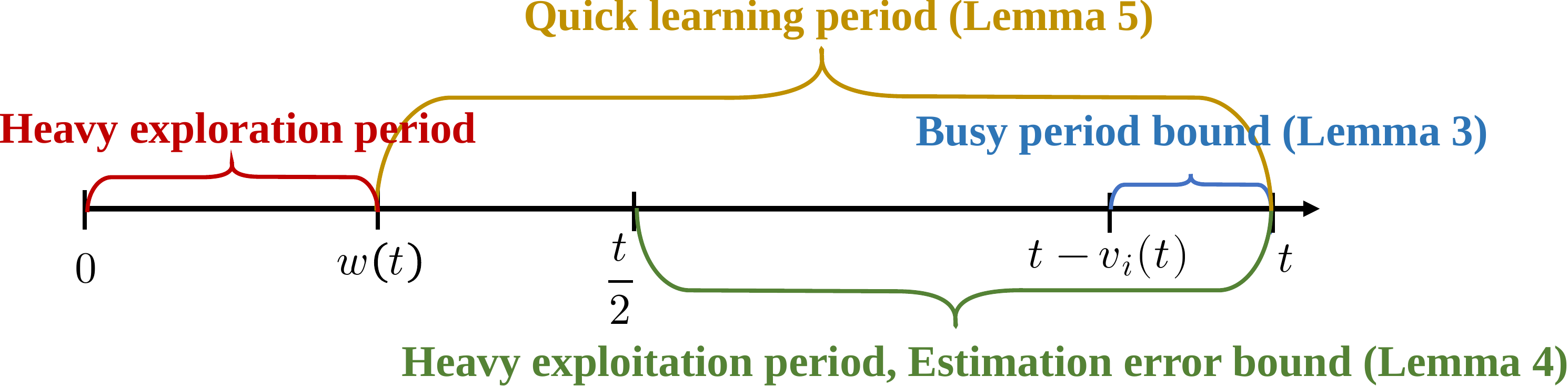}
    \caption{Time structure of lemmas in proof of Theorem~\ref{thm:main}.}
    \label{fig:dia_timeline}
\end{figure}
With Lemma~\ref{lem:inst-regret}, to bound the expected regret, now it suffices to bound the probabilities $\probability\left(\left(\eventfive\right)^c\right)$ and $\probability\left(\left(\eventsix\right)^c\right)$, which are established in Lemmas~\ref{lem:final_busy_period} and \ref{lem:tight_bound_p} below.  We demonstrate the time structure of the lemmas in Figure~\ref{fig:dia_timeline}.
\begin{lem}[Busy Period Bound]\label{lem:final_busy_period}
There exist a constant $k_2$ and a $t_0$ such that for any $t\ge t_0$, the event $\eventfive$ defined in \eqref{eqn:event5} satisfies
\begin{equation}
\probability\left(\left(\eventfive\right)^c\right) \leq 4\numofserver\left( \frac{1}{t^{7}} + \frac{k_2 + 1}{t^3} + \frac{1}{t^4}\right).
\end{equation}
\end{lem}
\begin{restatable}[Estimation Error Bound]{lem}{lemERRORestimate}
\label{lem:tight_bound_p}
There exist a constant $k_2$ and a $t_0$ such that for any $t\ge t_0$, the event $\eventsix$ defined in \eqref{eqn:event6} satisfies  
\begin{multline}
\probability\left(\left(\eventsix\right)^c\right) \leq \numofserver\left(\frac{1}{t^7} + \frac{k_2 + 1}{t^3}\right) \\+\sum_{i: p_i^*>0}\left(\frac{1}{t^3} + t \exp\left(-\frac{p_i^*\lambda t}{128}\right) +
t\exp\left(-\frac{\freecapacity_i^2 t}{4}\right)\right).
\end{multline}
\end{restatable}

Finally, we choose a common $t_0$ for Lemmas~\ref{lem:inst-regret}--\ref{lem:tight_bound_p} and a common $k_2$ for Lemmas~\ref{lem:final_busy_period} and \ref{lem:tight_bound_p}, and put Lemmas~\ref{lem:inst-regret}--\ref{lem:tight_bound_p} together to get
\begin{align}
&\mspace{21mu}\expectation\left[\regret_{\policy \policy^*}(t)\right]\nonumber\\
&= \sum_{\tau = 1}^{t_0-1}\sum_{i=1}^K \expectation\left[\queue_i(\tau)-\queue^*_i(\tau)\right] + \sum_{\tau = t_0}^{t-1} \sum_{i=1}^K \expectation\left[\queue_i(\tau)-\queue^*_i(\tau)\right]\nonumber\\
&\le t_0^2 + \sum_{\tau = t_0}^{t-1} \sum_{i=1}^K\frac{2v_i(\tau)\ln \tau}{\tau}
+\sum_{\tau = t_0}^{t-1}\sum_{i:p_i^*>0} k_1v_i(\tau) \sqrt{\frac{\ln \tau}{\tau}}\nonumber\\
&\mspace{21mu}+\sum_{\tau = t_0}^{t-1}\tau\cdot \numofserver\left(\frac{4}{\tau^{7}} + \frac{4k_2 + 4}{\tau^3} + \frac{4}{\tau^4}\right)+\sum_{\tau = t_0}^{t-1}2\tau\cdot \numofserver\left(\frac{1}{\tau^{7}} + \frac{k_2 + 1}{\tau^3}\right)\nonumber\\
&\mspace{21mu}+\sum_{\tau = t_0}^{t-1}2\tau\cdot\sum_{i: p_i^*>0}\left(\frac{1}{\tau^3} + \tau\exp\left(-\frac{p_i^*\lambda \tau}{128}\right) +\tau\exp\left(-\frac{\freecapacity_i^2 \tau}{4}\right)\right)\\
& = t_0^2 + \sum_{\tau = t_0}^{t-1} \sum_{i=1}^{\numofserver}\frac{132 \ln^2 \tau}{\freecapacity_i^2 \tau} + \sum_{\tau = t_0}^{t-1}\sum_{i: p_i^* > 0} \frac{66k_1\ln \tau}{\freecapacity_i^2}\sqrt{\frac{\ln \tau}{\tau}}  \nonumber\\
&\mspace{21mu} + \numofserver\sum_{\tau = t_0}^{t-1} \left( \frac{6}{\tau^{6}} + \frac{6k_2 + 6}{\tau^2} + \frac{4}{\tau^3}\right)\nonumber\\
&\mspace{21mu} + 2\sum_{\tau = t_0}^{t-1}\sum_{i: p_i^*>0}\left(\frac{1}{\tau^2} + \tau^2\exp\left(-\frac{p_i^*\lambda \tau}{128}\right) + \tau^2\exp\left(-\frac{\freecapacity_i^2 \tau}{4}\right)\right)\\
&=\sum_{\tau = t_0}^{t-1}\left(\sum_{i: p_i^* > 0} \frac{66k_1\ln \tau}{\freecapacity_i^2}\sqrt{\frac{\ln \tau}{\tau}}+\sum_{i=1}^{\numofserver}\frac{132 \ln^2 \tau}{\freecapacity_i^2 \tau}\right) +  \order\left(1\right),\nonumber
\end{align}
which completes the proof of Theorem~\ref{thm:main}.\qed

\begin{remark*}
Our proof techniques used the \emph{absolute} difference in the routing probabilities to analyze the difference in queue lengths.
We comment that it might be possible for one to prove a tighter regret upper bound by considering the actual difference in routing probabilities, but the analysis will become much more challenging.  Specifically, inaccurate routing probabilities can actually instantaneously benefit the queues whose $\hat{p}_i(t)$ is smaller than the optimal $p^*_i$.  This is a phenomenon not seen in multi-armed bandit problems, and it is worth further investigation.
\end{remark*}

\subsection{Proof Sketches of Lemmas~\ref{lem:inst-regret}--\ref{lem:tight_bound_p}}\label{subsec:proof-sketch-lemmas}
\begin{figure*}
    \centering
    \begin{subfigure}{0.33\linewidth}
        \centering
        \includegraphics[scale = 0.35]{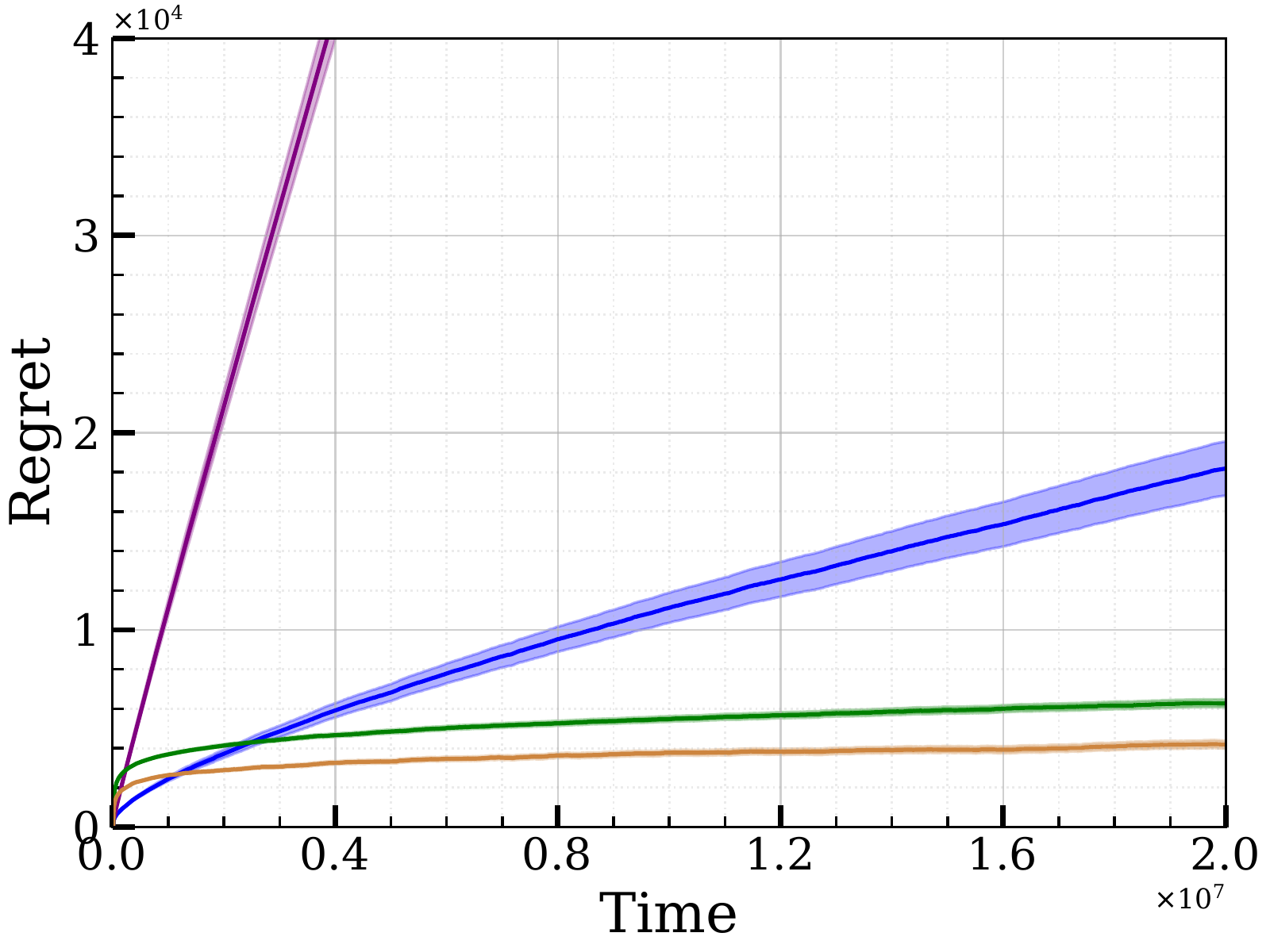}
        \caption{$\lambda = 0.1$}
    \end{subfigure}
    \begin{subfigure}{0.33\linewidth}  
        \centering 
        \includegraphics[scale = 0.35]{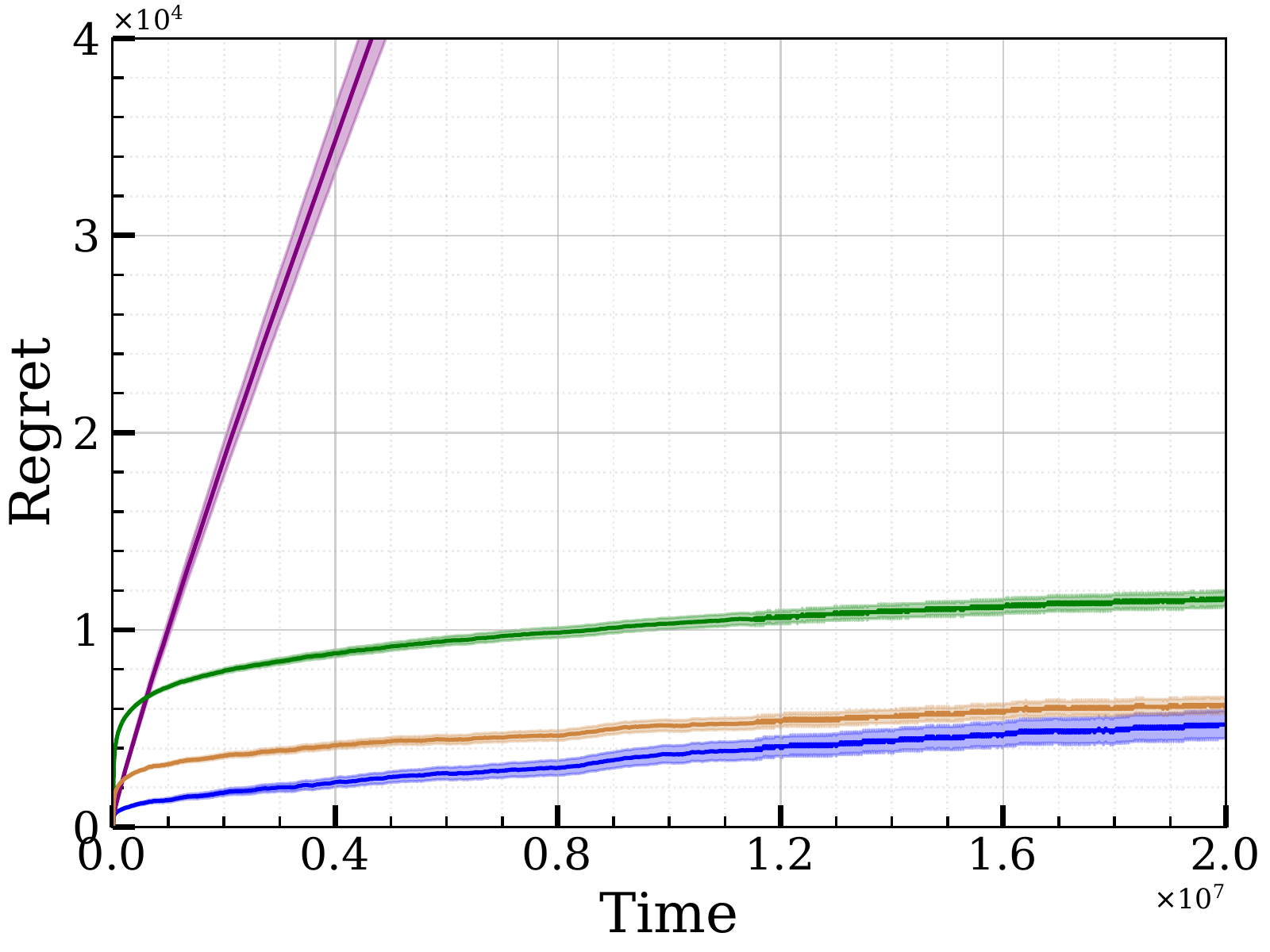}
        \caption{$\lambda = 0.2$}
    \end{subfigure}
    \begin{subfigure}{0.33\linewidth}  
        \centering 
        \includegraphics[scale = 0.35]{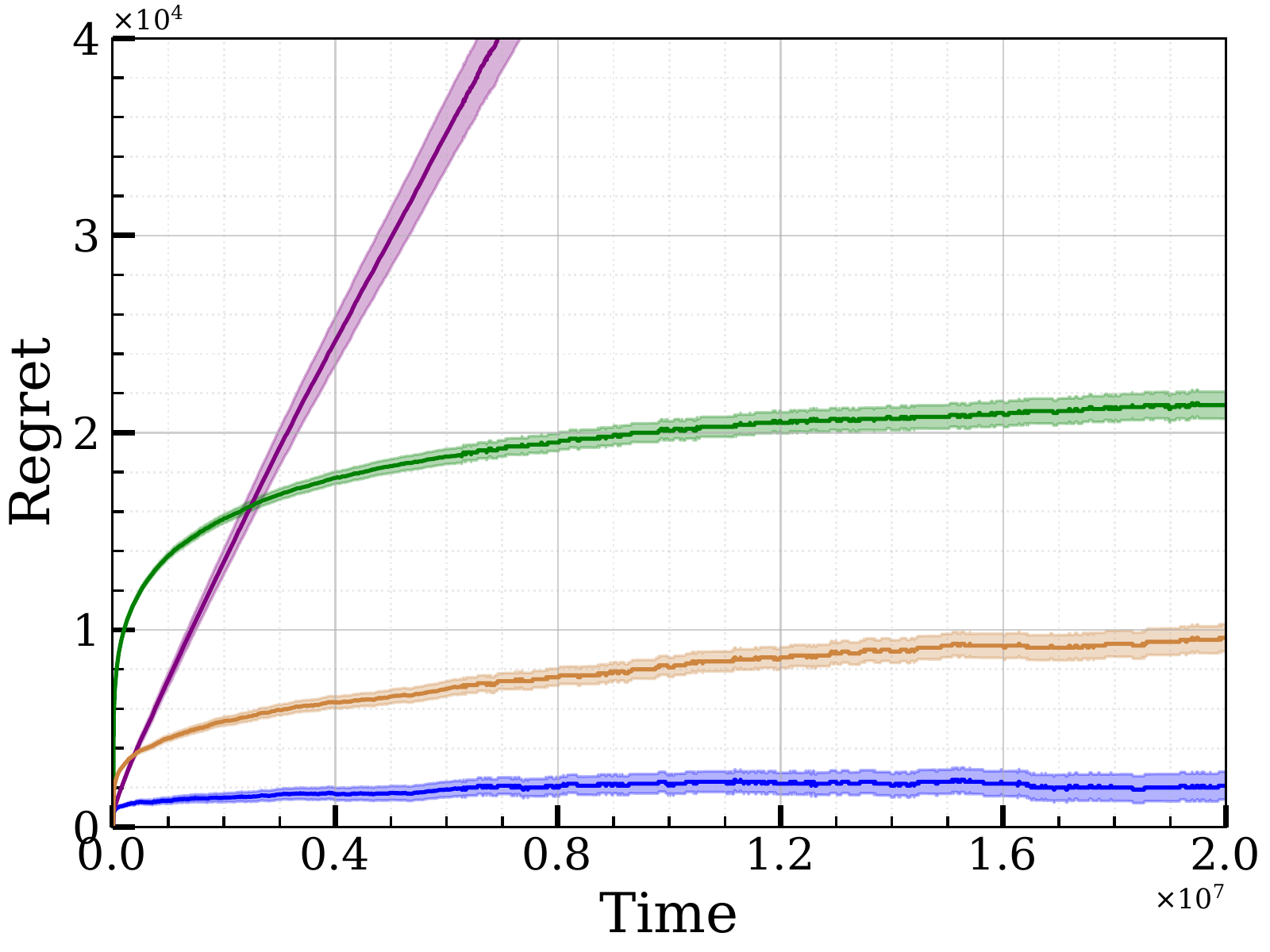}
        \caption{$\lambda = 0.3$}
    \end{subfigure}
        \centering
    \begin{subfigure}{0.33\linewidth}
        \centering
        \includegraphics[scale = 0.35]{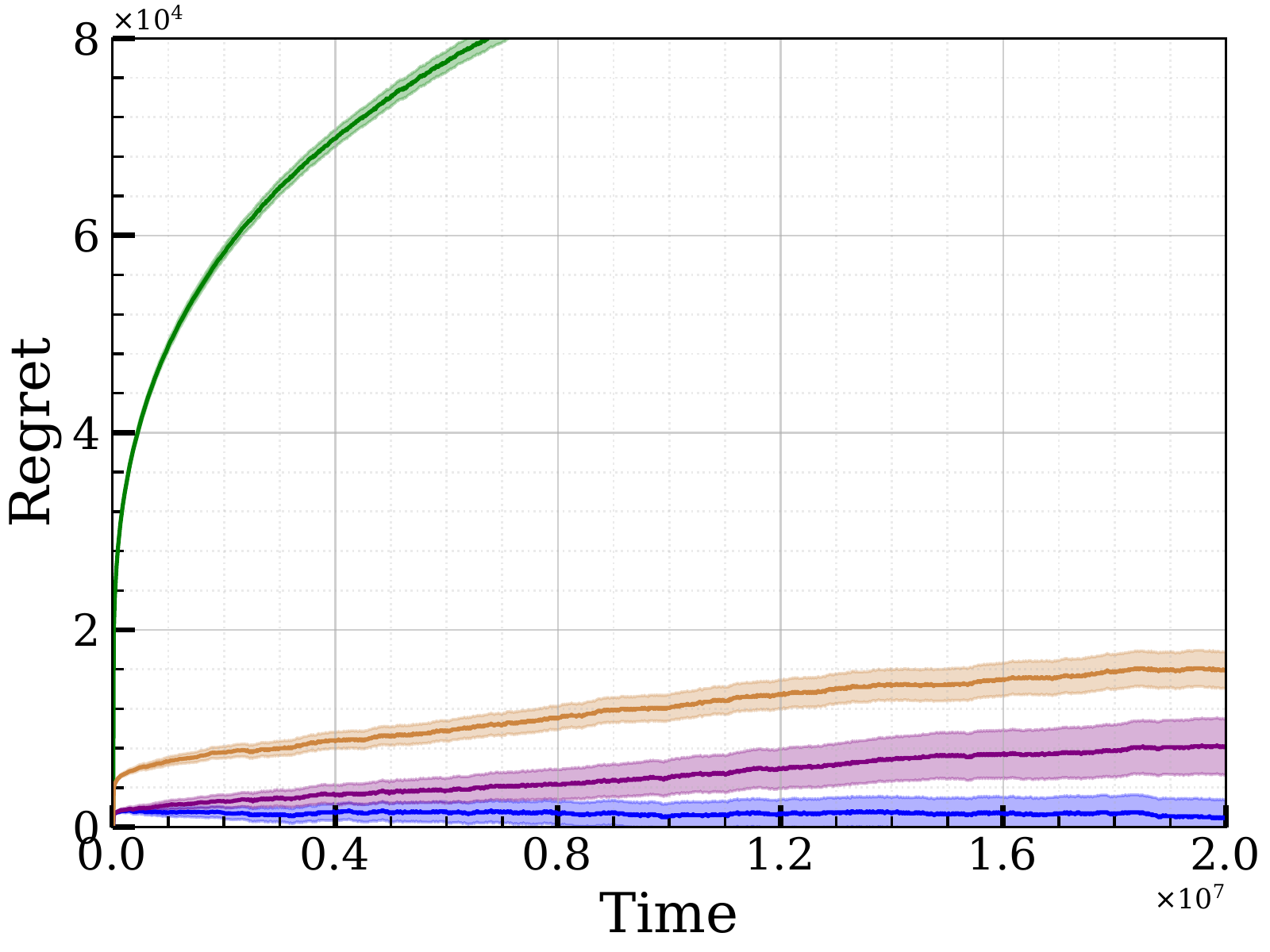}
        \caption{$\lambda = 0.5$}
    \end{subfigure}
    \centering
    \begin{subfigure}{0.33\linewidth}  
        \centering 
        \includegraphics[scale = 0.35]{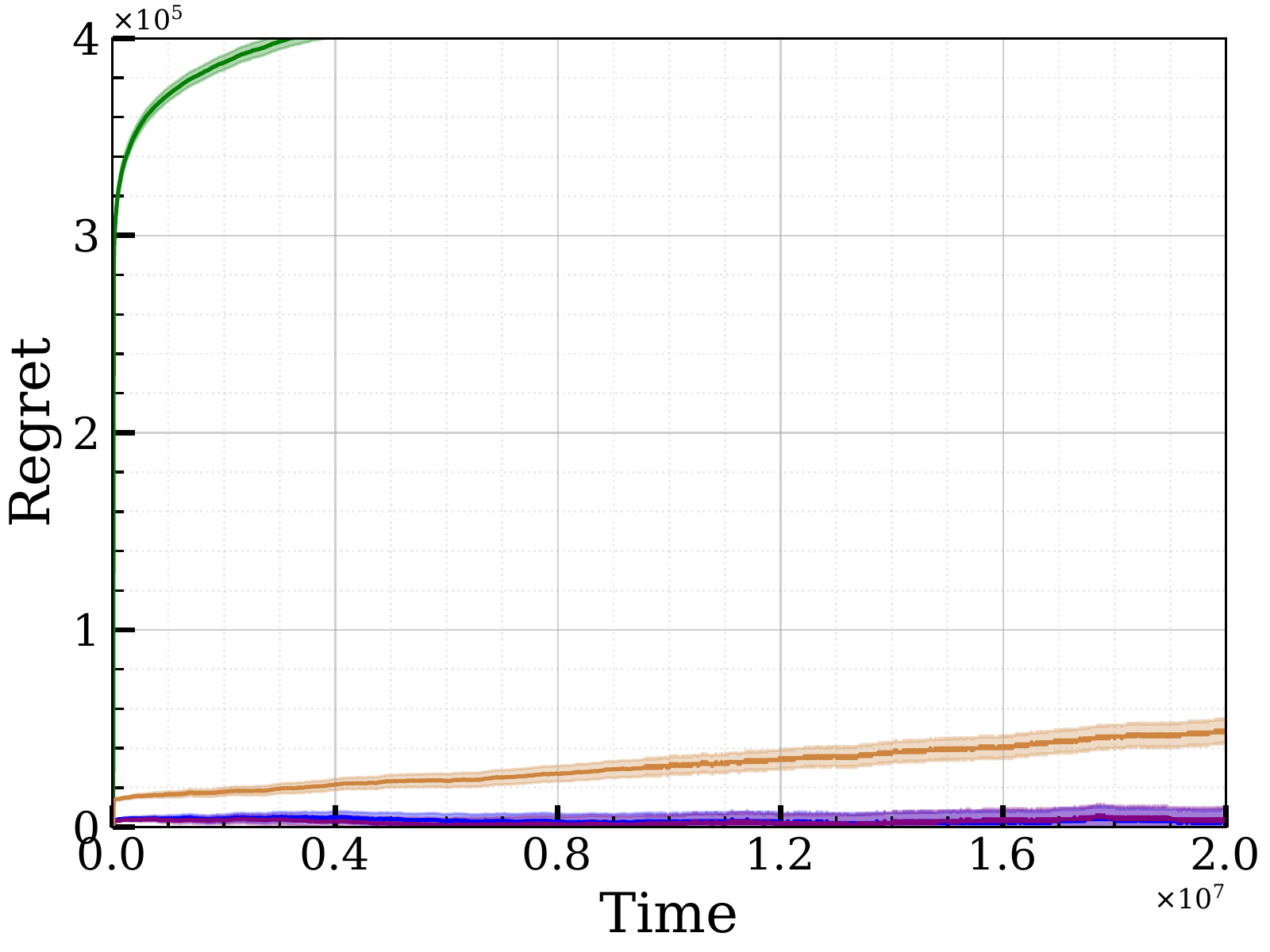}
        \caption{$\lambda = 0.7$}
    \end{subfigure}
    \centering
    \begin{subfigure}{0.33\linewidth}  
        \centering 
        \raisebox{1cm}{\includegraphics[scale = 0.9, trim={2mm 2mm 1mm 2mm}, clip]{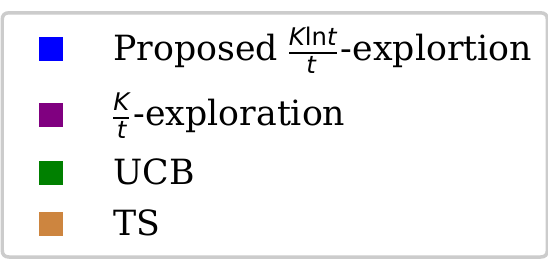}}
    \end{subfigure}
    \caption{Regret vs time for various policies. The shaded region represent $\pm2\sigma$ boundary of the mean regret. TS and UCB performs well in load load regime, while $K/t$-greedy exploration policy performs well in higher traffic regimes. Our policy moderately in very low load regime and performs well in higher traffic regimes. 
    }
    \label{fig:regret_policies}
\end{figure*}


The detailed proofs of Lemmas~\ref{lem:inst-regret}, \ref{lem:final_busy_period}, \ref{lem:tight_bound_p} and \ref{lem:p_wt} are presented in \secondversion{Appendices~\ref{prf_lem:inst-regret}, \ref{prf_lem:final_busy_period}, \ref{prf_lem:tight_bound_p} and \ref{prf_lem:p_wt} respectively.}{\cite{choudhury2020job}.}

\Cref{lem:inst-regret} can be proven using a series of conditioning on event $\eventfive$, event $\eventsix$ and also the estimated  $\hat{\bm{p}}(t)$.  Then the construction of $\eventfive$ gives an upper bound on the busy period, \Cref{lem:mismatch} helps translate the number of mismatches into the error in estimation, and finally the construction of $\eventsix$ upper bounds the estimation error.

Next for Lemmas~\ref{lem:final_busy_period} and \ref{lem:tight_bound_p}, we will just highlight a key lemma used in their proofs, presented as Lemma~\ref{lem:p_wt}.  
Lemma~\ref{lem:p_wt} below states that for any large enough time $t$, the estimated optimal support set, $\support\left(\lambda, \hat{\bm{\mu}}(t)\right)$, is correct (i.e., equal to the true optimal support set $\support\left(\lambda, \bm{\mu}\right)$) for a period of time $[w(t),t]$ with high probability, where $w(t)=2\exp\left(\Theta\left(\sqrt{\ln t}\right)\right)$.  Moreover, during this period of time, the rate at which we dispatch jobs to each server~$i$ lies between $\lambda p^*_i/2$ and $\mu_i-\freecapacity_i/2$; i.e., the arrival rate to each server is no smaller than half of the rate under the optimal weighted random routing, but still leaves at least half of residual capacity under the optimal weighted random routing.  We call the period of time $[w(t),t]$ the \emph{quick learning period} since we have ``locked'' the correct support set and spend all exploitation jobs on learning the service rates of servers in the correct support set.  This time structure of Lemma~\ref{lem:p_wt} is also illustrated in Figure~\ref{fig:dia_timeline}.

\begin{restatable}[Quick Learning Period]{lem}{restateLEMMApWT}
\label{lem:p_wt}
Define the event $\eventone$ as 
\begin{equation}\label{event1}
    \eventone := \eventoneone \cap \eventonetwo, \text{ where}
\end{equation} 
\begin{align*}
\eventoneone&:=\left\{\support\left(\lambda, \hat{\bm{\mu}}(\tau)\right) = \support(\lambda,\bm{\mu}),\forall \tau \in \left[w(t) + 1, t\right]\right\},\\
\eventonetwo&:= \left\{\frac{\lambda p_i^*}{2} \leq
\expectation\left(\arrival_i(\tau) \mmiddle \hat{p}_i(\tau)\right) \leq \mu_i - \frac{\freecapacity_i}{2},\forall i,\forall \tau \in \left[w(t) + 1, t\right]\right\}.
\end{align*}
Then there exist a constant $k_2$ and $t_0$ such that for all $t\geq t_0$,
\begin{equation*}
    \probability\left(\left(\eventone\right)^c\right)\leq \numofserver\left(\frac{1}{t^{7}} + \frac{k_2 + 1}{t^3}\right).
\end{equation*}
\end{restatable}
Based on Lemma~\ref{lem:p_wt}, Lemmas~\ref{lem:final_busy_period} and \ref{lem:tight_bound_p} can be proven through the following outline.  The bound on the busy period in Lemma~\ref{lem:final_busy_period} relies on the property that $\expectation\left(\arrival_i(\tau) \mmiddle \hat{p}_i(\tau)\right) \leq \mu_i - \frac{\freecapacity_i}{2}$ in the event $\eventonetwo(t)$, which leads to a negative drift in the queue length.  For the bound on the estimation error in Lemma~\ref{lem:tight_bound_p}, the property that $\expectation\left(\arrival_i(\tau) \mmiddle \hat{p}_i(\tau)\right)\ge \frac{\lambda p_i^*}{2}$ in the event $\eventonetwo$ guarantees that the expected number of jobs we dispatch to each server in the optimal support set is at least linear in time, resulting in enough samples for estimating the service rates of these servers.  For servers outside of the optimal support set, event $\eventoneone$ ensures that we do not dispatch exploitation jobs to those servers.

%% file: newversion/simulation.tex
\section{Simulation Results}
\label{sec:simulation}

In this section we compare the expected regret of our proposed $K \ln t/t$-exploration policy with three other policies that are also based on multi-armed bandits: (i) an $\epsilon_t$-exploration with a faster decaying exploration probability $\epsilon_t=\numofserver/t$, (ii) a variant of the upper confidence bound (UCB) policy \citep{auer2002finite}, and (iii) a variant of Thompson sampling \citep{thompson1933likelihood, agrawalgoyal2012thompson}, described in more detail below. Our simulation set-up consists of a system of $6$ servers with service rates $\mu_i$ such that $\mu_i = 2^{i-1} \mu_1$, and $\sum_{i=1}^{6} \mu_i = 0.99$. We consider $5$ different job arrival rates $\lambda = 0.1, 0.2, 0.4, 0.5 \text{ and } 0.7$. To compute the regret, we find the cumulative queue length $\sum_{\tau=1}^{t} \sum_{i=1}^{\numofserver} \queue_i(\tau)$ for $t \in [0,2\times 10^7]$ for each of the policies and the optimal weighted random routing policy. The regret $\regret_{\policy \policy^*}(t) = \sum_{\tau = 1}^t \sum_{i = 1}^{\numofserver}( \queue_i(\tau) - \queue_i^{*}(\tau))$ is then averaged over $20000$ simulation runs.
%
%
We compare the regret of our proposed policy with that of the three other policies in \Cref{fig:regret_policies}.

\vspace{0.1cm} \noindent  \textbf{$\numofserver/t$-exploration:} Instead of the $\epsilon_t = \numofserver \ln t/t$ probability of exploration used in our proposed policy, this policy sets $\epsilon_t = K/t$, which decays much faster. Because of the aggressive exploitation, this policy can exclude servers from the optimal support set, similar to the situation described in \Cref{subsec:exploration}. As a result, we observe linearly increasing regret in \Cref{fig:regret_policies} and it is clearly outperformed by our proposed $\numofserver \ln t /t$-exploration policy, especially for small $\lambda$. For larger $\lambda$, only a small amount of exploration is required to ensure that none of the servers in the optimal support set is excluded and hence the performance of the policy improves. 
    
    
\vspace{0.1cm} \noindent  \textbf{Upper Confidence Bound (UCB) variant}: This is a variant of the UCB policy \citep{auer2002finite}, where in each time slot, we compute the routing probability vector $f(\lambda, \bm{\mu}^{UCB}(t))$ using optimistic estimates of the service rates $\mu_i^{UCB}(t) = \hat{\mu}_i(t) + \frac{1}{\sqrt{N_i(t)}}$, where $N_i(t)$ is the number of jobs that have departed from server $i$ till time $t$. Using optimistic service rate estimates induces more exploration of slower servers by including them in the support set more often. 
As a result, UCB explores more aggressively than our proposed $\numofserver \ln t /t$-exploration policy and therefore, UCB performs well for small $\lambda$. However, as $\lambda$ increases, the additional exploration results in a higher regret. 
    
   
\vspace{0.1cm} \noindent  \textbf{Thompson Sampling (TS) variant}: This is a variant of the Thompson sampling \citep{thompson1933likelihood, agrawalgoyal2012thompson}. At each time slot, we compute the routing probability vector $f(\lambda, \bm{\mu}^{TS}(t))$ by sampling the service rates $\mu_i^{TS}(t)$ from a Beta distribution with parameters $\hat{\mu}_i(t)N_i(t) + 1$ and $(1 - \hat{\mu}_i(t))N_i(t) + 1$. The variance of the Beta distribution is roughly $O(1/N_i(t))$. The exploration in this policy comes from the fact $\mu_i^{TS}(t)$ lies within $\hat{\mu}_i(t) \pm O(1/\sqrt{N_i(t)})$ region. While similar to UCB, TS performs less exploration of slow servers because $\mu_i^{TS}(t)$ can be lower than the optimistic estimates $\hat{\mu}_i(t)$. Thus, we observe in \Cref{fig:regret_policies} that the regret of TS is similar to, but better than UCB.


A common trend in these results is that 
we need more exploration in the low $\lambda$ regime and less exploration for larger $\lambda$. In the low $\lambda$ regime, the optimal weighted random routing usually sends the job to the fastest server, which essentially reduces to a typical MAB setting. Hence, UCB and TS perform well in very low $\lambda$ regime. However, their performance worsens as $\lambda$ increases due to over-exploration. Unlike traditional MAB problems where the user either explores or exploits at each time, in queueing bandits every exploitation also acts as an exploration. As long as the servers in the optimal support set has a non-zero probability of assignment associated with it, there would be a steady flow of jobs to those servers which in turn will improve their service rate estimates. 



%% file: newversion/conclusion.tex
\section{Concluding Remarks}
In this paper, we study the problem of job dispatching policies in a system with unknown service rates and unknown queue length information. We propose a bandit-based $K \ln t/t$-exploration policy, which uses online estimate of the service rates to dispatch jobs, and asymptotically converges to the optimal weighted random (OWR) routing policy. We characterize the finite-time regret of this policy and present simulation results to demonstrate that it performs well in all load regimes. 

There are substantial open directions for future work. 
An immediate open challenge is to prove a matching lower bound on the regret. Unlike typical bandit problem where every wrong decision incurs a penalty, characterization of this penalty is difficult in our queueing setting. 
Another open direction is extending the work to characterize of regret for classes of policies that has access to the queue length information like JSQ, SED etc. The analysis of these policies is far more complicated than random routing policies, where the difference in the queue length can be characterized by the difference in the routing probabilities.


%% file: newversion/appendix.tex
\appendix
\addtocontents{toc}{\protect\contentsline{chapter}{Appendix:}{}}
\section{Constants and Standard results}
We define the constant $t_0$ below. 
\begin{align}
        t_0  = \inf_\tau &\Bigg\{\tau \in \mathbb{R}:\frac{\lambda \ln w(\tau)}{w(\tau)} \leq \min_i\frac{\freecapacity_i}{4},\label{condition_t0:arrival_condition_on_probab}\\
        &\qquad \qquad \min_i\left\{\frac{\mu_i w(\tau)}{2} - \frac{4\ln \tau}{c_g \Delta_0^2} - \sqrt{8\mu_i w(\tau)\ln \tau}\right\} \geq 0 ,\label{condition_t0:lemma_p_wt}\\
        &\qquad \qquad \min_i \freecapacity_{i} w(\tau) \geq 48 \ln \tau, \label{condition_t0:lemma_queue_length_bound}\\
        &\qquad \qquad \frac{k_{1}}{c}\sqrt{\frac{ \ln \tau}{\tau}} \leq \min\left\{\min_{i:p_i^*> 0} p_i,  \Delta\right\},\label{condition_t0:lemma_estimation_error}\\
        &\qquad \qquad w(\tau) \leq \frac{\tau}{4},\label{condition_t0:bound_w(t)}\\
        &\qquad \qquad\max_i \frac{66 \ln \tau}{\freecapacity_i^2} + \frac{6 w(\tau)}{ \freecapacity_i} \leq \frac{\tau}{2}\Bigg\} ,\label{condition_t0:bound_vi(t)}
\end{align}
where
\begin{equation}\label{eq:w(t)}
    w(\tau) = 2\exp\left(\frac{1}{\Delta_0}\sqrt{\frac{16 \ln \tau}{c_g\lambda}}\right)
\end{equation}
and $c_g$, $\Delta_0$ and $\Delta$ are constants depending on system parameters given by 
\begin{equation}\label{eq:c_g}
    c_g = \min_i\min\left\{\frac{1}{8\mu_i^2(1 - \mu_i)},  \frac{1}{6\mu_i^2(1 - \mu_i)(3 - \mu_i)}\right\},
\end{equation}
\begin{equation}\label{eq:delta0}
    \Delta_0 = \min\left\{\Delta, \min_i\{\mu_i(1 - \mu_i)\}, \frac{\min_{i:p_i^* > 0}p_i^*}{3c}\right\},
\end{equation}
\begin{equation}\label{eq:delta}
    \Delta = \min \left\{\frac{\tilde{\mu}}{2}, \frac{\sum_{j\in \support(\lambda, \bm{\mu})}\freecapacity_j}{|\support(\lambda, \bm{\mu})|}, \frac{\min_{j\in \support(\lambda, \bm{\mu})}\freecapacity_j}{4c\lambda}, \Delta_{\support}\right\},
\end{equation}
\begin{equation}
    \tilde{\mu} = \min_{j \in \support(\lambda, \bm{\mu})} \{\min \{\mu_j, 1- \mu_j\}\},
\end{equation}
\begin{equation}
    \freecapacity_{\min} = \min_i \freecapacity_i,
\end{equation}
\begin{equation}
  c = \max\left\{\frac{1}{\lambda}\left(1 + 4\frac{ \sum_{j\in \support(\lambda, \bm{\mu})}\freecapacity_i}{\tilde{\mu}} + \lvert \support(\lambda, \bm{\mu})\rvert \right), \frac{1}{\lambda} + \frac{30 \sum_{j\in \support(\lambda, \bm{\mu})}\freecapacity_j}{\lambda \tilde{\mu}} + \frac{16\lvert \support(\lambda, \bm{\mu})\rvert}{\lambda}\right\}
\end{equation}
and 
\begin{equation}
\Delta_{\support} := \sup\left\{\delta \geq 0\colon \support\left(\lambda,\bm{\mu}'\right) = \support(\lambda,\bm{\mu}), \forall\bm{\mu}'\mathrm{ s.t.\ } \left|\mu'_i-\mu_i\right|\leq \delta,\forall i\right\}
\end{equation}
The requirement of \eqref{condition_t0:arrival_condition_on_probab}-\eqref{condition_t0:bound_w(t)} is explained in the proofs of the lemmas. \eqref{condition_t0:bound_vi(t)} is to ensure that for that any $t\geq t_0$, $v_i(t) \leq t/2$, which is used in the proof of \Cref{thm:main}. Below, we state some standard results that we will use in the proofs.
\begin{itemize}
    \item \textbf{Chernoff's Inequality}: For i.i.d.\ Bernoulli random variables $X_1, X_2, \cdots, X_n$, and $\delta\in (0, 1)$
     \begin{equation}\label{ineq:Chernoff}
         \probability\left(\sum_iX_i \leq (1- \delta)\expectation\left[\sum_i X_i\right]\right) \leq \exp \left(-\frac{\delta^2\expectation\left[\sum_i X_i\right]}{2}\right)
     \end{equation}
     \item \textbf{Hoeffding's Inequality}: For i.i.d. random variables $X_1, X_2, \cdots, X_n$ such that $a \leq X_i \leq b$ for all $i$, we have, 
        \begin{equation}\label{ineq:Hoeffding}
            \probability\left(\sum_{i=1}^n X_i - \expectation\left(\sum_{i=1}^n X_i\right) \leq  - t\right) \leq \exp\left(-\frac{2t^2}{n(b-a)^2}\right).
        \end{equation}
\end{itemize}
\section{Proof of the Optimal Weighted Random Routing Policy} \label{prf_optimal_p}
The Lagrangian of the optimization problem is given by 
\begin{equation}
    \mathcal{L} = \sum_i \frac{\lambda p_i(1 - \mu_i)}{\mu_i - \lambda p_i} + a\left(1 - \sum_i p_i\right) + \sum_i b_i\left(\lambda p_i - \mu_i\right) - \sum_i c_ip_i.
\end{equation}
Now 
\begin{equation}\label{eq:langragian}
    \frac{d\mathcal{L}}{d p_i} = \frac{\lambda \mu_i (1-\mu_i)}{(\mu_i - \lambda p_i)^2} - a + b_i\lambda - c_i
\end{equation}
Let $p^*\left(\lambda, \bm{\mu}\right) = \left(p_1^*, p_2^*, \cdots, p_\numofserver^*\right)$ be the optimal primal solution which is also the optimal routing vector. Let $a^*, \bm{b}^* = (b_1^*, b_2^*, \cdots, b_\numofserver^*)$ and $\bm{c}^* = (c_1^*, c_2^*, \cdots, c_\numofserver^*)$ be the optimal dual solution. Since, objective function is convex, inequality constraints are convex and equality constraint is affine, the dual gap is zero. From the complementary slackness,
\begin{equation}
    c_i^* p_i^* = 0,
\end{equation}
\begin{equation}
    b_i^* (\lambda p_i^* - \mu_i) = 0.
\end{equation}
Since, $\lambda p_i^* < \mu_i$ for all $i$, $b_i^* = 0$ for all $i$. As defined already, $\support(\lambda, \bm{\mu})$ is the optimal support set, i.e., $i \in \support(\lambda, \bm{\mu})$ if and only if $p_i^* > 0$. Then, for all $i \in \support(\lambda, \bm{\mu})$, we have
\begin{equation}
    \frac{\lambda \mu_i (1-\mu_i)}{(\mu_i - \lambda p_i^*)^2} - a^*  = 0
\end{equation}
Simplifying the expression, we get
\begin{equation}\label{rel:prf_opt_dist:1}
    \mu_i - \lambda p_i^* = a's_i
\end{equation}
where $a' = \sqrt{\frac{\lambda}{a^*}}$, $s_i = \sqrt{\mu_i(1-\mu_i)}$. Summing the above expression for all $i\in \support(\lambda, \bm{\mu})$, we get
\begin{equation}\label{rel:prf_opt_dist:2}
    \sum_{j\in\support(\lambda, \bm{\mu})} \mu_j - \lambda = a'\left(\sum_{j \in \support(\lambda, \bm{\mu})}s_j\right).
\end{equation}
Using the value of $a'$ from \eqref{rel:prf_opt_dist:1} and \eqref{rel:prf_opt_dist:2}, for any $i\in \support(\lambda, \bm{\mu})$
\begin{equation}
    p_i^* = \frac{\mu_i}{\lambda} -  \frac{s_i}{\sum_{j\in\support(\lambda, \bm{\mu})} s_j}\left(\frac{\sum_{j\in \support(\lambda, \bm{\mu})}\mu_j}{\lambda} - 1\right).
\end{equation}
Hence, the task to find the optimal routing vector essentially boils down to finding the optimal support set. We will introduce the definition of valid support set and two lemmas below. 
\begin{definition}\label{propty:support_positive}
A set is a valid support set if the routing vector calculated using \eqref{optimal_p} for the set satisfies the constraints of the optimization problem and the corresponding routing probability for any server in the set is strictly positive.
\end{definition}
\begin{lemma}\label{lem:support_monotonic}
If $\mu_i \geq \mu_j$, then $p_i^* \geq p_j^*$.
\end{lemma}
\begin{lemma}\label{lem:support_maximal}
There exists no valid support set $V \supset \support(\lambda, \bm{\mu})$.
\end{lemma}
We will now prove the correctness of our algorithm using the two lemmas mentioned above. Without loss of generality, assume
\begin{equation}
    \mu_1 \geq \mu_2 \geq \cdots\geq \mu_{\numofserver}.
\end{equation}
Define the sets $V_1, V_2, \cdots, V_{\numofserver}$ such that $i\leq j$ implies $i \in V_j$. Observe that $V_i$ for $i\in [1, \numofserver]$ are the only possible sets that can be both valid set and satisfy \Cref{lem:support_monotonic}. Hence, $\support(\lambda, \bm{\mu})$ has to be one of the $V_i$'s. Assume that $\support(\lambda, \bm{\mu}) = V_q$. We want to argue that algorithm to find $\support(\lambda, \bm{\mu})$ converges to $V_q$. Since $V_q$ is a valid set, clearly the algorithm will not converge to any $V_{q'}$ where $q' < q$. To prove the validity of our algorithm, it is enough to show that for all $q'' > q$, either the routing vector calculated using $V_{q''}$ is not non-negative or some of the routing probabilities corresponding to the servers in the support set is zero. By \Cref{lem:support_maximal}, for any $q'' > q$, $V_{q''}$ is not a valid set. One way this is possible is if the routing vector corresponding to the servers in the support set is zero which is handled by the algorithm. The only other way for the set to be not valid is if it violates the optimization constraints. There are mainly three constraints for a set to be valid. The first constraint is that the sum of routing probabilities $\sum_{i = 1}^{\numofserver} p_i^*$ should equal $1$ which is always satisfied because of $\eqref{optimal_p}$. The second constraint ensures that for any $i$, $\mu_i - \lambda p_i^*$ should be positive. Clearly this is true for any $i$ which is not in the support set. For any $i \in V_{q''}$, we have
\begin{align}
    \mu_i - \lambda p_i^* = \frac{s_i}{\sum_{j\in V_{q''}}s_j}\left(\sum_{j \in V_{q''}}\mu_j - \lambda\right) \geq 0, 
\end{align}
where the last step follows from the fact that $\left(\sum_{j \in V_{q''}}\mu_j - \lambda\right) \geq \left(\sum_{j \in \support(\lambda, \bm{\mu})}\mu_j - \lambda\right) \geq 0$. Hence, the only other way $V_{q''}$ could not be a valid set, is if it violates the third constraint which ensures that the routing probabilities are non-negative.

Next we will prove the above lemmas.
\begin{enumerate}
    \item 
    \begin{proof}[
    Proof of \Cref{lem:support_monotonic}]
    Consider a feasible routing vector $\bm{p}^A = (p_1, p_2, \cdots, p_\numofserver)$ such that there exists $i, j$ such that $p_i< p_j$ and $\mu_i> \mu_j$. Consider another feasible routing vector $\bm{p}^B$ with the exact same routing probabilities as $\bm{p}^A$ except the $p_i, p_j$ are interchanged, i.e., $\bm{p}_i^A = \bm{p}_j^B$ and $\bm{p}_i^B = \bm{p}_j^A$. Now the difference of the value of objective function is given by, 
    \begin{align}
        &\expectation\left[\queue_A\right] - \expectation\left[\queue_B\right] \\
        =& \frac{\lambda p_i (1- \mu_i)}{\mu_i - \lambda p_i} + \frac{\lambda p_j (1- \mu_j)}{\mu_j - \lambda p_j} - \frac{\lambda p_j (1- \mu_i)}{\mu_i - \lambda p_j} - \frac{\lambda p_i (1- \mu_j)}{\mu_j - \lambda p_i}\\
        =&\lambda p_j\frac{(\mu_i - \mu_j)(1-\lambda p_j)}{(\mu_j - \lambda p_j)(\mu_i - \lambda p_j)} + \lambda p_i\frac{(\mu_j - \mu_i)(1-\lambda p_i)}{(\mu_i - \lambda p_i)(\mu_j - \lambda p_i)}\\
        =&\lambda \left(\mu_i - \mu_j\right) \left(p_j - p_i\right)\left(\frac{\mu_i\mu_j + \lambda^2 p_i p_j(\mu_i + \mu_j) - \lambda^2 p_ip_j - \lambda\mu_i\mu_j(p_i + p_j)}{(\mu_i - \lambda p_i)(\mu_j - \lambda p_i)(\mu_j - \lambda p_j)(\mu_i - \lambda p_j)}\right)\\
        =& \frac{\lambda\left(\mu_i - \mu_j\right) \left(p_j - p_i\right)}{(\mu_i - \lambda p_i)(\mu_j - \lambda p_i)(\mu_j - \lambda p_j)(\mu_i - \lambda p_j)}\Big(\left(\mu_j - \lambda p_j\right)\lambda p_i\left(1- \lambda p_i\right) + \left(\mu_i - \lambda p_i\right)\lambda p_j\left(1- \lambda p_j\right) + \nonumber\\
        &\qquad + \left(\mu_i - \lambda p_i\right)\left(\mu_j - \lambda p_j\right)\left(1 - \lambda p_i - \lambda p_j\right)\Big)\\
        \geq& 0
    \end{align}
    where the last inequality is non-negative since all the terms in the expression are non-negative. This implies that if $\mu_i \geq \mu_j$, then $p_i^* \geq p_j^*$.
    \end{proof}
    \item \begin{proof}[Proof of \Cref{lem:support_maximal}]
    Let $V_m$ and $V_{m+n}$ be valid support sets. We want to argue that $V_m$ cannot be the optimal support set. Assume that $V_m$ is the optimal support set. Since the solution is optimal, the dual variables corresponding to optimal support set $V_m$ has to be non-negative.
    \begin{align}
        a^* & = \frac{\lambda s_1^2}{(\mu_1 - \lambda p_1)}\\
        & = \frac{\lambda s_1^2}{\left(\frac{s_1}{\sum_{j = 1}^m s_j}\left(\sum_{j = 1}^m \mu_j - \lambda\right)\right)^2}\\
        & = \lambda\left(\frac{ \sum_{j = 1}^m s_j}{\sum_{j = 1}^m \mu_j - \lambda}\right)^2,
    \end{align}
    which is always non-negative. Since, $V_m$ is a valid set, $b_i^* = 0$ for all $i \in \{1, 2, 3, \cdots, \numofserver\}$. Using this along with \eqref{eq:langragian}, we have
    \begin{equation}\label{eq:kkt}
        \frac{\lambda \mu_i (1-\mu_i)}{(\mu_i - \lambda p_i^*)^2} - a^* - c_i^* = 0
    \end{equation}
    Since, dual variables has to be non-negative, for all $i>m$, 
    \begin{equation}\label{eq:dual_non_negative}
        c_i^* \geq 0.
    \end{equation}
    Using \eqref{eq:kkt} and \eqref{eq:dual_non_negative}, we have
    \begin{align}
        a^* & \leq \frac{\lambda \mu_i (1-\mu_i)}{(\mu_i)^2}\\
        \lambda\left(\frac{ \sum_{j = 1}^m s_j}{\sum_{j = 1}^m \mu_j - \lambda}\right)^2 & \leq \frac{\lambda \mu_i (1-\mu_i)}{(\mu_i)^2} = \lambda \frac{s_i^2}{\mu_i^2}\\
        \frac{ \sum_{j = 1}^m s_j}{\sum_{j = 1}^m \mu_j - \lambda} &\leq \frac{s_i}{\mu_i}\label{rel:proof_alg:lem2:1}
    \end{align}
    Using \eqref{rel:proof_alg:lem2:1} with the property that $s_i/\mu_i$ is monotonically increasing in $i$, we have 
    \begin{equation}\label{eq:alg:monotone:1}
       \frac{ \sum_{j = 1}^m s_j}{\sum_{j = 1}^m \mu_j - \lambda} \leq \frac{s_{m + 1}}{\mu_{m + 1}} \leq \frac{s_{m + 2}}{\mu_{m + 2}}\leq \cdots \leq \frac{s_{m + n}}{\mu_{m + n}}.
    \end{equation}
    Hence, 
    \begin{align}
        \frac{ \sum_{j = 1}^{m + n} s_j}{\sum_{j = 1}^{m + n} \mu_j - \lambda} &= \frac{ \sum_{j = m + 1}^{m + n}s_{j} + \sum_{j = 1}^{m } s_j}{ \sum_{j = m + 1}^{m + n}\mu_{j} + \sum_{j = 1}^{m} \mu_j - \lambda}\\
        &\leq \frac{ \sum_{j = m + 1}^{m + n}\frac{s_{m + n}}{\mu_{m + n}}\mu_{j} + \frac{s_{m + n}}{\mu_{m + n}} \left(\sum_{j = 1}^{m} \mu_j - \lambda\right)}{ \sum_{j = m + 1}^{m + n}\mu_{j} + \sum_{j = 1}^{m} \mu_j - \lambda}\\
        & = \frac{s_{m + n}}{\mu_{m + n}}.
    \end{align}
    This implies 
    \begin{equation}
        \frac{\mu_{m + n}}{\lambda} - \frac{s_{m + n}}{\sum_{j = 1}^{m + n} s_j}\left(\frac{\sum_{j = 1}^{m + n} \mu_j - \lambda}{\lambda}\right) \leq 0.
    \end{equation}
    But this is the routing probability to server $(m + n)$ if the support set is $V_{m + n}$. Since the routing probability is either negative or zero, $V_{m + n}$ is not a valid support set, which is a contradiction. This implies $V_m$ cannot be the optimal support set.
    
    
    
    \end{proof}
\end{enumerate}

\section{Proofs of Lemmas required for Theorem \ref{thm:main}}
\subsection{Proof of Lemma \ref{lem:mismatch}}\label{prf_lem:mismatch}
\begin{proof}
When $\exploit_i(t) = 1$ and $\arrival_i^*(t) = 0$, the $\epsilon_t$-exploration policy must have chosen to exploit and generated a dispatching decision $\dispdecision(t)=i$ and the optimal weighted random routing must have generated a dispatching decision $\dispdecision^*(t)\neq i$.
Therefore,
\begin{align}
\probability\left[\exploit_i(t) = 1,  \arrival_i^*(t) = 0 \Bigm| \hat{p}_i(t)\right] & = \probability\left[\explorerandom(t)=0,\dispdecision(t)=i,\dispdecision^*(t)\neq i\Bigm| \hat{p}_i(t)\right] \\
&\le \probability\left[\dispdecision(t)=i,\dispdecision^*(t)\neq i\Bigm| \hat{p}_i(t)\right] \\
&=\frac{\hat{p}_i(t)-\min\{\hat{p}_i(t),p^*_i\}}{d_{TV}(\hat{\bm{p}}(t), \bm{p}^*)}\sum_{j\ne i}\left(p^*_j(t)-\min\left\{\hat{p}_j(t),p^*_j\right\}\right) \\
& = \hat{p}_i(t)-\min\{\hat{p}_i(t),p^*_i\} \label{eqn:p_mismatch_lemma_4} \\
& \le \left|\hat{p}_i(t) - p_i^*\right|, 
\end{align}
where in \eqref{eqn:p_mismatch_lemma_4} we use the fact that the total variation distance $d_{TV}(\hat{\bm{p}}(t), \bm{p}^*) =\sum_{j=1}^K\left(p^*_j(t)-\min\left\{\hat{p}_j(t),p^*_j\right\}\right)$.
\end{proof}
\subsection{Proof of Lemma \ref{lem:inst-regret}}\label{prf_lem:inst-regret}
Recall from \eqref{rel:regret_equation}, we have
\begin{equation}\label{rel:regret_equation2}
    \queue_i(t) - \queue_i^*(t)
    \leq \sum_{\ell = t- \busy_i(t)}^{t-1} \explore_i(\ell) + \sum_{\ell = t- \busy_i(t)}^{t-1} \indicatorofevents_{ \left\{ \exploit_i(\ell)=1, \arrival_i^*(\ell) = 0\right\}}.
\end{equation}
We first utilize the event $\eventfive$ to further upper bound the queue length difference based on the upper bound in \eqref{rel:regret_equation2}. Specifically, recall that $\eventfive$ is the event where  $B_i(t)\le v_i(t)=\frac{66\ln t}{\freecapacity_{i}^2}$ for all server $i$.  Then based on \eqref{rel:regret_equation2}, we have
\begin{align}
    \sum_{i=1}^K\expectation\left[\queue_i(t) - \queue_i^*(t)\right]&=\expectation\left[\left(\sum_{i=1}^K\left(\queue_i(t) - \queue_i^*(t)\right)\right)\cdot (\indicatorofevents_{\eventfive} + \indicatorofevents_{\left(\eventfive\right)^c} ) \right]\\
    &\le \expectation\left[\left(\sum_{i=1}^K\sum_{\ell = t- \busy_i(t)}^{t-1} \explore_i(\ell)\right)\cdot\indicatorofevents_{\eventfive}\right] + \expectation\left[\sum_{i=1}^K\left(\sum_{\ell = t- \busy_i(t)}^{t-1} \indicatorofevents_{\left\{ \exploit_i(\ell)=1, \arrival_i^*(\ell) = 0\right\}}\right)\cdot\indicatorofevents_{\eventfive}\right] + t\probability\left(\left(\eventfive\right)^c\right)\label{eq:bound-eventfive-temp1}\\
    &\le \expectation\left[\sum_{i=1}^K\sum_{\ell = t- v_i(t)}^{t-1} \explore_i(\ell)\right] + \expectation\left[\sum_{i=1}^K\sum_{\ell = t- v_i(t)}^{t-1} \indicatorofevents_{\left\{ \exploit_i(\ell)=1, \arrival_i^*(\ell) = 0\right\}}\right]+t\probability\left(\left(\eventfive\right)^c\right)\label{eq:bound-eventfive-temp2}\\
    &\le \sum_{i=1}^K\frac{2v_i(t)\ln t}{t}+\sum_{i=1}^K\sum_{\ell = t- v_i(t)}^{t-1}\probability\left(\exploit_i(\ell)=1, \arrival_i^*(\ell) = 0\right) + t\probability\left(\left(\eventfive\right)^c\right),\label{eq:bound-eventfive-temp3}
\end{align}
where \eqref{eq:bound-eventfive-temp1} uses \eqref{rel:regret_equation2} and the fact that $\sum_{i=1}^K\left(\queue_i(t)-\queue^*_i(t)\right)\le t$ since there are at most $t$ arrivals before time $t$; \eqref{eq:bound-eventfive-temp2} is due to the property $\busy_i(t)\le v_i(t)$ given by the indicator $\indicatorofevents_{\eventfive}$; and \eqref{eq:bound-eventfive-temp3} is because $v_i(t)\le t/2$ for large enough $t$.

Next, it suffices to bound $\probability\left(\exploit_i(\ell)=1, \arrival_i^*(\ell) = 0\right)$ using event $\eventsix$. Recall $\eventsix$ is the event where the estimation error $\lvert\hat{p}_i(\tau) - p_i^*\rvert \leq k_1\min\{p_i^*, \sqrt{\ln t/t}\}$ for any $\tau \in [t/2, t]$. Using Lemma~\ref{lem:mismatch} for $\hat{\bm{p}}(t)=f(\lambda, \hat{\bm{\mu}(t)})$, the routing probability vector computed from the estimated service rates, we have
\begin{align}
\sum_{i=1}^K\sum_{\ell = t- v_i(t)}^{t-1}\probability\left(\exploit_i(\ell)=1, \arrival_i^*(\ell) = 0\right) &=\sum_{i=1}^K\sum_{\ell = t- v_i(t)}^{t-1}\expectation\left[\probability\left(\exploit_i(\ell)=1, \arrival_i^*(\ell) = 0\mmiddle\hat{p}_i(t)\right)\right]\\
&\le \sum_{i=1}^K\sum_{\ell = t- v_i(t)}^{t-1}\expectation\left[\left|\hat{p}_i(t)-p_i^*\right|\right]\label{eq:bound-eventsix-temp1}\\
&=\expectation\left[\left(\sum_{i=1}^K\sum_{\ell = t- v_i(t)}^{t-1}\left|\hat{p}_i(t)-p_i^*\right|\right)\cdot\indicatorofevents_{\eventsix}\right] + \expectation\left[\left(\sum_{i=1}^K\sum_{\ell = t- v_i(t)}^{t-1}\left|\hat{p}_i(t)-p_i^*\right|\right)\cdot\indicatorofevents_{\left(\eventsix\right)^c}\right]\\
&\le \sum_{i:p_i^*>0} k_1v_i(t) \sqrt{\frac{\ln t}{t}} + 2t\probability\left(\left(\eventsix\right)^c\right),\label{eq:bound-eventsix-temp2}
\end{align}
where \eqref{eq:bound-eventsix-temp1} is due to Lemma~\ref{lem:mismatch}; \eqref{eq:bound-eventsix-temp2} uses the definition of the event $\eventsix$, the fact that $v_i(t)\le t/2$ for large enough $t$, and the fact that $\sum_{i=1}^K\left|\hat{p}_i(t)-p_i^*\right|\le 2$ for the second summand.  Inserting \eqref{eq:bound-eventsix-temp2} back to \eqref{eq:bound-eventfive-temp3} completes the proof.\qed
\subsection{Proofs of preliminary results (Lemma \ref{lem:conc_geoemtric} and Lemma \ref{lem:relation_p_mu})}
From the service times of the departed jobs, the system can learn the $\mu_i$'s. Since $\expectation\left(\hat{\mu}_i(t)\right) \neq \mu_i$ and geometric random variable is unbounded, standard Hoeffding bounds cannot be used to bound the estimation error. \Cref{lem:conc_geoemtric} provides a relation to bound the error in the estimate using Chernoff bound. We also need to bound the error in the estimate of $p_i^*$. \Cref{lem:relation_p_mu} provides a relation between the estimation error of $|\hat{\mu}_i(\tau) - \mu_i|$ and $|\hat{p}_i(\tau) - p_i^*|$, for all $i$. 
\begin{lem}
\label{lem:conc_geoemtric}
For any $\delta \in [0, \mu_i(1 - \mu_i)]$, $n$ and the estimate $\mu_i^{(n)}$ of $\mu_i$ using $n$ i.i.d samples, ,
\begin{equation}
    \probability\left(\left|\mu_i^{(n)} - \mu_i\right| \geq \delta\right) \leq  \exp\left(-n c_{g} \delta^2\right),
\end{equation}
where $c_g = \min_i\min\left\{\frac{1}{8\mu_i^2(1 - \mu_i)},  \frac{1}{6\mu_i^2(1 - \mu_i)(3 - \mu_i)}\right\}$.
\end{lem}
\begin{proof}[Proof of \Cref{lem:conc_geoemtric}]\label{prf_lem:conc_geoemtric}
Let $X_{i1}, X_{i2}, \cdots, X_{in}$ be the $n$ i.i.d samples of a geometric random variable with parameter $\mu_i$. Then the estimate $\mu_i^{(n)}$ can be given as
\begin{equation}
    \mu_i^{(n)} = \frac{n}{\sum_{j = 1}^n X_{ij}}.
\end{equation}
Now, $\left|\frac{\sum_{j = 1}^n X_{ij}}{n} - \frac{1}{\mu_i} \right| \leq \delta'$ implies  
\begin{align}
    \frac{1}{\mu_i} - \delta'&\leq \frac{\sum_{j = 1}^n X_{ij}}{n} \leq \frac{1}{\mu_i} + \delta'\\
    \frac{\mu_i}{1+\delta'\mu_i} &\leq \mu_i^{(n)} \leq \frac{\mu_i}{1-\delta'\mu_i}\\
    \frac{-\delta'\mu_i^2}{1+\delta'\mu_i} &\leq \mu_i^{(n)} - \mu_i \leq \frac{\delta'\mu_i^2}{1-\delta'\mu_i}
\end{align}
implies $\left|\mu_i^{(n)} - \mu_i\right| \leq 2 \delta' \mu_i^2$ for any $ 0 \leq \delta' \leq 1/(2\mu_i)$. We want to prove that 
\begin{equation}
    \probability\left(\left|\frac{\sum_{j = 1}^n X_{ij}}{n} - \frac{1}{\mu_i}\right| \geq \delta'\right) \leq  \exp\left(-n c_{g}' \delta'^2\right).
\end{equation}
Then, for any $\delta \leq \mu_i$
\begin{equation}\label{eq:prf_conc_geom:rel1}
     \probability\left(\left|\mu_i^{(n)} - \mu_i\right| \geq \delta\right)  \leq \probability\left(\left|\frac{\sum_{j = 1}^n X_{ij}}{n} - \frac{1}{\mu_i}\right| \geq \frac{\delta}{2\mu_i^2}\right) \leq \exp\left( - n\frac{c_g'}{4\mu_i^4}\delta^2\right).
\end{equation}
Hence, to prove the lemma, we will first prove that
\begin{equation}
    \probability\left(\left|\frac{\sum_{j = 1}^n X_{ij}}{n} - \frac{1}{\mu_i}\right| \geq \delta'\right) \leq  \exp\left(-n c_{g}' \delta'^2\right).
\end{equation}
We will bound the upper confidence and lower confidence interval now. 
\begin{enumerate}
    \item \textbf{Bounding the upper confidence interval}:\\
    Now, for any $s$ such that $0 \leq s \leq -\ln(1 - \mu_i)$
        \begin{align}
        \probability\left(\frac{\sum_{j = 1}^n X_{ij}}{n} \geq \frac{1}{\mu_i} + \delta'\right) & \leq \left(\frac{\expectation e^{sX_{i1}}}{e^{s(\frac{1}{\mu_i} + \delta')}}\right)^n\\
        &= \left(\frac{\frac{\mu_i e^s}{1-(1-\mu_i)e^s}}{e^{s(\frac{1}{\mu_i} + \delta')}}\right)^n\\
        &= \exp\left(n\left(\ln \mu_i +s - \ln(1-(1-\mu_i)e^s)- s\left(\frac{1}{\mu_i} + \delta'\right)\right)\right).
    \end{align}
    Since, the above function is true for any $s \in [0, -\ln(1 - \mu_i)]$, we will try to find the $s$ that minimizes the probability. Now, define the function $f$ such that
    \begin{equation}
        f(s) = \ln \mu_i +s - \ln(1-(1-\mu_i)e^s)- s\left(\frac{1}{\mu_i} + \delta'\right).
    \end{equation}
    \begin{equation}
        f'(s) = 1 +\frac{(1-\mu_i)e^s}{1-(1-\mu_i)e^s}- \left(\frac{1}{\mu_i} + \delta'\right) = \frac{1}{1-(1-\mu_i)e^s}- \left(\frac{1}{\mu_i} + \delta'\right).
    \end{equation}
    \begin{equation}
        f''(s) = \frac{(1-\mu_i)e^s}{\left(1-(1-\mu_i)e^s\right)^2}\geq 0.
    \end{equation}
    Hence, $f(s)$ is convex w.r.t $s$. This implies that the minima of the function is at the point $s = s^*$ where derivative is $f'(s^*) = 0$. Thus
    \begin{align}
         &\frac{1}{1-(1-\mu_i)e^{s^*}}- \left(\frac{1}{\mu_i} + \delta'\right) =0,\\
         e^{s^*} &= \frac{1+\mu_i \delta'-\mu_i}{1+\mu_i \delta'-\mu_i-\mu_i^2\delta'} \leq \frac{1}{1-\mu_i},
    \end{align}
    where the last inequality satisfies the condition that $s^* \in [0, -\ln(1 - \mu_i)]$. Substituting the value of $s^*$ in $f(s)$, we get
    \begin{equation}
        f(s^*) = \ln \mu_i + \ln \left(\frac{1 + \mu_i\delta' - \mu_i}{\mu_i(1 - \mu_i)}\right) - \left(\frac{1}{\mu_i} + \delta'\right)\ln \left(\frac{1+\mu_i \delta'-\mu_i}{1+\mu_i \delta'-\mu_i-\mu_i^2\delta'}\right)
    \end{equation}
    Now, we want to argue that $\exists c_{g_i^+} > 0$ such that $f(s^*)\leq -c_{g_i^+} \delta'^2$. Consider the function 
    \begin{equation}
        h(x) = \ln \mu_i + \ln \left(\frac{1 + \mu_ix - \mu_i}{\mu_i(1 - \mu_i)}\right) - \left(\frac{1}{\mu_i} + x\right)\ln \left(\frac{1+\mu_i x-\mu_i}{1+\mu_i x-\mu_i-\mu_i^2x}\right) + c_{g_i^+}x^2
    \end{equation} 
    Now,
    \begin{align}
        h'(x) &=  \frac{\mu_i}{1 - \mu_i + \mu_i x} - \frac{1 + \mu_i x}{\mu_i}\left[\frac{\mu_i}{1 - \mu_i + \mu_i x} - \frac{\mu_i - \mu_i^2}{1+\mu_i x-\mu_i-\mu_i^2x}\right] \nonumber\\
        & \qquad - \ln\left(\frac{1-\mu_i-\mu_i x}{1+\mu_i x-\mu_i-\mu_i^2x}\right) + 2c_{g_i^+}x,\\
        h''(x) &= -\frac{\mu_i^2}{(1 - \mu_i + \mu_i x)^2} - \frac{1 + \mu_i x}{\mu_i}\left[-\frac{\mu_i^2}{(1 - \mu_i + \mu_i x)^2} + \frac{(\mu_i - \mu_i^2)^2}{(1+\mu_i x-\mu_i-\mu_i^2 x)^2}\right] \nonumber\\
        & \qquad - 2\left[\frac{\mu_i}{1 - \mu_i + \mu_i x} - \frac{\mu_i - \mu_i^2}{1+\mu_i x-\mu_i-\mu_i^2x}\right] + 2c_{g_i^+}\\
        &= -\frac{\mu_i^2}{(1 - \mu_i + \mu_i x)(1 + \mu_i x)} + 2c_{g_i^+}.
    \end{align}
    Hence, for all $0 \leq x \leq \frac{1 - \mu_i}{2\mu_i}$,
    \begin{equation}
        h''(x) \leq -\frac{4\mu_i^2}{3(1 - \mu_i)(3 - \mu_i)} + 2c_{g_i^+}.
    \end{equation}
    Choosing $c_{g_i^+} = \frac{2\mu_i^2}{3(1 - \mu_i)(3 - \mu_i)}$ implies $h''(0) < 0$. Also,  $h'(0) = 0, h(0) = 0$ Hence, for all $0 \leq x\leq \frac{1-\mu_i}{2\mu_i}$, 
    \begin{equation}
        h(x) \leq 0.
    \end{equation}
    Hence, for any $\delta' \in [0, (1 - \mu_i)/(2\mu_i)]$ and $c_{g_i^+} = 2\mu_i^2/(3(1 - \mu_i)(3 - \mu_i))$
    \begin{equation}\label{rel:upper_confidence}
        \probability\left(\frac{\sum_i X_i}{n} \geq \frac{1}{\mu_i} + \delta'\right) \leq \exp\left(-nc_{g_i^+} \delta'^2\right).
    \end{equation}
\item \textbf{Bounding the lower confidence interval}:\\
    Now, for any $s$ such that $0 \leq s \leq -\ln(1 - \mu_i)$
    \begin{align}
        \probability\left(\frac{\sum_{j = 1}^n X_{ij}}{n} \leq \frac{1}{\mu_i} - \delta'\right) &\leq \left(\frac{\expectation e^{-sX_{i1}}}{e^{-s(\frac{1}{\mu_i} - \delta')}}\right)^n\\
        &\leq \left(\frac{\frac{\mu_i e^{-s}}{1-(1-\mu_i)e^{-s}}}{e^{-s(\frac{1}{\mu_i} - \delta')}}\right)^n\\
        &\leq \exp\left(n\left(\ln \mu_i -s - \ln(1-(1-\mu_i)e^{-s})+ s\left(\frac{1}{\mu_i} - \delta'\right)\right)\right).
    \end{align}
    Since, the above function is true for any $s \in [0, -\ln(1 - \mu_i)]$, we will try to find the $s$ that minimizes the probability. Now, define the function $g$ such that
    \begin{equation}
        g(s) = \ln \mu_i -s - \ln(1-(1-\mu_i)e^{-s})+ s\left(\frac{1}{\mu_i} - \delta'\right).
    \end{equation}
    \begin{equation}
        g'(s) = -1 -\frac{(1-\mu_i)e^{-s}}{1-(1-\mu_i)e^{-s}}+ \left(\frac{1}{\mu_i} - \delta'\right) = -\frac{1}{1-(1-\mu_i)e^{-s}}+ \left(\frac{1}{\mu_i} - \delta'\right).
    \end{equation}
    \begin{equation}
        g''(s) = \frac{(1-\mu_i)e^{-s}}{\left(1-(1-\mu_i)e^{-s}\right)^2}\geq 0.
    \end{equation}
    Hence, $g(s)$ is convex w.r.t $s$. This implies that the minima of the function is at the point $s = s^*$ where derivative is $g'(s^*) = 0$. Thus
    \begin{align}
         &\frac{1}{1-(1-\mu_i)e^{-s^*}}- \left(\frac{1}{\mu_i} - \delta'\right) =0,\\
         e^{-s^*} &= \frac{1-\mu_i \delta'-\mu_i}{1-\mu_i \delta'-\mu_i+\mu_i^2\delta'} \leq 1.
    \end{align}
    To ensure that the minima is valid $e^{-s^*}\geq 0$ which implies $\delta'\leq \frac{1-\mu_i}{\mu_i}$. Substituting $s^*$ in g(s), we get  
    \begin{equation}
        g(s^*) = \ln \left(1-\frac{\mu_i \delta'}{1-\mu_i}\right)- \left(\frac{1}{\mu_i} - \delta'\right)\ln \left(1 - \frac{\mu_i^2\delta'}{(1-\mu_i)(1-\mu_i \delta')}\right).
    \end{equation}
    Now, we want to argue that $\exists c_{g_i^-} > 0$ such that $g(s^*)\leq -c_{g_i^-} \delta'^2$. Consider the function 
    \begin{equation}
        h(x) = \ln \left(1-\frac{\mu_i x}{1-\mu_i}\right)- \left(\frac{1}{\mu_i} - x\right)\ln \left(1 - \frac{\mu_i^2x}{(1-\mu_i)(1-\mu_i x)}\right) + c_{g_i^-}x^2
    \end{equation}
    Now,
    \begin{align}
        h'(x) &=  \frac{\mu_i}{\mu_i x + \mu_i - 1} - \frac{1-\mu_i x}{\mu_i}\left[\frac{\mu_i}{\mu_i x + \mu_i - 1} - \frac{\mu_i^2-\mu_i}{1 - \mu_i - \mu_i x + \mu_i^2x}\right] + \nonumber\\
        &\qquad + \ln\left(\frac{1-\mu_i-\mu_i x}{1 - \mu_i - \mu_i x + \mu_i^2x}\right) + 2c_{g_i^-}x\\
        h''(x) &= -\frac{\mu_i^2}{(\mu_i x + \mu_i - 1)^2} - \frac{1-\mu_i x}{\mu_i}\left[-\frac{\mu_i^2}{(\mu_i x + \mu_i - 1)^2} + \frac{(\mu_i^2-\mu_i)^2}{(1 - \mu_i - \mu_i x + \mu_i^2 x)^2}\right] + \nonumber\\
        &\qquad + 2\left[\frac{\mu_i}{\mu_i x + \mu_i - 1} - \frac{\mu_i^2-\mu_i}{1 - \mu_i - \mu_i x + \mu_i^2 x}\right] + 2c_{g_i}\\
        =& -\frac{\mu_i^2}{(1 - \mu_i x)(1 - \mu_i - \mu_i x)} + 2c_{g_i}.
    \end{align}
    Hence, for all $0 \leq x \leq \frac{1 - \mu_i}{2\mu_i}$, 
    \begin{equation}
        h''(x) \leq -\frac{\mu_i^2}{1 - \mu_i} + 2c_{g_i^-}.
    \end{equation}
    Choose $c_{g_i^-} = \mu_i^2/(2(1 - \mu_i))$. Now, $h(0) = 0, h'(0) = 0$. Hence, for all $ 0 \leq x\leq \frac{1-\mu_i}{2\mu_i}$,
    \begin{equation}
        h(x) \leq 0.
    \end{equation}
    Hence, for any $\delta' \in [0, (1 - \mu_i)/(2\mu_i)]$ and $c_{g_i^-} = \mu_i^2/(2(1 - \mu_i))$
    \begin{equation}\label{rel:lower_confidence}
        \probability\left(\frac{\sum_i X_i}{n} \leq \frac{1}{\mu_i} - \delta'\right) \leq \exp\left(-nc_{g_i^-} \delta'^2\right).
    \end{equation}
\end{enumerate}
Using \eqref{rel:upper_confidence} and \eqref{rel:lower_confidence}, we have for any $\delta' \in [0, (1 - \mu_i)/(2\mu_i)]$, 
\begin{equation}\label{eq:prf_conc_geom:rel2}
    \probability\left(\left|\frac{\sum_i X_i}{n} - \frac{1}{\mu_i}\right| \geq \delta'\right) \leq  \exp\left(-n c_{g_i}' \delta'^2\right).
\end{equation}
where $c_{g_i}' = \min\left\{\frac{\mu_i^2}{2(1 - \mu_i)},  \frac{2\mu_i^2}{3(1 - \mu_i)(3 - \mu_i)}\right\}$. Using \eqref{eq:prf_conc_geom:rel1} and \eqref{eq:prf_conc_geom:rel2}, we have, for any $0 \leq \delta \leq \mu_i(1 - \mu_i)$, 
\begin{equation}
     \probability\left(\left|\mu_i^{(n)} - \mu_i\right| \geq \delta\right)   \leq \exp\left( - nc_g\delta^2\right).
\end{equation}
where $c_g = \min_i\min\left\{\frac{1}{8\mu_i^2(1 - \mu_i)},  \frac{1}{6\mu_i^2(1 - \mu_i)(3 - \mu_i)}\right\}$
\end{proof}

\begin{restatable}{lem}{restateLEMrelationPmu}
    \label{lem:relation_p_mu}
    For any time $\tau \geq 0$ and any $\delta \in (0, \Delta)$, $\> \lvert\hat{\mu}_i(\tau) - \mu_i\rvert \leq \delta$ for all $i$, implies
    \begin{equation}
        \lvert\hat{p}_i(\tau) - p_i^*\rvert \leq \min \left\{c\delta, \frac{\freecapacity_i}{4\lambda}\right\}
    \end{equation}
    where $c\geq 1$ is some positive constant and $\Delta = \min \left\{\frac{\tilde{\mu}}{2}, \frac{\sum_{j\in \support(\lambda, \bm{\mu})}\freecapacity_j}{|\support(\lambda, \bm{\mu})|}, \frac{\min_{j\in \support(\lambda, \bm{\mu})}\freecapacity_j}{4c\lambda}, \Delta_{\support}\right\}, \tilde{\mu} = \min_{j\in \support(\lambda, \bm{\mu})} \{\min \{\mu_j, 1- \mu_j\}\}$.
\end{restatable}
\begin{proof}\label{prf_lem:relation_p_mu}
We will mainly use the following set on inequalities to prove the lemma.
\begin{equation}\label{ineq1}
    1 - x \leq \sqrt{1-x} \leq 1 - \frac{x}{2}, \>\> x \in [0, 1]
\end{equation}
\begin{equation}\label{ineq2}
    1 + \frac{x}{3} \leq \sqrt{1+x} \leq 1 + x, \>\> x \in [0, 1]
\end{equation}
\begin{equation}\label{ineq3}
    1 + x \leq \frac{1}{1 - x}, \>\> x \in [0, 1]
\end{equation}
\begin{equation}\label{ineq4}
    \frac{1}{1 - x} \leq 1 + 2x, \>\> x \in [0, \frac{1}{2}]
\end{equation}
From the definition of $\Delta$, it is clear that if $\lvert\hat{\mu}_i(\tau) - \mu_i\rvert \leq \Delta$ for all $i$, then $\lvert\hat{\mu}_i(\tau) - \mu_i\rvert \leq \Delta_\support$ which implies that the optimal support set will remain unchanged. Hence, $\hat{p}_i(\tau) = 0$ for all $i \notin \support(\lambda, \bm{\mu})$. For any $i \in \support(\lambda, \bm{\mu})$
\begin{align}
    \hat{p}_i(\tau) &= \frac{\hat{\mu}_i(\tau)}{\lambda} - \frac{\sqrt{\hat{\mu}_i(\tau)(1-\hat{\mu}_i(\tau))}}{\sum_{j\in \support(\lambda, \bm{\mu})}\sqrt{\hat{\mu}_j(\tau)(1-\hat{\mu}_j(\tau))}} \left(\frac{\sum_{j\in \support(\lambda, \bm{\mu})}\hat{\mu}_j(\tau)}{\lambda} - 1\right)\\
    &\leq \frac{\mu_i + \delta}{\lambda} - \frac{\sqrt{(\mu_i - \delta)(1-\mu_i - \delta)}}{\sum_{j\in \support(\lambda, \bm{\mu})} \sqrt{(\mu_j + \delta)(1 - \mu_j + \delta)}} \left(\frac{\sum_{j\in \support(\lambda, \bm{\mu})}(\mu_j - \delta)}{\lambda} - 1\right)\\
    &\leq \frac{\mu_i + \delta}{\lambda} - \frac{\sqrt{\mu_i(1-\mu_i)}}{\sum_{j\in \support(\lambda, \bm{\mu})}\sqrt{\mu_j(1-\mu_j)}}\frac{\sqrt{(1 - \frac{\delta}{\mu_i})(1 - \frac{\delta}{1-\mu_i})}}{\sqrt{(1 + \frac{\delta}{\mu_{\min}})(1 + \frac{\delta}{1-\mu_{\max}})}} \left(\frac{\sum_{j\in \support(\lambda, \bm{\mu})}(\mu_j - \delta)}{\lambda} - 1\right)\\
    &\leq \frac{\mu_i + \delta}{\lambda} - \frac{\sqrt{\mu_i(1-\mu_i)}}{\sum_{j\in \support(\lambda, \bm{\mu})}\sqrt{\mu_j(1-\mu_j)}}\frac{(1 - \frac{\delta}{\mu_i})(1 - \frac{\delta}{1-\mu_i})}{(1 + \frac{\delta}{\mu_{\min}})(1 + \frac{\delta}{1-\mu_{\max}})} \left(\frac{\sum_{j\in \support(\lambda, \bm{\mu})}(\mu_j - \delta)}{\lambda} - 1\right)\label{rel:p_mu_upper_error:1},
\end{align}
where 
\begin{itemize}
    \item $\mu_{\min} = \min_{i \in \support(\lambda, \bm{\mu})} \mu_i$ and $\mu_{\max} = \max_{i \in \support(\lambda, \bm{\mu})} \mu_i$ .
    \item \eqref{rel:p_mu_upper_error:1} uses \eqref{ineq1}, \eqref{ineq2}.
\end{itemize}
Hence, if $\tilde{\mu} = \min (\mu_{\min}, 1- \mu_{\max})$
\begin{equation}
    \delta \leq \min \left\{\tilde{\mu}, \frac{\sum_{j\in \support(\lambda, \bm{\mu})}\mu_j - \lambda}{|\support(\lambda, \bm{\mu})|}\right\},
\end{equation}
then, combining \eqref{rel:p_mu_upper_error:1} with \eqref{ineq3} and the above conditions, we get
\begin{equation}
    \hat{p}_i(\tau) \leq \frac{\mu_i + \delta}{\lambda} - \frac{\sqrt{\mu_i(1-\mu_i)}}{\sum_{j\in \support(\lambda, \bm{\mu})}\sqrt{\mu_j(1-\mu_j)}}\left(\frac{\sum_{j\in \support(\lambda, \bm{\mu})}\mu_j}{\lambda} - 1\right)\left(1 - \frac{\delta}{\tilde{\mu}}\right)^4 \left(1 - \frac{\lvert \support(\lambda, \bm{\mu})\rvert \delta}{\sum_{j\in \support(\lambda, \bm{\mu})}\mu_j - \lambda}\right)\label{rel:p_mu_upper_error:2}
\end{equation}
Hence,
\begin{align}
    \hat{p}_i(\tau) - p_i^* &\leq  \frac{\delta}{\lambda} + \frac{\sqrt{\mu_i(1-\mu_i)}}{\sum_{j\in \support(\lambda, \bm{\mu})}\sqrt{\mu_j(1-\mu_j)}}\left(\frac{\sum_{j\in \support(\lambda, \bm{\mu})}\mu_j}{\lambda} - 1\right)\left(1 - \left(1 - \frac{\delta}{\tilde{\mu}}\right)^4 \left(1 - \frac{\lvert \support(\lambda, \bm{\mu})\rvert \delta}{\sum_{j\in \support(\lambda, \bm{\mu})}\mu_j - \lambda}\right)\right)\\
    &\leq  \frac{\delta}{\lambda} + \left(\frac{\sum_{j\in \support(\lambda, \bm{\mu})}\mu_j}{\lambda} - 1\right)\left(1 - \left(1 - 4\frac{\delta}{\tilde{\mu}}\right) \left(1 - \frac{\lvert \support(\lambda, \bm{\mu})\rvert \delta}{\sum_{j\in \support(\lambda, \bm{\mu})}\mu_j - \lambda}\right)\right)\\
    &\leq  \frac{\delta}{\lambda} + \left(\frac{\sum_{j\in \support(\lambda, \bm{\mu})}\mu_j}{\lambda} - 1\right)\left(\frac{4\delta}{\tilde{\mu}} + \frac{\lvert \support(\lambda, \bm{\mu})\rvert \delta}{\sum_{j\in \support(\lambda, \bm{\mu})}\mu_j - \lambda}\right)\\
    &\leq  \frac{\delta}{\lambda}\left(1 + 4\frac{ \sum_{j\in \support(\lambda, \bm{\mu})}\mu_j - \lambda}{\tilde{\mu}} + \lvert \support(\lambda, \bm{\mu})\rvert \right)\\
    &=  \frac{\delta}{\lambda}\left(1 + 4\frac{ \sum_{j\in \support(\lambda, \bm{\mu})}\freecapacity_i}{\tilde{\mu}} + \lvert \support(\lambda, \bm{\mu})\rvert \right)\\
    &=  c_1 \delta\label{eq:p_rel:mu:1}
\end{align}
where $c_1 = \frac{1}{\lambda}\left(1 + 4\frac{ \sum_{j\in \support(\lambda, \bm{\mu})}\freecapacity_i}{\tilde{\mu}} + \lvert \support(\lambda, \bm{\mu})\rvert \right)$. 
To prove a lower bound, we will use a similar argument like upper bound. ,
\begingroup
\allowdisplaybreaks
\begin{align}
    \hat{p}_i(\tau) &= \frac{\hat{\mu}_i(\tau)}{\lambda} - \frac{\sqrt{\hat{\mu}_i(\tau)(1-\hat{\mu}_i(\tau))}}{\sum_{j\in \support(\lambda, \bm{\mu})}\sqrt{\hat{\mu}_j(\tau)(1-\hat{\mu}_j(\tau))}} \left(\frac{\sum_{j\in \support(\lambda, \bm{\mu})}\hat{\mu}_j(\tau)}{\lambda} - 1\right)\\
    &\geq\frac{\mu_i - \delta}{\lambda} - \frac{\sqrt{(\mu_i + \delta)(1-\mu_i + \delta)}}{\sum_{j\in \support(\lambda, \bm{\mu})} \sqrt{(\mu_j - \delta)(1 - \mu_j - \delta)}} \left(\frac{\sum_{j\in \support(\lambda, \bm{\mu})}(\mu_j + \delta)}{\lambda} - 1\right)\\
    &\geq \frac{\mu_i - \delta}{\lambda} - \frac{\sqrt{\mu_i(1-\mu_i)}}{\sum_{j\in \support(\lambda, \bm{\mu})}\sqrt{\mu_j(1-\mu_j)}}\frac{\sqrt{(1 + \frac{\delta}{\mu_i})(1 + \frac{\delta}{1-\mu_i})}}{\sqrt{(1 - \frac{\delta}{\mu_{\min}})(1 - \frac{\delta}{1-\mu_{\max}})}} \left(\frac{\sum_{j\in \support(\lambda, \bm{\mu})}(\mu_j + \delta)}{\lambda} - 1\right)\\
    &\geq\frac{\mu_i - \delta}{\lambda} - \frac{\sqrt{\mu_i(1-\mu_i)}}{\sum_{j\in \support(\lambda, \bm{\mu})} \sqrt{\mu_j(1-\mu_j)}} \frac{(1 + \frac{\delta}{\mu_i})(1 + \frac{\delta}{1-\mu_i})}{(1 - \frac{\delta}{\mu_{\min}})(1 - \frac{\delta}{1-\mu_{\max}})} \left(\frac{\sum_{j\in \support(\lambda, \bm{\mu})}(\mu_j + \delta)}{\lambda} - 1\right)\label{rel:p_mu_lower_error:1}
\end{align}
\endgroup
where \eqref{rel:p_mu_lower_error:1} uses \eqref{ineq1}, \ref{ineq2}. If 
\begin{equation}
    \delta \leq \min \left\{\frac{\tilde{\mu}}{2}, \frac{\sum_{j\in \support(\lambda, \bm{\mu})}\mu_j - \lambda}{|\support(\lambda, \bm{\mu})|}\right\},
\end{equation}
then combining \eqref{rel:p_mu_lower_error:1} with \eqref{ineq4}, we get
\begin{equation}\label{rel:p_mu_lower_error:2}
    \hat{p}_i(\tau) \geq \frac{\mu_i - \delta}{\lambda} - \frac{\sqrt{\mu_i(1-\mu_i)}}{\sum_{j\in \support(\lambda, \bm{\mu})}\sqrt{\mu_j(1-\mu_j)}}\left(\frac{\sum_{j\in \support(\lambda, \bm{\mu})}\mu_j}{\lambda} - 1\right)\left(1 + \frac{2\delta}{\tilde{\mu}}\right)^4 \left(1 + \frac{\lvert \support(\lambda, \bm{\mu})\rvert \delta}{\sum_{j\in \support(\lambda, \bm{\mu})}\mu_j - \lambda}\right).
\end{equation}
\begin{align}
    \hat{p}_i(\tau) - p_i^* &\geq  -\frac{\delta}{\lambda} + \frac{\sqrt{\mu_i(1-\mu_i)}}{\sum_{j\in \support(\lambda, \bm{\mu})}\sqrt{\mu_j(1-\mu_j)}}\left(\frac{\sum_{j\in \support(\lambda, \bm{\mu})}\mu_j}{\lambda} - 1\right)\left(1 - \left(1 + \frac{2\delta}{\tilde{\mu}}\right)^4 \left(1 + \frac{\lvert \support(\lambda, \bm{\mu})\rvert \delta}{\sum_{j\in \support(\lambda, \bm{\mu})}\mu_j - \lambda}\right)\right)\\
    &\geq  -\frac{\delta}{\lambda} + \left(\frac{\sum_{j\in \support(\lambda, \bm{\mu})}\mu_j}{\lambda} - 1\right)\left(1 - \left(1 + 30\frac{\delta}{\tilde{\mu}}\right) \left(1 + \frac{\lvert \support(\lambda, \bm{\mu})\rvert \delta}{\sum_{j\in \support(\lambda, \bm{\mu})}\mu_j - \lambda}\right)\right)\\
    &\geq  -\frac{\delta}{\lambda} - \left(\frac{\sum_{j\in \support(\lambda, \bm{\mu})}\mu_j}{\lambda} - 1\right)\left(\frac{30\delta}{\tilde{\mu}} + \frac{\lvert \support(\lambda, \bm{\mu})\rvert \delta}{\sum_{j\in \support(\lambda, \bm{\mu})}\mu_j - \lambda} + \frac{30\delta^2 \lvert \support(\lambda, \bm{\mu})\rvert}{(\sum_{j\in \support(\lambda, \bm{\mu})}\mu_j - \lambda)\tilde{\mu}}\right)\\
    &\geq  -\delta\left(\frac{1}{\lambda} + \frac{30\sum_{j\in \support(\lambda, \bm{\mu})}\freecapacity_i}{\lambda \tilde{\mu}} + \frac{16\lvert \support(\lambda, \bm{\mu})\rvert}{\lambda}\right)\\
    &=  -c_2\delta\label{eq:p_rel:mu:2}
\end{align}
where $c_2 = \left(\frac{1}{\lambda} + \frac{30 \sum_{j\in \support(\lambda, \bm{\mu})}\freecapacity_j}{\lambda \tilde{\mu}} + \frac{16\lvert \support(\lambda, \bm{\mu})\rvert}{\lambda}\right)$. Hence, using \eqref{eq:p_rel:mu:1} and \eqref{eq:p_rel:mu:2} if $\delta \leq \Delta = \min \left\{\frac{\tilde{\mu}}{2}, \frac{\sum_{j\in \support(\lambda, \bm{\mu})}\freecapacity_j}{|\support(\lambda, \bm{\mu})|}, \frac{\min_{j\in \support(\lambda, \bm{\mu})}\freecapacity_j}{4c\lambda}\right\}$ and $c = \max (c_1, c_2)$
\begin{equation}
    \lvert \hat{\mu}_i - \mu_i \rvert \leq  \delta \mbox{ for all $i$, } \mbox{ implies } \lvert \hat{p}_i - p_i^* \rvert \leq  \min\left\{c \delta, \frac{\freecapacity_i}{4\lambda}\right\}\mbox{ for all $i$.}
\end{equation}
\end{proof}

\subsection{Proof of Lemma \ref{lem:p_wt}} \label{prf_lem:p_wt}
We restate Lemma~\ref{lem:p_wt} below for ease of reference. \restateLEMMApWT*

\begin{proof}
We first show that if the service rates estimates are within $\Delta_0$ ball of the true parameter for all time after $w(t)$, i.e., $|\hat{\mu}_i(\tau) - \mu_i| \leq \Delta_0$ for all $i$ and all $\tau \in \left[w(t) + 1, t\right]$, then $\eventoneone$ and $\eventonetwo$ hold true, where $w(t)  = 2\exp \left(\frac{1}{\Delta_0} \sqrt{\frac{16 \ln t}{c_g\lambda}}\right)$.  Recall that the definitions of $\Delta_0$, $\Delta$ and $\Delta_{\support}$ in \eqref{eq:delta0}, \eqref{eq:delta} and \eqref{eq:gap} imply that $\Delta_0\le \Delta\le \Delta_{\support}$.
\begin{enumerate}
    \item For any $\tau$, $|\hat{\mu}_i(\tau) - \mu_i| \leq \Delta_0$  for all $i$ implies $|\hat{\mu}_i(\tau) - \mu_i| \leq \Delta_{\support}$ for all $i$, which from the definition of $\Delta_{\support}$ implies that $\support(\lambda, \bm{\mu}) = \support(\lambda, \hat{\bm{\mu}}(\tau))$. Hence, $ |\hat{\mu}_i(\tau) - \mu_i| \leq \Delta_0$, for all $i$, for all $\tau \in \left[w(t) + 1, t\right]$ implies $\support(\lambda, \bm{\mu}) = \support(\lambda, \hat{\bm{\mu}}(\tau))$, for all $\tau \in \left[w(t) + 1, t\right]$, which is essentially $\eventoneone$.
    \item To prove the lower bound in $\eventonetwo$, we will use mainly \Cref{lem:relation_p_mu}. We restate the \Cref{lem:relation_p_mu} below. \restateLEMrelationPmu*
    Now, for any $\tau$, $|\hat{\mu}_i(\tau) - \mu_i| \leq \Delta_0$, for all $i$ implies $|\hat{\mu}_i(\tau) - \mu_i|\leq \min\left\{\Delta, \frac{\min_{i:p_i^*>0}p_i^*}{3c}\right\}$ for all $i$, which by \Cref{lem:relation_p_mu}  further implies $|\hat{p}_i(\tau) - p_i^*|\leq c\min\left\{\Delta, \frac{\min_{i:p_i^*>0}p_i^*}{3c}\right\}$. Hence, 
    \begin{align}
        |\hat{p}_i(\tau) - p_i^*| &\leq c\min\left\{\Delta, \frac{\min_{i:p_i^*>0}p_i^*}{3c}\right\}\\
        &\leq \frac{\min_{i:p_i^*>0}p_i^*}{3}\\
        &\leq \frac{p_i^*}{3}\\
        \hat{p}_i(\tau) &\geq \frac{2p_i^*}{3}\\
        \lambda\left(\frac{\ln \tau}{\tau} + \left(1 - \frac{\ln \tau}{\tau}\right)\hat{p}_i(\tau)\right) & \geq \lambda\left(\frac{\ln \tau}{\tau} + \left(1 - \frac{\ln \tau}{\tau}\right)\frac{2p_i^*}{3}\right)\\
        & \geq \frac{\lambda p_i^*}{2}\label{eq:lemma_p_wt:lowerbound_arrival_rate},
    \end{align}
    where \eqref{eq:lemma_p_wt:lowerbound_arrival_rate} is true for any $\tau \geq 1$. Hence, for any $t\geq t_0$ and and for all $\tau \in [w(t) + 1, t]$, $|\hat{\mu}_i(\tau) - \mu_i| \leq \Delta_0$  for all $i$, implies for all $i$,
    \begin{equation}
        \expectation\left(\arrival_i(\tau)\Big|\hat{p}_i(\tau) \right) = \lambda\left(\frac{\ln \tau}{\tau} + \left(1 - \frac{\ln \tau}{\tau}\right)\hat{p}_i(\tau)\right) \geq \frac{\lambda p_i^*}{2},
    \end{equation}
    which is the lower bound in $\eventonetwo$.
    \item For the upper bound, like previous argument, we can show that for any $\tau$, $|\hat{\mu}_i(\tau) - \mu_i| \leq \Delta_0$, for all $i$ implies $|\hat{p}_i(\tau) - p_i^*|\leq \freecapacity_i/4\lambda$. Hence, for all $i$,
    \begin{align}
        \expectation\left(\arrival_i(\tau)\Big|\hat{p}_i(\tau) \right) &= \lambda\left(\frac{\ln \tau}{\tau} + \left(1 - \frac{\ln \tau}{\tau}\right)\hat{p}_i(\tau)\right)\\
        &\leq \lambda\left(\frac{\ln \tau}{\tau} + \hat{p}_i(\tau)\right)\\
        &\leq \lambda\frac{\ln \tau}{\tau} + \lambda(\hat{p}_i(\tau) - p_i^*) + \lambda p_i^*\\
        &\leq \lambda\frac{\ln \tau}{\tau} + \lambda\frac{\freecapacity_i}{4\lambda} + (\mu_i - \freecapacity_i)\label{eq:lemma_p_wt:upperbound_arrival_rate:1}\\
        &\leq \frac{\freecapacity_i}{4} + \lambda\frac{\freecapacity_i}{4\lambda} + (\mu_i - \freecapacity_i)\label{eq:lemma_p_wt:upperbound_arrival_rate:2}\\
        &\leq \mu_i - \frac{\freecapacity_i}{2},
    \end{align}
    where
    \begin{itemize}
        \item \eqref{eq:lemma_p_wt:upperbound_arrival_rate:1} uses the fact that $\freecapacity_i = \mu_i - \lambda p_i^*$ and $|\hat{p}_i(\tau) - p_i^*|\leq \freecapacity_i/4\lambda$.
        \item \eqref{eq:lemma_p_wt:upperbound_arrival_rate:2} uses \eqref{condition_t0:arrival_condition_on_probab} that ensures that for any $\tau \geq w(t)$ and $t\geq t_0$, $\lambda \ln \tau/\tau \leq \min_{i} \freecapacity_i/4$.
    \end{itemize}
\end{enumerate}
Hence, we proved that for any $t \geq t_0$,
\begin{equation}
    \eventoneone \cup \eventonetwo \supseteq \underset{\tau = w(t) + 1}{\overset{t}{\cap}}\left\{\underset{i = 1}{\overset{\numofserver}{\cap}}\left\{|\hat{\mu}_i(\tau) - \mu_i| \leq \Delta_0\right\}\right\}.
\end{equation}
Hence, 
\begin{equation}
    \probability\left((\eventone)^c\right)\leq \sum_{\tau = w(t) + 1}^t\sum_{i = 1}^\numofserver\probability\left(|\hat{\mu}_i(\tau) - \mu_i| \geq \Delta_0\right).
\end{equation}
Hence, to prove the lemma it is sufficient to prove that for any $t \geq t_0$, 
\begin{equation}
    \sum_{\tau = w(t) + 1}^t\sum_{i = 1}^\numofserver\probability\left(|\hat{\mu}_i(\tau) - \mu_i| \geq \Delta_0\right) \leq \numofserver\left(\frac{2}{t^{7}} + \frac{k_2}{t^3}\right).
\end{equation}
The main idea behind proving the above bound is that because of exploration, each server will observe a sufficient number of departures such that the estimates are within the $\Delta_0$ ball of the true parameter. Consider any $\tau \in [ w(t) + 1, t]$. Now, for any $n_0(t)$
\begin{align}
    \probability\left(\left|\hat{\mu}_i(\tau) - \mu_i\right|> \Delta_0\right) &=  \probability\left( \left|\hat{\mu}_i(\tau) - \mu_i\right|> \Delta_0, \sum_{\ell=1}^{\tau - 1} \departure_i(\ell) < n_0(t)\right) + \sum_{n=n_0(t)}^{\infty}\probability\left(\left|\hat{\mu}_i(\tau) - \mu_i\right|> \Delta_0, \sum_{\ell=1}^{\tau - 1} \departure_i(\ell) = n\right)\\
    &\leq \probability\left(\sum_{\ell=1}^{\tau - 1} \departure_i(\ell) < n_0(t)\right) + \sum_{n=n_0(t)}^{\infty}\probability\left(\left|\hat{\mu}_i^{(n)} - \mu_i\right|> \Delta_0, \sum_{\ell=1}^{\tau - 1} \departure_i(\ell) = n\right)\\
    &\leq \probability\left(\sum_{\ell=1}^{\tau - 1} \departure_i(\ell) < n_0(t)\right) + \sum_{n=n_0(t)}^{\infty}\probability\left(\left|\hat{\mu}_i^{(n)} - \mu_i\right|> \Delta_0 \right),\label{eq:lemma_p_wt:difference:0}
\end{align}
where $\hat{\mu}_i^{(n)}$ is the estimate of $\mu_i$ using $n$ independent samples of i.i.d. geometric random variable with parameter $\mu_i$.
Now, the term $\probability\left(\left|\hat{\mu}_i^{(n)} - \mu_i\right|> \Delta_0\right)$ can be upper bounded using \Cref{lem:conc_geoemtric}. Using \Cref{lem:conc_geoemtric} with \eqref{eq:lemma_p_wt:difference:0}, we have
\begin{align}
    &\probability\left(\left|\hat{\mu}_i(\tau) - \mu_i\right|> \Delta_0\right) \\
    \leq &\probability\left(\sum_{\ell=1}^{\tau - 1} \departure_i(\ell) < n_0(t)\right) + \sum_{n=n_0(t)}^{\infty}\exp\left( -n c_g\Delta_0^2 \right)\\
    \leq &\probability\left( \left\{\sum_{\ell=1}^{\tau - 1}\departure_i(\ell) < n_0(t)\right\}\cap \left\{ \sum_{\ell=1}^{(\tau - 1)/2}\arrival_i(\ell) \geq n_0(t)\right\}\right) + \probability \left(\sum_{\ell=1}^{(\tau - 1)/2} \arrival_i(\ell) < n_0(t)\right) + \sum_{n=n_0(t)}^{\infty}\exp\left( -n c_g\Delta_0^2 \right)\label{eq:lemma_p_wt:difference:1}\\
    \leq& \probability\left( \left\{\sum_{\ell=1}^{\tau - 1}\departure_i(\ell) < n_0(t)\right\}\cap \left\{ \sum_{\ell=1}^{(\tau - 1)/2}\arrival_i(\ell) \geq n_0(t)\right\}\right) + \probability \left(\sum_{\ell=1}^{w(t)/2} \arrival_i(\ell) < n_0(t)\right) + \frac{\exp\left( -n_0(t) c_g\Delta_0^2 \right)}{1 - \exp\left( -c_g\Delta_0^2 \right)}\\
    \leq& \probability\left( \left\{\sum_{\ell=1}^{\tau - 1}\departure_i(\ell) < n_0(t)\right\}\cap \left\{ \sum_{\ell=1}^{(\tau - 1)/2}\arrival_i(\ell) \geq n_0(t)\right\}\cap \left\{\sum_{\ell=(\tau - 1)/2}^{\tau - 1}\service_i(\ell) < n_0(t)\right\}\right) + \nonumber\\
    &\qquad + \probability\left( \left\{\sum_{\ell=1}^{\tau - 1}\departure_i(\ell) < n_0(t)\right\}\cap \left\{ \sum_{\ell=1}^{(\tau - 1)/2}\arrival_i(\ell) \geq n_0(t)\right\}\cap \left\{\sum_{\ell=(\tau - 1)/2}^{\tau - 1}\service_i(\ell) \geq n_0(t)\right\}\right) + \probability \left(\sum_{\ell=1}^{w(t)/2} \explore_i(\ell) < n_0(t)\right) + \frac{\exp\left( -n_0(t) c_g\Delta_0^2 \right)}{1 - \exp\left( -c_g\Delta_0^2 \right)}\label{eq:lemma_p_wt:difference:20}.
\end{align}
Now, \begin{equation}\label{eq:lemma_p_wt:difference:21}
    \probability\left( \left\{\sum_{\ell=1}^{\tau - 1}\departure_i(\ell) < n_0(t)\right\}\cap \left\{ \sum_{\ell=1}^{(\tau - 1)/2}\arrival_i(\ell) \geq n_0(t)\right\}\cap \left\{\sum_{\ell=(\tau - 1)/2}^{\tau - 1}\service_i(\ell) < n_0(t)\right\}\right)\leq \probability\left(\sum_{\ell=(\tau - 1)/2}^{\tau - 1}\service_i(\ell) < n_0(t)\right).
\end{equation}
Also, 
\begin{equation}\label{eq:lemma_p_wt:difference:22}
    \probability\left( \left\{\sum_{\ell=1}^{\tau - 1}\departure_i(\ell) < n_0(t)\right\}\cap \left\{ \sum_{\ell=1}^{(\tau - 1)/2}\arrival_i(\ell) \geq n_0(t)\right\}\cap \left\{\sum_{\ell=(\tau - 1)/2}^{\tau - 1}\service_i(\ell) \geq n_0(t)\right\}\right) = 0,
\end{equation}
because, if the server $i$ sees at least $n_0(t)$ jobs till time $(\tau - 1)/2$ and the total offered service in time $(\tau - 1)/2$ to $(\tau - 1)$ exceeds $n_0(t)$, then the total departures till time $\tau$ should be at least $n_0(t)$. Using \eqref{eq:lemma_p_wt:difference:20},  \eqref{eq:lemma_p_wt:difference:21} and \eqref{eq:lemma_p_wt:difference:22} we have,
\begin{align}
    \probability\left(\left|\hat{\mu}_i(\tau) - \mu_i\right|> \Delta_0\right) &\leq \probability\left(\sum_{\ell=(\tau - 1)/2}^{\tau - 1}\service_i(\ell) < n_0(t)\right)  + \probability \left(\sum_{\ell=1}^{w(t)/2} \explore_i(\ell) < n_0(t)\right) + \frac{\exp\left( -n_0(t) c_g\Delta_0^2 \right)}{1 - \exp\left( -c_g\Delta_0^2 \right)}\label{eq:lemma_p_wt:difference:2}\\
    &= \probability\left( \sum_{\ell = 1}^{(\tau - 1)/2}\service_i(\ell) < n_0(t)\right)  + \probability \left(\sum_{\ell=1}^{w(t)/2} \explore_i(\ell) < n_0(t)\right) + \frac{\exp\left( -n_0(t) c_g\Delta_0^2 \right)}{1 - \exp\left( -c_g\Delta_0^2 \right)}\label{eq:lemma_p_wt:difference:3}\\
    &\leq \underbrace{\probability\left( \sum_{\ell = 1}^{w(t)/2}\service_i(\ell) < n_0(t)\right)}_{T_1}  + \underbrace{\probability \left(\sum_{\ell=1}^{w(t)/2} \explore_i(\ell) < n_0(t)\right)}_{T_2} + \underbrace{\frac{\exp\left( -n_0(t) c_g\Delta_0^2 \right)}{1 - \exp\left( -c_g\Delta_0^2 \right)}}_{T_3},\label{eq:lemma_p_wt:difference:4}
\end{align}
where \eqref{eq:lemma_p_wt:difference:3} uses the fact that the service process is i.i.d across time.
Take $n_0(t) = \frac{4\ln t}{c_g \Delta_0^2}$. Now,
\begin{equation}
    \probability\left(\explore_i(\ell) = 1\right) = \frac{\lambda\ln \ell}{\ell}.
\end{equation}
Hence, 
\begin{equation}
    \expectation\left(\sum_{\ell=1}^{w(t)/2}\explore_i(\ell)\right) \geq \frac{3}{8}\lambda\ln^2\left(\frac{w(t)}{2}\right)= \frac{6\ln t}{c_g \Delta_0^2}.
\end{equation}
Bounding the term $T_2$ in \eqref{eq:lemma_p_wt:difference:4}, we get
\begin{align}
    \probability\left(\sum_{\ell=1}^{w(t)/2}\explore_i(\ell) \leq \frac{4\ln t}{c_g \Delta_0^2}\right)\label{rel:bound_arrival_sup_lemma1} & \leq \probability\left(\sum_{\ell=1}^{w(t)/2}\explore_i(\ell)  \leq \left(1-\frac{1}{3}\right)\expectation\left(\sum_{\ell=1}^{w(t)/2}\explore_i(\ell) \right)\right)\\
    &\leq \exp\left(-\frac{1}{18}\expectation\left(\sum_{\ell=1}^{w(t)/2}\explore_i(\ell) \right)\right)\label{rel:bound_arrival_sup_lemma2}\\
    &\leq \exp\left(-\frac{\ln t}{3\Delta_0^2 c_g}\right)\\
    &\leq \frac{1}{t^{4}}\label{rel:bound_arrival_sup_lemma3}
\end{align}
where, 
\begin{itemize}
    \item \eqref{rel:bound_arrival_sup_lemma2} uses Chernoff bound given in \eqref{ineq:Chernoff}.
    \item \eqref{rel:bound_arrival_sup_lemma3} uses the fact that $\Delta_0 \leq \frac{1}{4}$ and $c_g \leq 1$.
\end{itemize}
Bounding the $T_1$ in \eqref{eq:lemma_p_wt:difference:4}, we get
\begingroup\allowdisplaybreaks
\begin{align}
    \probability\left( \sum_{\ell= 1}^{w(t)/2}\service_i(\ell) < \frac{4\ln t}{c_g \Delta_0^2}\right) &= \probability\left( \sum_{\ell = 1}^{w(t)/2}\service_i(\ell) < \frac{\mu_i w(t)}{2}\frac{\frac{4\ln t}{c_g \Delta_0^2}}{\frac{\mu_i w(t)}{2}}\right)\\
    &\leq \exp\left( - \frac{1}{2}\left(1 - \frac{\frac{4\ln t}{c_g \Delta_0^2}}{\frac{\mu_i w(t)}{2}}\right)^2\frac{\mu_i w(t)}{2}\right)\label{rel:departure_arrival_service1:1}\\
    &\leq \exp\left(-\frac{\left(\frac{\mu_i w(t)}{2} - \frac{4\ln t}{c_g \Delta_0^2}\right)^2}{\mu_i w(t)}\right)\\
    & \leq \frac{1}{t^{8}}\label{rel:departure_arrival_service1:2}
\end{align}
\endgroup
where
\begin{itemize}
    \item \eqref{rel:departure_arrival_service1:1} uses the Chernoff bound given in \eqref{ineq:Chernoff}. 
    \item \eqref{rel:departure_arrival_service1:2} uses $\eqref{condition_t0:lemma_p_wt}$ that ensures that for any $t \geq t_0$, $\min_i\left\{\frac{\mu_i w(t)}{2} - \frac{4\ln t}{c_g \Delta_0^2} - \sqrt{8\mu_i w(t)\ln t}\right\} \geq 0$.
\end{itemize}
Finally, to bound the $T_3$ in \eqref{eq:lemma_p_wt:difference:4}, we substitute the value of $n_0(t) = \frac{4\ln t}{c_g \Delta_0^2}$ in the expression and hence get
\begin{equation}
    \frac{\exp\left( -n_0(t) c_g\Delta_0^2 \right)}{1 - \exp\left( -c_g\Delta_0^2 \right)} \leq \frac{k_2}{t^4}.
\end{equation}
where $k_2 = \frac{1}{1 - \exp\left( -c_g\Delta_0^2 \right)}$. Using bounds on $T_1, T_2$ and $T_3$ we have, for any $i$ and for any $\tau \in [w(t) + 1, t]$
\begin{equation}
    \probability\left(\left|\hat{\mu}_i(\tau) - \mu_i\right|> \Delta_0\right) \leq \frac{1}{t^8} + \frac{k_2 + 1}{t^4},\\
\end{equation}
Hence,  
\begin{equation}\label{rel:mu_wt}
    \probability((\eventone)^c) \leq\sum_{\tau = w(t) + 1}^t\sum_{i = 1}^\numofserver\probability\left(\left\{|\hat{\mu}_i(\tau) - \mu_i| \geq \Delta_0\right\}\right) \leq \numofserver\left(\frac{1}{t^{7}} + \frac{k_2 + 1}{t^3}\right),
\end{equation}
which completes the proof.
\end{proof}

\subsection{Implication of Lemma \ref{lem:p_wt}}
An immediate consequence of \Cref{lem:p_wt} is the \Cref{lem:neg_drift} which would be instrumental in proving the future lemmas.  Recall that $w(t) = 2\exp\left(\frac{1}{\Delta_0}\sqrt{\frac{16 \ln t}{c_g\lambda}}\right)$ and $\freecapacity_i$ is the residual capacity at server~$i$.
\begin{lem}
    \label{lem:neg_drift}
    For any $t_1, t_2$ such that $w(t)\leq t_1 \leq t_2 \leq t$, 
    \begin{enumerate}
        \item for $\tau \leq (t_2 - t_1 + 1)\frac{\freecapacity_i}{2}$, we have
        \begin{equation}
            \probability\left(\left\{\sum_{\ell=t_1}^{t_2}\left( \service_i(\ell) - \arrival_i(\ell)\right)\leq \tau\right\}\cap \eventone\right) \leq \exp\left( - \frac{\left(\tau - (t_2 - t_1 + 1)\frac{\freecapacity_i}{2}\right)^2}{2(t_2 - t_1 + 1)}\right).
        \end{equation}
        \item for $0 \leq \tau \leq (t_2 - t_1 + 1)\frac{\lambda p_i^*}{2}$, we have
        \begin{equation}
            \probability\left(\left\{\sum_{\ell=t_1}^{t_2} \arrival_i(\ell)\leq \tau\right\}\cap \eventone\right) \leq \exp\left(- \frac{1}{2}\left(1 - \frac{\tau}{\frac{(t_2 - t_1 + 1)\lambda p_i^*}{2}}\right)^2\frac{(t_2 - t_1 + 1)\lambda p_i^*}{2}\right).
        \end{equation}
    \end{enumerate}
\end{lem}
\begin{proof}\label{prf_lem:neg_drift}
    \begin{enumerate}
        \item 
        \begingroup
        \allowdisplaybreaks
        \begin{align}
                 &\probability\left(\left\{\sum_{\ell=t_1}^{t_2}\left( \service_i(\ell) - \arrival_i(\ell)\right)\leq \tau\right\}\cap \eventone\right) \\
                 =  & \int_{p_i(t_2)} \probability \left(\left\{\sum_{\ell=t_1}^{t_2}\left( \service_i(\ell) - \arrival_i(\ell)\right)\leq \tau\right\}\cap \eventone \Big| \hat{p}_i(t_2) = p_i(t_2)\right)f_{\hat{p}_i(t_2)}(p_i(t_2))dp_i(t_2)\\
                 =& \int_{p_i(t_2)}\probability\left(\left\{\sum_{\ell=t_1}^{t_2}\left( \service_i(\ell) - \arrival_i(\ell)\right)\leq \tau\right\}\cap \eventone \cap\left\{\expectation[\arrival_i(t_2)\Big| \hat{p}_i(t_2)] \leq \mu_i - \frac{\freecapacity_i}{2}\right\}\Big|\hat{p}_i(t_2) = p_i(t_2)\right)\times \nonumber\\
                 &\qquad\times f_{\hat{p}_i(t_2)}(p_i(t_2))dp_i(t_2)\\
                 =& \int_{p_i(t_2):\expectation[\arrival_i(t_2)| \hat{p}_i(t_2) = p_i(t_2)] \leq \mu_i - \frac{\freecapacity_i}{2}}\probability\left(\left\{\sum_{\ell=t_1}^{t_2}\left( \service_i(\ell) - \arrival_i(\ell)\right)\leq \tau\right\}\cap \eventone \Big| \hat{p}_i(t_2) = p_i(t_2)\right)\times \nonumber\\
                 &\qquad \times f_{\hat{p}_i(t_2)}(p_i(t_2))dp_i(t_2) \\
                 = & \int_{p_i(t_2):\expectation[\arrival_i(t_2)| \hat{p}_i(t_2) = p_i(t_2)] \leq \mu_i - \frac{\freecapacity_i}{2}}\probability\Bigg(\bigg\{\service_i(t_2) - Ber\left(\lambda\frac{\ln t_2}{t_2} + \lambda\left(1 - \frac{\ln t_2}{t_2}\right) p_i(t_2)\right) + \nonumber \\
                 &\qquad + \sum_{ \ell=t_1}^{t_2 - 1} \left(\service_i(\ell) - \arrival_i(\ell)\right)\leq \tau\bigg\}\cap \eventone \Big|\hat{p}_i(t_2) = p_i(t_2)\Bigg)f_{\hat{p}_i(t_2)}(p_i(t_2))dp_i(t_2)\label{rel:negative_drift_departure:1}\\
                 \leq & \int_{p_i(t_2):\expectation[\arrival_i(t_2)| \hat{p}_i(t_2) = p_i(t_2)] \leq \mu_i - \frac{\freecapacity_i}{2}}\probability\Bigg(\left\{\service_i(t_2) - \xi_i'(t_2)+\sum_{ \ell=t_1}^{t_2 - 1} \left(\service_i(\ell) - \arrival_i(\ell)\right)\leq \tau\right\}\cap \nonumber\\
                 &\qquad \cap\eventone \Big|\hat{p}_i(t_2) = p_i(t_2)\Bigg) \times f_{\hat{p}_i(t_2)}(p_i(t_2))dp_i(t_2)\label{rel:negative_drift_departure:2}\\
                 \leq &\probability\left(\left\{\service_i(t_2) - \xi'_i(t_2)+\sum_{ \ell=t_1 }^{t_2 - 1} \left(\service_i(\ell) - \arrival_i(\ell)\right)\leq \tau\right\}\cap \eventone \right)\\
                 \leq &\probability\left(\left\{\sum_{ \ell=t_1 }^{t_2} \left(\service_i(\ell) - \xi_i'(\ell)\right)\leq \tau\right\}\cap \eventone \right)\label{rel:negative_drift_departure:3}\\
                 \leq  &\probability\left(\sum_{\ell=t_1}^{t_2} \left(\service_i(\ell) - \xi_i'(\ell)\right) \leq \tau \right)\\
                 \leq  &\exp\left( - \frac{\left(\tau - (t_2 - t_1 + 1)\frac{\freecapacity_i}{2}\right)^2}{2(t_2 - t_1 + 1)}\right)\label{rel:negative_drift_departure:4}
            \end{align}
        \endgroup
        where
        \begin{itemize}
            \item \eqref{rel:negative_drift_departure:1} uses the fact that conditioned on $\hat{p}_i(t_2)$, $\arrival_i(t_2)$ is bernoullli random variable independent of past arrivals and services.
            \item \eqref{rel:negative_drift_departure:2} uses the fact that we can have a Bernoulli random variables $\xi_i'(t_2)$ with mean $\mu_i - \frac{\freecapacity_i}{2}$ independent of everything else, that stochasticallly dominates the random variable
            $Ber\left(\lambda\frac{\ln t_2}{t_2} + \lambda\left(1 - \frac{\ln t_2}{t_2}\right)p_i(t_2)\right)$.
            \item \eqref{rel:negative_drift_departure:3} follows repeating the same argument as above.
            \item \eqref{rel:negative_drift_departure:4} uses the Hoeffding inequality given in \eqref{ineq:Hoeffding}.
        \end{itemize}
        \item 
        \begingroup
        \allowdisplaybreaks
        \begin{align}
                 &\probability\left(\left\{\sum_{\ell=t_1}^{t_2} \arrival_i(\ell)\leq \tau\right\}\cap \eventone\right) \\
                 =  & \int_{p_i(t_2)} \probability \left(\left\{\sum_{\ell=t_1}^{t_2} \arrival_i(\ell)\leq \tau\right\}\cap \eventone \Big| \hat{p}_i(t_2) = p_i(t_2)\right)f_{\hat{p}_i(t_2)}(p_i(t_2))dp_i(t_2)\\
                 =& \int_{p_i(t_2)}\probability\left(\left\{\sum_{\ell=t_1}^{t_2} \arrival_i(\ell)\leq \tau\right\}\cap \eventone \cap\left\{\expectation[\arrival_i(t_2)\Big| \hat{p}_i(t_2)] \geq \frac{\lambda p_i^*}{2}\right\}\Big|\hat{p}_i(t_2) = p_i(t_2)\right)\times \nonumber\\
                 &\qquad \times f_{\hat{p}_i(t_2)}(p_i(t_2))dp_i(t_2)\\
                 =& \int_{p_i(t_2):\expectation[\arrival_i(t_2)| \hat{p}_i(t_2) = p_i(t_2)] \geq \frac{\lambda p_i^*}{2}}\probability\left(\left\{\sum_{\ell=t_1}^{t_2} \arrival_i(\ell)\leq \tau\right\}\cap \eventone \Big| \hat{p}_i(t_2) = p_i(t_2)\right)\times \nonumber\\
                 &\qquad \times f_{\hat{p}_i(t_2)}(p_i(t_2))dp_i(t_2) \\
                 = & \int_{p_i(t_2):\expectation[\arrival_i(t_2)| \hat{p}_i(t_2) = p_i(t_2)] \geq \frac{\lambda p_i^*}{2}}\probability\Bigg(\bigg\{ Ber \left( \lambda\frac{\ln t_2}{t_2} + \lambda\left(1 - \frac{\ln t_2}{t_2}\right) p_i(t_2)\right) + \nonumber\\
                 &\qquad + \sum_{ \ell=t_1}^{t_2 - 1}  \arrival_i(\ell)\leq \tau\bigg\}\cap \eventone \Big|\hat{p}_i(t_2) = p_i(t_2)\Bigg)f_{\hat{p}_i(t_2)}(p_i(t_2))dp_i(t_2)\label{rel:negative_drift_arrival:1}\\
                 \leq & \int_{p_i(t_2):\expectation[\arrival_i(t_2)| \hat{p}_i(t_2) = p_i(t_2)] \geq \frac{\lambda p_i^*}{2}}\probability\Bigg(\left\{ \chi_i(t_2)+\sum_{ \ell=t_1}^{t_2 - 1} \arrival_i(\ell)\leq \tau\right\}\cap\eventone \Big|\hat{p}_i(t_2) = p_i(t_2)\Bigg) \times \nonumber\\
                 &\qquad \times f_{\hat{p}_i(t_2)}(p_i(t_2))dp_i(t_2) \label{rel:negative_drift_arrival:2}\\
                 \leq &\probability\left(\left\{\chi_i(t_2)+\sum_{ \ell=t_1 }^{t_2 - 1}  \arrival_i(\ell)\leq \tau\right\}\cap \eventone \right)\\
                 \leq &\probability\left(\left\{\sum_{ \ell=t_1 }^{t_2} \chi_i(\ell)\leq \tau\right\}\cap \eventone \right)\label{rel:negative_drift_arrival:3}\\
                 \leq  &\probability\left(\sum_{\ell=t_1}^{t_2} \chi_i(\ell) \leq \tau \right)\\
                 \leq  &\exp\left( - \frac{1}{2}\left(1 - \frac{\tau}{\frac{(t_2 - t_1 + 1)\lambda p_i^*}{2}}\right)^2\frac{(t_2 - t_1 + 1)\lambda p_i^*}{2}\right)\label{rel:negative_drift_arrival:4}
            \end{align}
        \endgroup
        where
        \begin{itemize}
            \item \eqref{rel:negative_drift_arrival:1} uses the fact that conditioned on $\hat{p}_i(t_2)$, $\arrival_i(t_2)$ is Bernoulli random variable independent of past arrivals.
            \item \eqref{rel:negative_drift_arrival:2} uses the fact that we can have a Bernoulli random variables $\chi_i(t_2)$ with mean $\frac{\lambda p_i^*}{2}$ independent of everything else, that is stochasticallly dominated by the random variable $Ber\left(\lambda\frac{\ln t_2}{t_2} + \lambda\left(1 - \frac{\ln t_2}{t_2}\right)p_i(t_2)\right)$.
            \item \eqref{rel:negative_drift_arrival:3} follows repeating the same argument as above.
            \item \eqref{rel:negative_drift_arrival:4} uses Chernoff inequality, given in \eqref{ineq:Chernoff}. 
        \end{itemize}
    \end{enumerate}
\end{proof}

\subsection{Proof of Lemma \ref{lem:final_busy_period}} \label{prf_lem:final_busy_period}
We will prove this using a sequence of lemmas where we iteratively bound the length of the busy period and the queue length. In \Cref{lem:bound_q_length}, we will first prove a coarse all time bound on the queue length. Using bounded queue length from \Cref{lem:bound_q_length} and negative drift from \Cref{lem:p_wt}, we will prove a coarse bound on the length of the busy period in \Cref{lem:bound_busy_period_1}. Using the coarse bound on the busy period length in \Cref{lem:bound_busy_period_1} and the negative drift in \Cref{lem:p_wt}, we will provide a tighter bound on the queue length in \Cref{lem:tighter_queue_length}. Finally, using \Cref{lem:tighter_queue_length} and \Cref{lem:p_wt}, we will prove \Cref{lem:final_busy_period}.

\Cref{lem:bound_q_length} is given below which proves a coarse all time bound on the queue length.
\begin{lem}\label{lm_queue_upper}
    \label{lem:bound_q_length}
    For any $t\ge t_0$ with $t_0$ defined in (\ref{condition_t0:arrival_condition_on_probab}--\ref{condition_t0:bound_vi(t)}) and $w(t)$ defined in \eqref{eq:w(t)} and the event $\eventtwo$ defined as 
    \begin{equation}\label{event2}
        \eventtwo := \overset{\numofserver}{\underset{i = 1}{\cap}}\overset{t}{\underset{\tau =1}{\cap}}\left\{\queue_i(\tau)\leq 2w(t)\right\},
    \end{equation}
    it holds that
    \begin{equation}
        \probability\left((\eventtwo)^c\right)\leq \numofserver\left( \frac{1}{t^4} + \frac{1}{t^{7}} + \frac{k_2 + 1}{t^3}\right).
    \end{equation}
\end{lem}
\begin{proof}\label{prf_lem:bound_q_length}
    Now, using the law of total probability and the union bound, we have
    \begin{equation}\label{rel:bound_q}
        \probability\left((\eventtwo)^c\right)  \leq \sum_i \sum_{\tau = 1}^{t} \probability\left(\left\{\queue_i(\tau)\geq 2w(t)\right\}\cap \eventone\right) + \probability\left((\eventone)^c\right).
    \end{equation}
    Hence, to bound the probability of $(\eventtwo)^c$, it is sufficient to bound $\probability\left(\left\{\queue_i(\tau)\geq 2w(t)\right\}\cap \eventone\right)$ for all $i$ and for any $\tau$. Clearly, the probability is zero for any $\tau \leq 2w(t)$
    Now, consider any $\tau > 2w(t)$. Define 
    \begin{equation}
        l_1 = \max\{\max\left\{\ell \leq \tau :\queue_i(\ell) = 0 \right\}, w(t)\}
    \end{equation}
    \begin{equation}
        l_2 = \max\left\{\ell \leq \tau - 1: \arrival_i(\ell) - \departure_i(\ell) \neq 0\right\}
    \end{equation}
    From the definition of $l_1, l_2$, $\queue_i(\ell)$ has to be positive in the duration $\ell \in [l_1 + 1, l_2]$. Now $\queue_i(\tau) \geq 2w(t)$ implies 
    \begin{equation}
        \queue_i(\tau) = \queue_i(l_1) + \sum_{\ell=l_1}^{l_2}(\arrival_i(\ell) - \service_i(\ell)) \geq 2w(t).
    \end{equation}
    If $l_1 > w(t)$, then $\queue_i(l_1) = 0$, else $\queue_i(l_1) = \queue_i(w(t)) \leq w(t)$. This implies 
    \begin{equation}
        \sum_{\ell=l_1}^{l_2}(\arrival_i(\ell) - \service_i(\ell)) \geq w(t).
    \end{equation}
    The above argument implies that if $\queue_i(\tau) \geq 2w(t)$, then there must exists some $l_1', l_2'$ such that $\sum_{\ell =l_1'}^{l_2'} \left( \arrival_i(\ell) - \service_i(\ell)\right) \geq w(t)$. This implies that the probability of the event $\left\{\left\{\queue_i(\tau) \geq 2w(t)\right\}\cap \eventone\right\}$ can be upper bounded by the probability of union of the the events $\left\{\left\{\sum_{\ell =l_1'}^{l_2'} \left( \arrival_i(\ell) - \service_i(\ell)\right) \geq w(t)\right\}\cap \eventone\right\}$, where the union is taken over all combination of $l_1', l_2'$. Hence, 
    \begin{align}
        \probability\left(\left\{\queue_i(\tau)\geq 2w(t)\right\}\cap \eventone\right) &\leq 
        \sum_{l_1', l_2'}\probability\left(\left\{\sum_{\ell=l_1'}^{l_2'} \left(  \service_i(\ell) - \arrival_i(\ell) \right)\leq  - w(t)\right\}\cap \eventone\right) \\
        &\leq \sum_{l_1', l_2'}\exp \left(-\frac{\left((l_2' - l_1' + 1)\frac{\freecapacity_i}{2} + w(t)\right)^2}{2(l_2' - l_1' + 1)}\right)\label{prfeqn:alltime_length2}\\
        &\leq \sum_{l_1', l_2'}\exp \left(-\frac{\freecapacity_i w(t)}{2}\right)\label{prfeqn:alltime_length3}\\
        &\leq \sum_{l_1', l_2'}\frac{1}{t^7}\label{prfeqn:alltime_length4}\\
        &\leq \frac{1}{t^5}
    \end{align}
    where,
    \begin{itemize}
        \item \eqref{prfeqn:alltime_length2} uses \Cref{lem:neg_drift}.
        \item \eqref{prfeqn:alltime_length3} uses the property that an arithmetic mean is greater than its corresponding geometric mean.
        \item \eqref{prfeqn:alltime_length4} uses \eqref{condition_t0:lemma_queue_length_bound} that ensures that for any $t\geq t_0$, $\min_i \freecapacity_{i} w(t) \geq 48 \ln t \geq 14 \ln t$.
    \end{itemize}
    Using \eqref{rel:bound_q} and \Cref{lem:p_wt} and taking an union bound over all $\tau$ and for all $i$, we get,
    \begin{equation}
        \probability\left((\eventtwo)^c\right) \leq \sum_i\sum_\tau \probability\left(\left\{\queue_i(\tau)\geq 2w(t)\right\}\cap \eventone\right) + \probability\left((\eventone)^c\right) \leq \numofserver \left( \frac{1}{t^4} + \frac{1}{t^{7}} + \frac{k_2 + 1}{t^3}\right).
    \end{equation}
\end{proof} 
Using the preceding lemma, we will now prove a coarse high probability bound on the busy period length.
\begin{lem}
    \label{lem:bound_busy_period_1}
    For any $t\ge t_0$ with $t_0$ defined in (\ref{condition_t0:arrival_condition_on_probab}--\ref{condition_t0:bound_vi(t)}) and $w(t)$ defined in \eqref{eq:w(t)} and  $\eventthree$ defined as
    \begin{equation}\label{event3}
        \eventthree := \overset{\numofserver}{\underset{i = 1}{\cap}}\left\{\busy_i(t - v_i(t)) \leq \frac{6w(t)}{\freecapacity_{i}}\right\},
    \end{equation}
    it holds that
    \begin{equation}
        \probability\left((\eventthree)^c\right)\leq \numofserver\left(\frac{2}{t^{7}} + \frac{2k_2 + 2}{t^3} + \frac{2}{t^4}\right).
    \end{equation}
\end{lem}
\begin{proof} \label{prf_lem:bound_busy_period_1}
    The main idea behind the lemma is that since the queue length at each server is bounded for all time and each server is experiencing a negative drift, the length of the busy period cannot be too large. Using law of total probability, we have
    \begin{equation}\label{rel:main_eqn:eventthree}
        \probability\left((\eventthree)^c\right) \leq \sum_i \probability\left(\left\{\busy_i(t - v_i(t)) \geq \frac{6w(t)}{\freecapacity_{i}}\right\}\cap \eventone\cap\eventtwo\right) + \probability\left((\eventone)^c\right) + \probability\left((\eventtwo)^c\right)
    \end{equation}
    Recall,
    \begin{equation*}
        \eventtwo := \overset{\numofserver}{\underset{i =1}{\cap}}\overset{t}{\underset{\tau =1}{\cap}}\left\{\queue_i(\tau)\leq 2w(t)\right\}
    \end{equation*}
    From its definition, $\eventtwo$ implies that $Q_i(t - v_i(t) - \frac{6w(t)}{\freecapacity_{i}}) \leq 2w(t)$. Now if $\busy_i(t - v_i(t)) \geq \frac{6w(t)}{\freecapacity_{i}}$, then $\eventtwo$ implies that $\queue_i(\ell) > 0$ for \\
    all $\ell \in \left[t - v_i(t) - \frac{6w(t)}{\freecapacity_{i}}, t - v_i(t)\right]$ and
    \begin{equation}
        2w(t) - \sum_{\ell=t_1}^{t_2 - 1}\left(\service_i(\ell) - \arrival_i(\ell)\right) \geq Q_i\left(t_1\right) - \sum_{\ell=t_1}^{t_2 - 1}\left(\service_i(\ell) - \arrival_i(\ell)\right) = Q_i(t_2) \geq 0,
    \end{equation}
    where $t_1 = t - v_i(t) - \frac{6w(t)}{\freecapacity_{i}}$ and $t_2 = t - v_i(t)$. Therefore
    \begin{equation}
        \left\{\busy_i(t_2) \geq \frac{6w(t)}{\freecapacity_{i}}\right\}\cap\left\{ \queue_i(t_1) \leq 2w(t)\right\}\subseteq \left\{\sum_{\ell=t_1}^{t_2 - 1} \left( \service_i(\ell) - \arrival_i(\ell)\right) \leq 2w(t)\right\}.
    \end{equation}
    Hence, 
    \begin{align}
        \probability\left(\left\{\busy_i(t - v_i(t)) \geq \frac{6w(t)}{\freecapacity_i}\right\}\cap \eventone\cap \eventtwo\right) &\leq \probability \left( \left\{\sum_{\ell=t_1}^{t_2 - 1}\left(\service_i(\ell) - \arrival_i(\ell)\right) \leq 2w(t)\right\}\cap \eventone\right)\label{rel_pref:eventthree:1}\\
        &\leq \exp \left(-\frac{\left(2w(t) - 3w(t) \right)^2}{\frac{12w(t)}{\freecapacity_i}}\right)\label{rel_pref:eventthree:2}\\
        &= \exp \left(-\frac{\freecapacity_i w(t)}{12}\right)\\
        &\leq \exp\left(-4\ln t\right)\label{rel_pref:eventthree:3}\\
        &= \frac{1}{t^4}\label{rel_pref:eventthree:4}
    \end{align}
    where,
    \begin{itemize}
        \item \eqref{rel_pref:eventthree:2} uses \Cref{lem:neg_drift}.
        \item \eqref{rel_pref:eventthree:3} uses  \eqref{condition_t0:lemma_queue_length_bound} that ensures that for any $t\geq t_0$, $\min_i \freecapacity_{i} w(t) \geq 48 \ln t$.
    \end{itemize}
    Using \eqref{rel_pref:eventthree:4} with \eqref{rel:main_eqn:eventthree} and substituting results from \Cref{lem:p_wt} and \Cref{lem:bound_q_length}, we have
    \begin{equation}
        \probability\left((\eventthree)^c\right) \leq  \numofserver\left(\frac{2}{t^{7}} + \frac{2k_2 + 2}{t^3} + \frac{2}{t^4}\right)
    \end{equation}
\end{proof}
Using the preceding lemma, we will now prove a tight high probability bound on the length of the queue.
\begin{lem}
    \label{lem:tighter_queue_length} 
    For any $t\ge t_0$ with $t_0$ defined in (\ref{condition_t0:arrival_condition_on_probab}--\ref{condition_t0:bound_vi(t)}) and $\eventfour$ defined as 
    \begin{equation}\label{event4}
        \eventfour := \overset{\numofserver}{\underset{i = 1}{\cap}}\left\{Q_i(t - v_i(t)) \leq \frac{10 \ln t}{\freecapacity_i}\right\}.
    \end{equation}
    it holds that
    \begin{equation}
    \probability\left((\eventfour)^c\right)\leq \numofserver\left(\frac{3}{t^{7}} + \frac{3k_2 + 3}{t^3} + \frac{3}{t^4}\right).
    \end{equation}
\end{lem}
\begin{proof}\label{prf_lem:tighter_queue_length}
    Recall that $\eventthree$ given by 
    \begin{equation}
        \eventthree = \overset{\numofserver}{\underset{i = 1}{\cap}}\left\{\busy_i(t - v_i(t)) \leq \frac{6w(t)}{\freecapacity_{i}}\right\},
    \end{equation}
    is a highly probable event. The main idea behind the proof is that since the $\busy_i(t - v_i(t))$ is bounded with high probability, the queue length at time $t - v_i(t)$ cannot be too large. Define $r_i(t) = t - v_i(t)$. Using the equation, 
    \begin{equation}
        \queue_i(r_i(t)) = \sum_{\ell = r_i(t) - \busy_i(r_i(t))}^{r_i(t) - 1} \arrival_i(\ell) - \service_i(\ell)
    \end{equation}
    we have,
    \begin{equation}\label{rel:simplifyfourproof}
        \probability\left((\eventfour)^c\right) \leq \sum_i\probability\left(\left\{\sum_{\ell = r_i(t) - \busy_i(r_i(t))}^{r_i(t) - 1} \left(\service_i(\ell) - \arrival_i(\ell)\right) \leq -\frac{10  \ln t}{\freecapacity_i}\right\}\cap \eventone\cap \eventthree\right) + \probability\left((\eventone)^c\right) + \probability\left((\eventthree)^c\right).
    \end{equation}
    Now, 
    \begingroup
    \allowdisplaybreaks
    \begin{align}
        &\probability\left(\left\{\sum_{\ell = r_i(t) - \busy_i(r_i(t))}^{r_i(t) - 1} \left(\service_i(\ell) - \arrival_i(\ell)\right) \leq -\frac{10  \ln t}{\freecapacity_i}\right\}\cap \eventone\cap \eventthree\right) \\
        =&\probability\left(\left\{\sum_{\ell = r_i(t) - \busy_i(r_i(t))}^{r_i(t) - 1} \left(\service_i(\ell) - \arrival_i(\ell)\right) \leq -\frac{10  \ln t}{\freecapacity_i}\right\}\cap \eventone\cap \eventthree\cap\left\{\cup_l\left\{\busy_i(r_i(t)) = b\right\}\right\}\right) \\
        \leq&\sum_{b}\probability\left(\left\{\sum_{\ell = r_i(t) - \busy_i(r_i(t))}^{r_i(t) - 1} \left(\service_i(\ell) - \arrival_i(\ell)\right) \leq -\frac{10  \ln t}{\freecapacity_i}\right\}\cap \eventone\cap \eventthree\cap\left\{\busy_i(r_i(t)) = b\right\}\right) \\
        =&\sum_{b}\probability\left(\left\{\sum_{\ell = r_i(t) - b}^{r_i(t) - 1} \left(\service_i(\ell) - \arrival_i(\ell)\right) \leq -\frac{10  \ln t}{\freecapacity_i}\right\}\cap \eventone\cap \eventthree\right) \\
        \leq &\sum_{b:b \leq \frac{6w(t)}{\freecapacity_i}}\probability\left(\left\{\sum_{\ell = r_i(t) - b}^{r_i(t) - 1} \left(\service_i(b) - \arrival_i(\ell)\right) \leq -\frac{10 \ln t}{\freecapacity_i}\right\}\cap \eventone\right)\label{rel_prf:eventfour:1}\\
        \leq &\sum_{b:b \leq \frac{6w(t)}{\freecapacity_i}}\exp\left(-\frac{\left(\frac{10 \ln t}{\freecapacity_i} + \frac{b \freecapacity_i}{2}\right)^2}{2b}\right)\label{rel_prf:eventfour:2} \\
        \leq &\sum_{b:b \leq \frac{6w(t)}{\freecapacity_i}}\exp\left(-5 \ln t\right)\label{rel_prf:eventfour:3}\\
        \leq &\frac{1}{t^4}
    \end{align}
    \endgroup
    where, 
    \begin{itemize}
        \item \eqref{rel_prf:eventfour:1} uses the fact that $\eventthree$ implies that $l \leq \frac{6w(t)}{\freecapacity_i}$.
        \item \eqref{rel_prf:eventfour:2} uses \Cref{lem:neg_drift}.
        \item \eqref{rel_prf:eventfour:3} uses the property that an arithmetic mean is greater than its corresponding geometric mean.
    \end{itemize}
    Taking an union bound over all $i$ and using \eqref{rel:simplifyfourproof} with \Cref{lem:p_wt} and \Cref{lem:bound_busy_period_1} completes the proof.
\end{proof}


Finally, using \Cref{lem:tighter_queue_length} we will prove \Cref{lem:final_busy_period}.
\begin{proof}[Proof of Lemma~\ref{lem:final_busy_period}]
Using the law of total probability and the union bound, we have
    \begin{equation}\label{rel:final_busy_period}
        \probability\left((\eventfive)^c\right) \leq \sum_i\probability\left(\left\{\busy_i(t) \geq v_i(t)\right\}\cap \eventone \cap \eventfour\right) +  \probability\left((\eventone)^c\right) +   \probability\left((\eventfour)^c\right)
    \end{equation}
    Recall that, 
    \begin{equation}
        \eventfour := \overset{\numofserver}{\underset{i = 1}{\cap}}\left\{Q_i(t - v_i(t)) \leq \frac{10 \ln t}{\freecapacity_i}\right\}.
    \end{equation} 
  $\eventfour$ implies that $Q_i\left(t - v_i(t)\right) \leq \frac{10 \ln t}{\freecapacity_i}$. Similar to arguments used in \Cref{lem:bound_busy_period_1}, we can show that if $\busy_i(t) \geq v_i(t)$, then $\queue_i(\ell) > 0$ for all $\ell \in \left[t - v_i(t), t\right]$ and
    \begin{equation}
        \frac{10 \ln t}{\freecapacity_i} - \sum_{\ell=t - v_i(t)}^{}\left(\service_i(\ell) - \arrival_i(\ell)\right) \geq Q_i\left(t - v_i(t)\right) - \sum_{\ell=t - v_i(t)}^{t}\left(\service_i(\ell) - \arrival_i(\ell)\right) = Q_i(t) \geq 0.
    \end{equation}
    Hence,
    \begin{equation}
        \eventfour \cap \left\{\busy_i(t)) \geq v_i(t)\right\} \subseteq \left\{\sum_{\ell=t - v_i(t)}^{t}\left(\service_i(\ell) - \arrival_i(\ell)\right) \leq  \frac{10 \ln t}{\freecapacity_i}\right\}
    \end{equation}
    Therefore, for any $i$,
    \begin{align}
        \probability\left(\left\{\busy_i(t) \geq v_i(t)\right\}\cap\eventone\cap\eventfour\right) &\leq \probability\left(\left\{\sum_{\ell = t - v_i(t)}^{t - 1} \left(\service_i(\ell) - \arrival_i(\ell)\right) \leq \frac{10\ln t}{\freecapacity_i}\right\}\cap \eventone \right)\label{rel_prf:eventfive:1} \\
        &\leq \exp\left(-\frac{\left(\frac{10\ln t}{\freecapacity_i} - \frac{66\ln t}{2\freecapacity_i}\right)^2 }{\frac{132\ln t}{\freecapacity_i^2}}\right)\label{rel_prf:eventfive:2}\\
        &\leq \frac{1}{t^4}
    \end{align}
    where \eqref{rel_prf:eventfive:2} uses \Cref{lem:neg_drift}. Finally, taking an union bound over all $i$ and using the results from \Cref{lem:tighter_queue_length} and \Cref{lem:p_wt} in \eqref{rel:final_busy_period} completes the proof.
    \end{proof}
\subsection{Proof of Lemma \ref{lem:tight_bound_p}}\label{prf_lem:tight_bound_p}
Restating the \Cref{lem:tight_bound_p} 

\newtheorem{innercustomlemma}{Lemma}
\newenvironment{customlemma}[1]
  {\renewcommand\theinnercustomlemma{#1}\innercustomlemma}
  {\endinnercustomlemma}

\begin{customlemma}{4}[Estimation error bound]
There exist a constant $k_2$ and a $t_0$ such that for any $t\ge t_0$, the event $\eventsix$ defined as 
\begin{equation}
\eventsix := \left\{\left| \hat{p}_i(\tau) - p_i^*\right| \leq k_1\min\left\{\sqrt{\frac{ \ln t}{t}}, p_i^*\right\},\forall \tau \in \left[\frac{t}{2} + 1, t\right],\forall i\right\},
\end{equation}
satisfies  
\begin{equation}
\probability\left(\left(\eventsix\right)^c\right) \leq \numofserver\left(\frac{1}{t^7} + \frac{k_2 + 1}{t^3}\right) + \sum_{i: p_i^*>0} \left(\frac{1}{t^3} + t \exp\left(-\frac{p_i^*\lambda t}{128}\right) +
t\exp\left(-\frac{\freecapacity_i^2 t}{4}\right)\right).
\end{equation}
\end{customlemma}
The proof idea of this lemma is similar to \Cref{lem:p_wt}. We will argue using \Cref{lem:p_wt} that there is $\order(t)$ number of samples to each of the servers in the support set which ensures with high probability that the estimated routing probabilities are within $\order(\sqrt{\ln t / t})$ ball of the true routing probabilities. Using the law of total probability, we have
\begin{align}
    \probability\left((\eventsix)^c\right) &= \probability\left(\overset{\numofserver}{\underset{i = 1}{\cup}}\overset{t}{\underset{\tau = \frac{t}{2} + 1}{\cup}}\left\{\left| \hat{p}_i(\tau) - p_i^*\right| > k_1\min\left\{\sqrt{\frac{ \ln t}{t}}, p_i^*\right\}\right\}\right)\\
    &\leq  \probability\left(\left\{\overset{\numofserver}{\underset{i = 1}{\cup}}\overset{t}{\underset{\tau = \frac{t}{2} + 1}{\cup}}\left\{\left| \hat{p}_i(\tau) - p_i^*\right| > k_1\min\left\{\sqrt{\frac{ \ln t}{t}}, p_i^*\right\}\right\}\right\}\cap \eventone\right) + \probability((\eventone)^c)\label{eq:tightbound_p:simplify1:1}.
\end{align}
Now, $\eventone$ implies that for all $\tau \geq w(t)$, $\support(\lambda, \hat{\bm{\mu}}(t))$ is the same $\support(\lambda, \bm{\mu})$ which implies for any $\tau \geq w(t)$ and for any $i$ such that $i \notin \support(\lambda, \bm{\mu})$, $\hat{p}_i(\tau) = p_i^* = 0$. Hence, for any $i \notin \support(\lambda, \bm{\mu})$ and for any $\tau \geq w(t)$, 
\begin{equation}\label{eq:use_supportset_in tight bound_p}
    \left\{\left| \hat{p}_i(\tau) - p_i^*\right| > k_1\min\left\{\sqrt{\frac{ \ln t}{t}}, p_i^*\right\}\right\} \cap \eventone = \phi,
\end{equation}
where $\phi$ is a zero probability event. Using \eqref{eq:tightbound_p:simplify1:1} with \eqref{eq:use_supportset_in tight bound_p}, we have
\begin{align}
    \probability\left((\eventsix)^c\right) &\leq  \probability\left(\left\{\underset{i:p_i^*>0}{\cup}\overset{t}{\underset{\tau = \frac{t}{2} + 1}{\cup}}\left\{\left| \hat{p}_i(\tau) - p_i^*\right| > k_1\min\left\{\sqrt{\frac{ \ln t}{t}}, p_i^*\right\}\right\}\right\}\cap \eventone\right) + \probability((\eventone)^c)\\
    &\leq \probability\left(\left\{\underset{i:p_i^*>0}{\cup}\overset{t}{\underset{\tau = \frac{t}{2} + 1}{\cup}}\left\{\left| \hat{p}_i(\tau) - p_i^*\right| > k_1\min\left\{\sqrt{\frac{ \ln t}{t}}, p_i^*\right\}\right\}\right\} \cap \left\{\underset{i:p_i^*>0}{\cap}\overset{t}{\underset{\tau = \frac{t}{2} + 1}{\cap}}\left\{\left| \hat{\mu}_i(\tau) - \mu_i\right| \leq k_{11}\min\left\{\sqrt{\frac{ \ln t}{t}}, p_i^*\right\}\right\}\right\}\cap \eventone\right) + \nonumber\\
    &\qquad + \probability\left(\left\{\overset{\numofserver}{\underset{i = 1}{\cup}}\overset{t}{\underset{\tau = \frac{t}{2} + 1}{\cup}}\left\{\left| \hat{\mu}_i(\tau) - \mu_i\right| > k_{11}\min\left\{\sqrt{\frac{ \ln t}{t}}, p_i^*\right\}\right\}\right\}\cap \eventone\right) + \probability((\eventone)^c)\label{eq:tightbound_p:simplify2:2},
\end{align}
where $k_{11} = k_1/c$. \eqref{condition_t0:lemma_estimation_error} ensures that for any $t\geq t_0$, $k_{11}\sqrt{\frac{ \ln t}{t}} \leq \min_{i:p_i^*> 0} p_i$. Using \Cref{lem:relation_p_mu}, we also have 
\begin{equation}\label{eq:use_supportset_in tight bound_p2}
    \left\{\underset{i:p_i^*>0}{\cup}\left\{\left| \hat{p}_i(\tau) - p_i^*\right| > k_1\min\left\{\sqrt{\frac{ \ln t}{t}}, p_i^*\right\}\right\}\right\} \cap \left\{\underset{i:p_i^*>0}{\cap}\left\{\left| \hat{\mu}_i(\tau) - \mu_i\right| \leq k_{11}\min\left\{\sqrt{\frac{ \ln t}{t}}, p_i^*\right\}\right\}\right\} = \phi.
\end{equation}
Using \eqref{eq:tightbound_p:simplify2:2} with \eqref{eq:use_supportset_in tight bound_p2} along with union bound, we have
\begin{equation}\label{eq:tightbound_p:main_relation}
    \probability\left((\eventsix)^c\right) \leq \sum_{i:p_i^* >0}\sum_\tau\probability\left(\left\{\left|\hat{\mu}_i(\tau) - \mu_i\right| \geq k_{11}\sqrt{\frac{\ln \tau}{\tau}}\right\}\cap \eventone\right) + \probability\left((\eventone)^c\right).
\end{equation}
For any $\tau \in [t/2 + 1, t]$ and any $n_0(t)$, we have 
\begingroup
\allowdisplaybreaks
\begin{align}
    &\probability\left(\left\{\left|\hat{\mu}_i(\tau) - \mu_i\right|> k_{11}\sqrt{\frac{\ln t}{t}}\right\}\cap \eventone \right) \\
    \leq &  \probability\left(\left\{\sum_{\ell=1}^{\tau - 1} \departure_i(\ell) < n_0(t)\right\}\cap \eventone\right) + \sum_{n=n_0(t)}^{t}\probability\left(\left\{\left|\hat{\mu}_i(\tau) - \mu_i\right|> k_{11}\sqrt{\frac{\ln t}{t}}\right\}\cap\left\{\departure_i(\tau) = n_0(t)\right\}\cap \eventone \right)\\
    \leq &  \probability\left(\left\{\sum_{\ell=1}^{\tau - 1} \departure_i(\ell) < n_0(t)\right\}\cap \eventone\right) + \sum_{n=n_0(t)}^{t}\probability\left(\left\{\left|\hat{\mu}_i^{(n_0(t))} - \mu_i\right|> k_{11}\sqrt{\frac{\ln t}{t}}\right\}\cap \eventone \right)\label{eq:tightbound_p:simplify3:1}.
\end{align}
The second term in \eqref{eq:tightbound_p:simplify3:1} can be simplified using \Cref{lem:conc_geoemtric} and \eqref{condition_t0:lemma_estimation_error} that ensures that for any $t\geq t_0$, $k_{11}\sqrt{\ln t/t} \leq \Delta$. Hence, \eqref{eq:tightbound_p:simplify3:1} can be further bounded as
\begin{align}
    &\probability\left(\left\{\left|\hat{\mu}_i(\tau) - \mu_i\right|> k_{11}\sqrt{\frac{\ln t}{t}}\right\}\cap \eventone \right)\\
    \leq& \probability\left( \left\{\sum_{\ell=1}^{\tau - 1}\departure_i(\ell) < n_0(t)\right\} \cap \left\{\sum_{\ell=1}^{\frac{3t}{8}}\arrival_i(\ell) \geq n_0(t)\right\}\cap  \eventone\right) + \probability \left(\left\{\sum_{\ell=1}^{\frac{3t}{8}} \arrival_i(\ell) < n_0(t)\right\}\cap \eventone\right) + \nonumber\\
    &\qquad + \sum_{n=n_0(t)}^{t}\exp\left( -\frac{k_{11}^2 n_0(t) c_g \ln t}{t}\right)\\
    \leq& \probability\left( \left\{\sum_{\ell=1}^{\tau - 1}\departure_i(\ell) < n_0(t)\right\} \cap \left\{\sum_{\ell=1}^{\frac{3t}{8}}\arrival_i(\ell) \geq n_0(t)\right\}\right) + \probability \left(\left\{\sum_{\ell=1}^{\frac{3t}{8}} \arrival_i(\ell) < n_0(t)\right\}\cap \eventone\right) + \nonumber\\
    &\qquad  + t\exp\left( -\frac{k_{11}^2 n_0(t) c_g \ln t}{t}\right)\\
    \leq& \probability\left( \left\{\sum_{\ell=1}^{\tau - 1}\departure_i(\ell) < n_0(t)\right\} \cap \left\{\sum_{\ell=1}^{\frac{3t}{8}}\arrival_i(\ell) \geq n_0(t)\right\}\cap\left\{ \sum_{\ell=\frac{3t}{8} + 1}^{\tau - 1}\service_i(\ell) < n_0(t)\right\}\right) + \probability \left(\left\{\sum_{\ell=1}^{\frac{3t}{8}} \arrival_i(\ell) < n_0(t)\right\}\cap \eventone\right) + \nonumber\\
    &\qquad + \probability\left( \left\{\sum_{\ell=1}^{\tau - 1}\departure_i(\ell) < n_0(t)\right\} \cap \left\{\sum_{\ell=1}^{\frac{3t}{8}}\arrival_i(\ell) \geq n_0(t)\right\}\cap\left\{ \sum_{\ell=\frac{3t}{8} + 1}^{\tau - 1}\service_i(\ell) \geq n_0(t)\right\}\right) +  t\exp\left( -\frac{k_{11}^2 n_0(t) c_g \ln t}{t}\right)\label{eq:tightbound_p:simplify3:2}.
\end{align}
Now, 
\begin{equation}\label{eq:tightbound_p:simplify3:3}
    \probability\left( \left\{\sum_{\ell=1}^{\tau - 1}\departure_i(\ell) < n_0(t)\right\} \cap \left\{\sum_{\ell=1}^{\frac{3t}{8}}\arrival_i(\ell) \geq n_0(t)\right\}\cap\left\{ \sum_{\ell=\frac{3t}{8} + 1}^{\tau - 1}\service_i(\ell) \geq n_0(t)\right\}\right) = 0
\end{equation}
because, if the server $i$ has $n_0(t)$ arrival till time $3t/8$ and total offered service in time $(3t/8 + 1)$ to $\tau$ exceed $n_0(t)$, then the total number of departures till time $\tau$ should be at least $n_0(t)$. Hence, using \eqref{eq:tightbound_p:simplify3:2} and \eqref{eq:tightbound_p:simplify3:3} we have
\begin{align}
    \probability\left(\left\{\left|\hat{\mu}_i(\tau) - \mu_i\right|> k_{11}\sqrt{\frac{\ln t}{t}}\right\}\cap \eventone \right) & \leq \probability\left( \sum_{\ell=\frac{3t}{8} + 1}^{\tau - 1}\service_i(\ell) < n_0(t) \right) + \probability \left(\left\{\sum_{\ell=1}^{\frac{3t}{8}} \arrival_i(\ell) < n_0(t)\right\}\cap \eventone\right) + t\exp\left( -\frac{k_{11}^2 n_0(t) c_g \ln t}{t} \right)\\
    & \leq \underbrace{\probability\left( \sum_{\ell=\frac{3t}{8} + 1}^{t/2}\service_i(\ell) < n_0(t) \right)}_{T_1} + \underbrace{\probability \left(\left\{\sum_{\ell=\frac{t}{4} + 1}^{\frac{3t}{8}} \arrival_i(\ell) < n_0(t)\right\}\cap \eventone\right)}_{T_2} + \underbrace{t\exp\left( -\frac{k_{11}^2 n_0(t) c_g \ln t}{t} \right)}_{T_3}.\label{rel:departure_arrival_service2:1}
\end{align}
\endgroup
Choosing $n_0(t) = \frac{p_i^* \lambda t}{32}$ and $k_{11} =\max_{i:p_i^* > 0} \sqrt{\frac{160}{c_g p_i^* \lambda}}$, bounds $T_3$ by
\begin{equation}
    t\exp\left( -\frac{k_{11}^2 n_0(t) c_g \ln t}{t} \right) \leq \frac{1}{t^4}
\end{equation}
Again, using the \eqref{condition_t0:bound_w(t)} that for any $t\geq t_0$, $w(t) \leq t/4$ for any $t \geq t_0$ and \Cref{lem:neg_drift}, we can bound $T_2$ by,
\begin{equation}
    \probability \left(\left\{\sum_{\ell=\frac{t}{4} + 1}^{\frac{3t}{8}} \arrival_i(\ell) < \frac{p_i^* \lambda t}{32}\right\}\cap \eventone\right) \leq \exp\left(-\frac{p_i^*\lambda t}{128}\right).
\end{equation}
 Similarly bounding $T_1$, we have
\begin{align}
    \probability\left(\sum_{\ell=\frac{3t}{8} + 1}^{t/2}\service_i(\ell) < \frac{p_i^* \lambda t}{32}\right) &\leq 
    \probability\left(\sum_{\ell=\frac{3t}{8} + 1}^{t/2}\service_i(\ell) < \frac{p_i^* \lambda t}{8}\right)\\
    &\leq \exp \left(-\frac{2\left(\frac{\freecapacity_it}{8}\right)^2} {\frac{t}{8}}\right)\label{rel:eventone:bound_service:1}\\
    &\leq \exp\left(-\frac{\freecapacity_i^2 t}{4}\right).
\end{align}
where \eqref{rel:eventone:bound_service:1} follows from Hoeffding's inequality given in \eqref{ineq:Hoeffding}.
Taking summation over all $\tau \in \left(\frac{t}{2} + 1, t\right)$ in \eqref{eq:tightbound_p:main_relation} and using \Cref{lem:p_wt}, we get
\begin{equation}
    \probability\left((\eventsix)^c\right)  \leq \numofserver\left(\frac{1}{t^7} + \frac{k_2 + 1}{t^3}\right) + \sum_{i: p_i^*>0}\left(\frac{1}{t^3} + t \exp\left(-\frac{p_i^*\lambda t}{128}\right) + t\exp\left(-\frac{\freecapacity_i^2 t}{4}\right)\right).
\end{equation}